%% file: Main_thesis.tex
\documentclass[PhD,binding=0.6cm]{sapthesis}

\usepackage{microtype}
\usepackage{bm}

\usepackage{hyperref}
\hypersetup{colorlinks=true,
	linkcolor=blue,
	anchorcolor=blue,
	citecolor=blue,
	urlcolor=blue,
	pdftitle={Sapthesis class example},
	pdfauthor={Andrea Cipollone}}

\usepackage{lipsum}
\usepackage{slashed}
\usepackage{curve2e}
\definecolor{gray}{gray}{0.4}

\newcommand{\be}{\begin{equation}}
\newcommand{\ee}{\end{equation}}
\newcommand{\bal}{\begin{align}}
\newcommand{\eal}{\end{align}}
\newcommand{\bea}{\begin{eqnarray}}
\newcommand{\eea}{\end{eqnarray}}
\newcommand{\rmd}{{\rm d}}

\def\g{\gamma}

\def\th{\theta}

\def\tt{\hat \tau}
\def\n{\nu}

\def\s{\sigma}

\def\O{\Omega}
\def\nn{\nonumber\\}

\def\up{\uparrow}

\def\down{\downarrow}
\def\ran{\rangle}
\def\lan{\langle}

\def\pd{\partial}

\def\bk{{\bf k}}

\def\bq{{\bf q}}
\def\br{{\bf r}}
\def\bp{{\bf p}}

\def\nn{\nonumber}

\def\sa{( \bm{\sigma}_1\cdot \hat{{\bf r}})}
\def\sb{( \bm{\sigma}_2\cdot \hat{{\bf r}})}
\def\st{(\bm{\sigma}_1\cdot  \bm{\sigma}_2)}
\def\tt{(\bm{\tau}_1\cdot  \bm{\tau}_2)}
\def\sab{3\sa\sb-\st}
\def\rh{ \hat{{\bf r}}}
\def\rr{\rangle}
\def\ll{\langle}

\def\<{\langle}
\def\>{\rangle}

\title{Neutrino Interactions in Neutron Matter}
\author{Andrea Cipollone}
\IDnumber{696555}
\course[Physics]{Fisica} 
\courseorganizer{Scuola di dottorato in Fisica "Vito Volterra"}
\cycle{XXV}
\submitdate{Sept 2012}
\copyyear{2012}
\advisor{Prof. Omar Benhar}
\authoremail{and.cipollone@gmail.com}

\begin{document}

\frontmatter

\maketitle




\tableofcontents







\mainmatter

\chapter*{Introduction}
\addcontentsline{toc}{chapter}{Introduction}
\input{Chap_Intro}


\chapter{Overview of nuclear many-body theory}\label{NMB}

\input{Chap_NMBT}

\chapter{Correlated basis functions formalism}\label{CB}

\input{Chap_CBF}


\chapter{Landau theory of normal Fermi liquids}\label{lab}
\input{Chap_Landau}

\chapter{Calculation of Landau parameters}\label{val}
\input{Chap_Value}

\chapter{Weak response of neutron matter}\label{Wea}
\input{Chap_Weak}

\chapter{Neutrino mean free path}\label{mf}
\input{Chap_Neutrino}

\chapter{Three-nucleon forces}\label{TBF}
\input{Chap_TBF}

\chapter*{Summary \& Outlook}\label{SO}
\addcontentsline{toc}{chapter}{Summary \& Outlook}
\input{Chap_SO}




\appendix

\chapter{Properties of the operators $O^n$}\label{On}
\input{Appendix1}
\chapter{Energy functional}\label{EnFun}
\input{appendix2}
\chapter{Euler-Lagrange equations}\label{EuLag}
\input{appendix3}

\chapter{Tensor operator in momentum space}\label{FourierTr}
\input{App_ex}

\backmatter
 \bibliography
\cleardoublepage
\phantomsection
\bibliographystyle{sapthesis} 


\end{document}

%% file: Chap_Intro.tex
Almost sixty years past the discovery of the neutrino \cite{Cowan}, the study of its physics is still one of the outstanding and  most exciting challenges to the scientific community. Thanks to the 
striking progress of particle physics in the last decades, the neutrino interaction mechanisms \emph{in vacuum} are now well-established. On the other hand, little is known about its nature (whether 
it is a Dirac or Majorana particle), its mass or the existence of sterile neutrino.
 Many constraints on its microscopic features, most notably the mass,  can be 
inferred from astrophysical data. Recently, neutrinos have been also used as astronomical probes, as they can carry unaltered information over cosmological distances \cite{Bigongiari}. 

Although several questions about neutrinos have been answered in the past few years, a number of issues are still obscure. In this context, it is worth mentioning the 
over one hundred papers that appeared within a couple of days after the abrupt announcement of the observation of superluminal neutrinos. 
This claim was later withdrawn, but in those days many new intriguing models and theoretical tools were developed.  
The neutrino \emph{"gold-rush"} seems to be never ending. 

The behavior of neutrinos in matter is far from being well understood, and should be regarded as a largely independent challenge, involving 
issues connected with both neutrino physics itself and the physics of the system neutrinos are interacting with.

Neutrino interactions with matter strongly depend on energy. In this Thesis we restrict our focus to low-energy neutrinos, whose interactions 
with many-body systems,  such as nuclei or, more generally, nuclear matter, 
turn out to be critical to the description of a variety of different scenarios, from supernov\ae $ \ $ explosion to neutron-star cooling. 

The supernova is one of the last stages of star evolution \cite{ReddyLong}: when the ignition of the natural elements is no longer exothermic, the star starts contracting
under the gravitational pressure. If the star is very heavy, its mass exceeding $4 M_{\odot}$, where $M_{\odot}$ denotes the solar mass, the electron degeneracy pressure is not 
able to balance gravity, and the core collapses, reaching the typical density of a nucleus. Now the core reacts elastically to a further compression, leading to a bounced 
shock wave that ejects the mantle of the star. When the expanding ejecta become gravitationally decoupled from the stellar residue, the unshocked core   
evolves into a proto-neutron star. For several seconds it is still optically thick, and neutrinos act as a very efficient mediator to transfer energy from the core to the colder external 
region\footnote{This high efficiency is reminescent of the fact that the impact of an elementary particle on the stellar structure is maximized when its mean free path is of the order of the size of the system.}. As a consequence, the temperature rises and the star gets smaller, reaching a radius of about $10-15$ km. After $\sim50$ s, the neutrino mean-free-path exceeds the radius of the star, and neutrino flow turns into in a cooling mechanism \cite{Pet}. It can remain the dominant cooling mechanism for a period of time that varies from weeks to several thousand years,  depending on the stiffness of the equation of state, driving in turn the onset of the direct Urca processes. 

Experimentally, the importance of neutrino-driven mechanisms has been shown by the analysis of the signal from the core collapse supernova SN1987A, carried
out by the KAMIOKANDE \cite{Hirata} and IBM \cite{Bionta} collaborations. Most of the emitted energy turns out to be carried by neutrinos, the relevant figures 
being  \cite{laganke} :
\begin{align}
\nonumber
&E_{grav} \approx 10^{53}\ \textrm{erg}\ ,  &E_{\nu}\approx 2.7\times 10^{53}\,\textrm{erg} \ , \\
\nonumber
&E_{rad} \approx 8\times10^{49} \,\textrm{erg}  &E_{kin}\approx 10^{51}\,\textrm{erg}  \ .
\end{align}
From a quantitative point of view, the understanding of neutrino emission rates and propagation in dense matter is regarded as one of the most critical issues, particularly in view of the large effort being made to carry out realistic large scale simulations of supernov\ae $ \ $ and neutron star 
evolution \cite{B1,B2}. The systematic uncertainty associated with these simulations depends heavily on the values of the neutrino-nucleon and neutrino-nucleus cross sections used as input. 
Many results discussed in the literature have been obtained using somewhat oversimplified models of nuclear dynamics, although this source of 
uncertainty has been recently reduced adopting more realistic models.

In this Thesis we explore the effects of nuclear correlations on neutrino interactions with the \emph{nuclear medium}. We will discuss the main coupling mechanisms  
at low-momentum transfer and the role of many-body excitations. 
In this regime, collective modes turn out to be important, and must be taken into account. As a final outcome, a quantitative estimate of the  
neutrino mean free path in dense neutron matter will be provided. 

Nuclear interactions are described within the {\em ab initio} approach based on realistic hamiltonians and the Correlated Basis Function (CBF) formalism. Within this 
scheme, nucleon-nucleon correlations are included in the nuclear wave functions, and one can define an \emph{effective} potential, modified by medium effects, 
suitable for use in standard perturbation theory. Unlike the effective potentials derived within mean field approaches, the CBF effective potential reduces to 
the \emph{bare} potential in the limit of vanishing density, and is therefore strongly constrained by the experimental information on the two-nucleon system. 

Many-body effects at low momentum transfer are treated within Landau theory of normal Fermi liquids,  suitable to describe both coherent and incoherent excitations on 
the same footing. The Landau parameters are derived from matrix elements of the CBF effective potential, and turn out to provide a description of static properties
of neutron matter in very good agreement with that obtained from CBF calculations. \\

The Thesis is structured as follows:

\begin{itemize}

\item
Chapter \ref{NMB} outlines the main features of nuclear dynamics and the many-body approaches commonly employed to study nuclear matter;

\item
Chapter \ref{CB} is devoted to the discussion of the approach based on the CBF formalism and the cluster expansion technique, as well as the 
derivation of the effective potential; 

\item
In Chapter \ref{lab} we briefly review Landau theory of normal Fermi liquids;

\item
Chapter \ref{val} reports the Landau parameters obtained from the effective potential  derived in Chapter \ref{CB}. 
The static properties of matter obtained from Landau theory and the CBF approach are also compared;

\item
In Chapter \ref{Wea} we discuss the dynamic form factors of neutron matter, entering the definition of the neutrino cross section in matter;

\item
Chapter \ref{mf} is devoted to the discussion of the mean free path of non degenerate neutrinos, including its dependence on matter density and
temperature;

\item
As the treatment of three-nucleon interactions is one of the critical issues in the definition of the effective interaction, in Chapter \ref{TBF}, we report 
the results of a recent study of their effects on the binding energies of the oxygen isotopes, carried out within a novel approach based on the Green 
function formalism. 

\end{itemize}

%% file: Chap_NMBT.tex
The main challenge of any many-body approach is to provide an  \emph{ab initio} description of the system under study. Starting from the "elementary" degrees of freedom and including 
few-body interactions, one should be able, at least in principle, to reach a full understanding of the physical behavior.  
So forth, the system should be modeled without the need of any approximations or the need to include new phenomenological parameters. However, this lucky chance is somewhat exceptional in physics. Even in vacuum, as in particle physics, this situation occurs at some energy scales only, depending on the interaction we are dealing with. Focusing on nuclear interactions, the celebrated Quantum Chromodynamics (QCD), universally accepted as the fundamental theory of strong interaction, provides an excellent scheme to deal with high-momentum transfer processes. On the other hand, it is clearly not the right approach to describe the low-energy region, where systems are highly non-perturbative and any perturbative expansion is designed to fail. 

\begin{figure}[htbp]
\begin{center}
\includegraphics[scale=1.3]{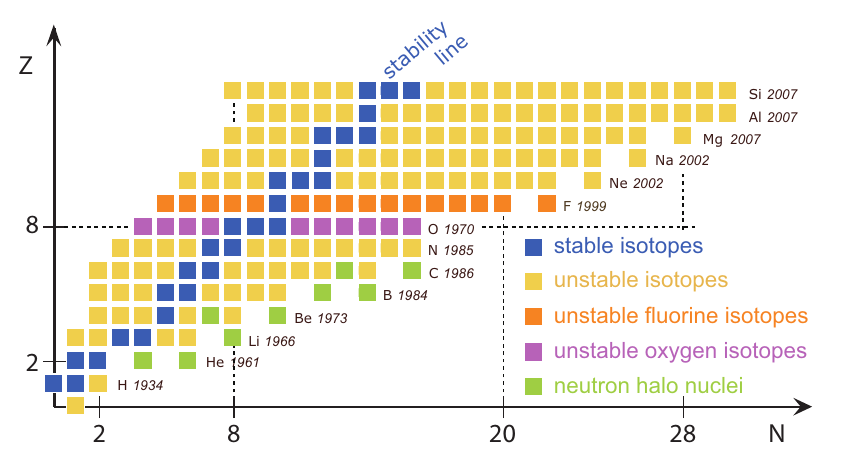}
\caption{Table of Nuclides, also referred to as "Segr\'e Chart". The figure shows only part of the table, hilighting the  different properties of the included nuclei \cite{Otsuka}}
\label{Nuc_im}
\end{center}
\end{figure}

\begin{figure}[htbp]
\begin{center}
\includegraphics[scale=1]{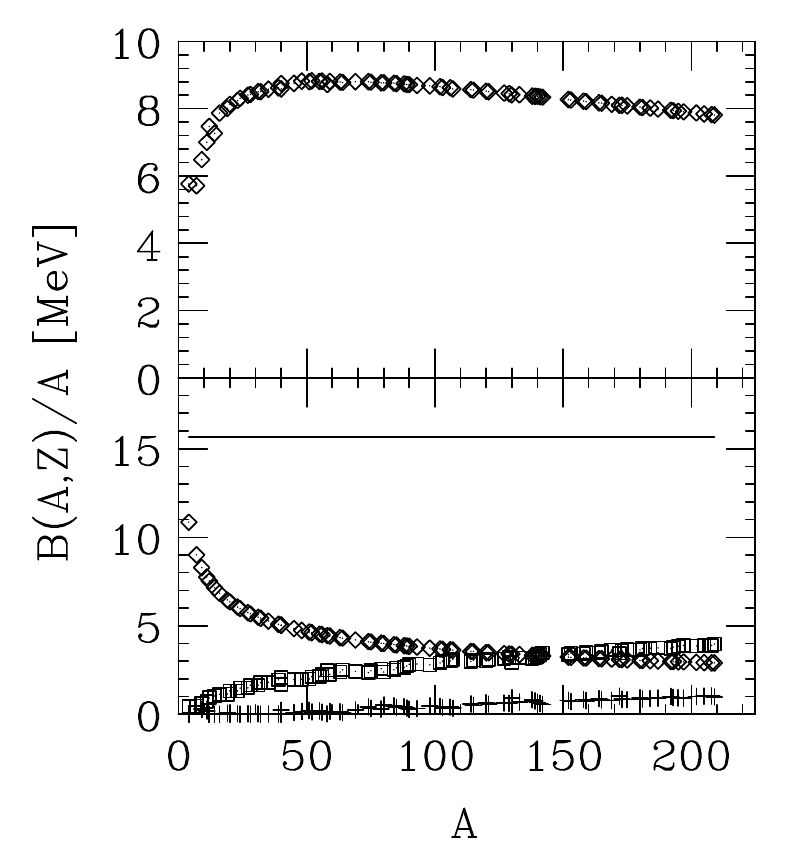}
\caption{Upper panel: Binding energy per nucleon for stable nuclei according to eq. (\ref{BinEn}), with $a\rb{V}=15.67 $ MeV,$a\rb{s}=17.23 $ MeV, $a\rb{c}=0.714 $ MeV, $a\rb{A}=93.15 $ MeV and $a\rb{p}=11.2 $ MeV. Lower panel: the solid line shows the volume contribution to
 the binding energy per nucleon, whereas the surface, Coulumb and symmetry contribution are represented by diamonds, square and crosses, respectively.}
\label{BEE}
\end{center}
\end{figure}

In this case, a \emph{coarse-graining} method is absolutely necessary. One usually averages on fast variables (like quarks), and defines new variables that are linked to a much larger 
time scale. In nuclear physics this coarse-grained (averaged) variables are already provided by nature: the nucleons. They still represent the actual basis for any realistic low-energy 
calculation, classification and prediction of new bound nuclei, see Segr\'e chart of fig. (\ref{Nuc_im}).
Several, quite important features of nuclear dynamics can be understood and {\em constrained} thanks to description of nuclei in terms of nucleons: binding energy, symmetries, saturation properties are only few of these. Moreover, thanks to the large amount of data available, the equilibrium properties of finite nuclei are well established, and consequently the properties of infinite system can be extrapolated with little error. To see how this can be possible, let us consider the celebrated \emph{semi-empirical} mass formula, yielding the binding energy as a function of the total number of nucleons $A$ and the number of protons $Z$,
\be\label{BinEn}
\frac{B(Z,A)}{A}=\frac{1}{A}\Big[a\rb{V}A-a\rb{s}A^{2/3}-a\rb{c}\frac{Z^2}{A^{1/3}}-a\rb{A}\frac{(A-2Z)^2}{4A}+\lambda a\rb{p}\frac{1}{A^{1/2}}\Big] \ .
\ee
The first term in square brackets, proportional to $A$, is called the \emph{volume term} and corresponds to the bulk energy. The second term, proportional to the squared nuclear radius, is associated with the surface energy, while the third accounts for the Coulomb repulsion between $Z$ protons uniformly distributed within the sphere of radius R. The fourth term, that goes under the name of \emph{symmetry energy} is required to describe the experimental observation that stable nuclei tend to have the same number of neutrons and protons. Moreover, even-even nuclei (i.e. nuclei having even $Z$ and even $A-Z$) tend to be more stable than even-odd or odd-odd ones. This property is taken into account by the last term in the above equation where $\lambda=-1,0,1$ for even-even, even-odd, odd-odd nuclei respectively. In fig. (\ref{BEE}) the value of $B/A$ for stable nuclei is shown together with the empirical values of the coefficients. 
By inspecting fig. \ref{BEE}, we can readily infer some important pieces of information. The nuclear interaction length is roughly the same as the radius of light nuclei, and smaller than the radius of heavy nuclei,  otherwise the binding energy would increase with increasing $A$. The properties of infinite matter can be explored in the limit $A\to \infty$, yielding
\begin{align} 
&E/A=a\rb{V}=15.7\,\textrm{MeV}\quad\textrm{for} \,\,A=2Z \quad \textrm{(symmetric matter, or SNM)}\nonumber \ ,\\
&E/A=a\rb{V}-\frac{a\rb{A}}{4}=-19.37\,\textrm{MeV}\quad\textrm{for}\,\, Z=0 \quad \textrm{(pure neutron matter, or PNM)} \ .\nonumber
\end{align}
The above values indicate that, while SNM is stable, PNM is not, since the system unbound $(E/A<0)$. Note that the fact that pure neutron matter alone is unstable does not
prevent the occurrence of PNM in a broad region in the interior of neutron stars. In an astrophysical object of mass comparable to the solar mass and radius $\sim 10 $ km has 
the gravitational pressure is strong enough confine PNM within the core.  
\begin{figure}[htbp]
\begin{center}
\includegraphics[scale=1]{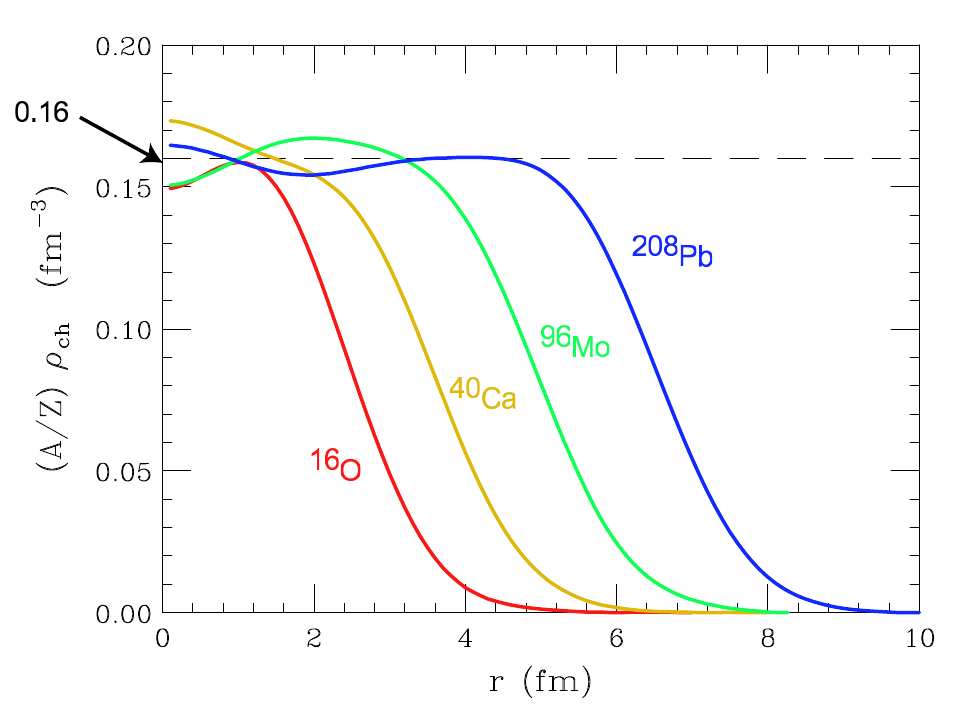}
\caption{Charge distribution as function of the radial distance, obtained from elastic electron scattering on nuclei.}
\label{fidens}
\end{center}
\end{figure}

Many other features of the nuclear interactions can be figured out studying the static properties of finite nuclei and extrapolating to infinite matter. 
The most important are:
\begin{description}
\item[Saturation of charge densities]
In fig.(\ref{fidens}) the charge density distributions of different nuclei are plotted as a function of the radial coordinate. 
It clearly appears that the density in the nucleus interior is nearly constant and independent of the mass number A. This means 
that nucleons cannot be packed too tightly, implying in turn that at short distance the nucleon-nucleon $(NN)$ force must be repulsive. Assuming that the interaction can be described by a non relativistic potential $v$ depending on the interparticle distance ${\bf r}$ we can write:
\[
v({\bf r})>0\quad,\quad |{\bf r}|< r_c \ ,
\]
where $r_c$ is the radius of repulsive core.
\item[Behavior of the binding energy per nucleon] The nuclear binding energy per nucleon is roughly the same for all nuclei with $A\ge 12$. This observation 
 suggests that $NN$ interaction has a finite range $r_{0}$, i.e. that 
\[
v({\bf r})\to 0\quad,\quad |{\bf r}|> r_0 \ .
\] 
\item[Properties of mirror nuclei] The spectra of the so-called mirror nuclei, i.e. pairs of nuclei having the same $A$ and charges differing by one unit (implying that the number of protons in a nucleus is the same as the number of neutrons in its mirror companion), e.g. $^{15}_7N$ (A=15, Z=7) and $^{15}_8 O$ (A=15, Z=8), exhibit striking similarities. The energy levels with the 
same parity and angular momentum are the same up to small electromagnetic corrections, showing that protons and neutrons have similar nuclear interactions, i.e. that nuclear forces 
are \emph{charge symmetric}.
\end{description}
The last property is a manifestation of the "isotopic invariance", associated with the approximated isospin symmetry of strong interactions\footnote{The isospin symmetry would be exact if the proton and neutron masses were equal. It is explicitly broken by small electromagnetic corrections and by the mass difference between the $u$ and $d$ quark.}. This mean that the proton and
the neutron can be viewed as different states of the same particle, described by the isodoublet field
\[
\Psi_N=
\left(
\begin{array}{c}
 p  \\
 n    
\end{array}
\right) \ .
\]
The isospin formalism allows one to describe free protons and neutrons using the lagrangian density
\[
\mathcal{L} = {\bar \Psi}_N ( \slashed{\partial} - m){\bar \Psi}_N \ ,
\]
invariant under the $SU(2)$ global phase transformation 
\[
U=\eu^{\iu\, \alpha_j\tau_j/2} \ ,
\]
where the $\alpha_j$ are constants and the $\tau_j$ denote Pauli matrices in acting in isospin space.
The proton and the neutron correspond to isospin projections $+1/2$ and $-1/2$, respectively. Proton-proton and neutron-neutron pairs always 
have total isospin $T=1$, whereas a proton-neutron pair may have either $T=0$ or $T=1$. 
The two-nucleon isospin states $|T,M_T\rangle$ can be summarized as follows:
\begin{eqnarray}
|1,1\rangle  &=&  |p,p\rangle  \ ,\nonumber \\
|1,0\rangle &=& \frac{1}{\sqrt{2}}(|p,n\rangle+|n,p\rangle) \ ,\nonumber     \\
|1,-1\rangle  &=&  |n,n\rangle  \ ,\nonumber\\
|0,0\rangle &=& \frac{1}{\sqrt{2}}(|p,n\rangle-|n,p\rangle) \ . \nonumber  
\end{eqnarray}

Isospin invariance implies that the interaction between two nucleons separated by a distance $r=|{\bf r}_1-{\bf r}_2|$ and having total spin $S$ depends on their total isospin $T$ but not on its projection $M_T$. For example, the potential $v({\bf r})$ acting between two protons with spins coupled to S=0, is the same as the potential acting between a proton and a neutron with spins and isospins coupled to $S=0$ and $T=1$.

\section{The two-nucleon system}
The two-body interaction is the basic ingredient to perform theoretical calculations in both finite nuclei and infinite nuclear matter. However, the derivation of this interaction from the 
fundamental theory, QCD, is still out of reach of the available theoretical approaches. As pointed out in the previous Section, the behavior of the $NN$ force in the 
${\bf r}\to 0,\infty$ limit can be infrrred and a qualitative trend can be drawn. This is of course not enough, since the same static properties can be predicted from different 
models of the microscopic dynamics. 

\begin{figure}[htbp]
\begin{center}
\includegraphics[scale=1.]{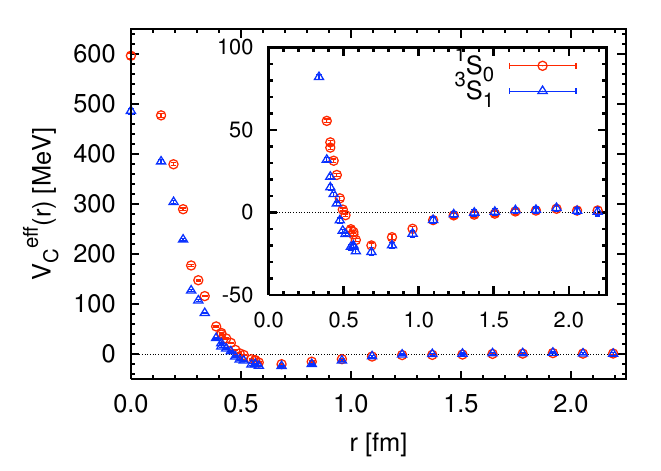}
\includegraphics[scale=1.]{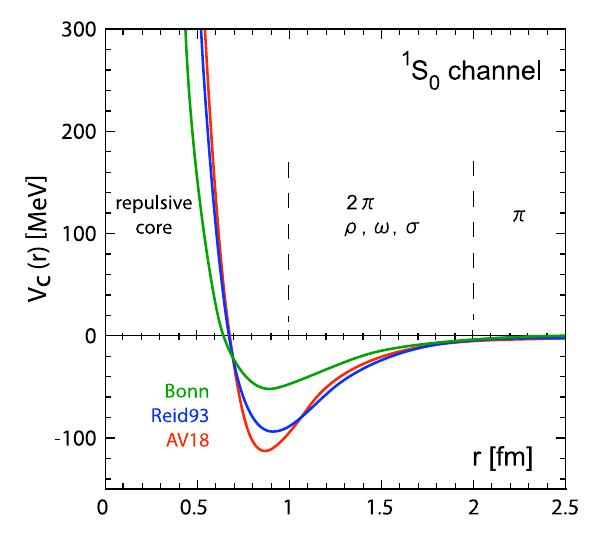}
\caption{\emph{Left panel}: Central componet of the $NN$ potential obtained from lattice QCD \cite{Ish,Ish1} setting the pion mass to $529$ MeV. It is apparent that, while the expected 
shape, exhibiting a repulsive core followed by an attractive region, is achieved, the attractive component is much weaker than that  predicted by the phenomenological potential showed in the (\emph{righ pannel})  }
\label{Ish}
\end{center}
\end{figure}

Several different approaches, described in the literature, have been employed to overcome this problem.
The left panel of fig. (\ref{Ish}) shows the preliminary results of the ongoing effort aimed at obtaining the $NN$ potential within the framework of lattice QCD
recently reported in Refs. \cite{Ish,Ish1}. The radial dependence of the potentials in the $^{1}S_0$ and $^{3}S_1$ channels have been obtained setting the pion mass 
to $m_{\pi}\simeq 529 $ MeV. It is apparent that the expected shape, exhibiting  a repulsive core followed by an attractive region, is reproduced at qualitative level.
On the other hand, comparison with the corresponding phenomenological potentials (to be discussed below) displayed in the the right panel, clearly shows that the attractive component is far too 
weak.  Note that the $NN$ potential reduces to the Yukawa one-pion exchange interaction at long distances, while the short-range repulsive core is to be ascribed to 
heavy-meson exchange or to more complicated mechanisms involving nucleon constituents and the intermediate range attaction is largely arising from two-pion exchange processes. 
Although the progress of QCD calculations has been impressive, the presently available potentials, as honestly stated by the authors of Refs. \cite{Ish,Ish1}, cannot be 
regarded as realistic. Use of these interactions would in fact lead to predict that a number of nuclei, including the deuterium, be unbound. 

Chiral perturbation theory $\chi PT$ \cite{Ecker} has been widely employed to obtain a theoretically sound model of the $NN$ interaction. 
This approach, originally proposed by Weinberg, exploits the Goldstone boson nature of the pion, arising from the breaking of chiral symmetry  \cite{WeinBook1}\footnote{The pion must 
be actually regarded as a {\em quasi} Goldstone boson, as chiral symmetry, besides being spontaneously broken, is also explicitly broken by the mass difference between the $u$ and $d$ quark.}. 
As a consequence, the interactions of low energy pions are weak, and can be treated in perturbation theory. $\chi PT$ has been first successfully applied to the pion-pion and pion-nucleon systems.
Many pion systems can be still be described within the perturbative approach, using the formalism of Ref. \cite{CCWZ}.
The extension to the nucleon-nucleon sector is conceptually more problematic, due to the nature of the interaction that may lead to bound states. However, Weinberg has shown that 
the non perturbative contributions to the $NN$ scattering amplitude are associated with diagrams involving purely nucleonic intermediate states, that can be summed up to all
orders in perturbation theory solving a Lippmann-Schwinger equation. 

Up to now, well-tested two-body potential based on $\chi PT$ have been obtained at $N^3LO$ (Next-to-Next-to-Next Leading Order).  They involve a relatively small number of 
free parameters, to be determined phenomenologically . 

\subsubsection{Phenomenological approach}
A different approach to attack the problem of the determination of the $NN$ potential relies on phenomenology only. For example, the observation that the only observed $NN$ bound state, 
the nucleus of deuterium, consists of one proton and one neutron coupled to total spin and isospin $S=1,T=0$, respsctively, is a clear indication of the spin-isospin dependence of nuclear forces. 
Furthermore the deuteron exhibits a non vanishing electric quadrupole moment, implying that its charge distribution is not spherically symmetric. Hence, the $NN$ potential 
in non-central.

Besides the properties of the two-nucleon bound state, the large data base of phase shifts measured in $NN$ scattering experiments  (the Nijmegen data base\cite{Nij} includes $\sim 4000$ data points, corresponding to the energies up to the pion threshold ($350 MeV$) provides valuable additional information on the nature of $NN$ forces. Phenomenological potentials are usually 
written as a sum of two terms:
\[
v=\tilde{v}_{\pi}+v_R,
\]
where $\tilde{v}_{\pi}$ takes into account the long range interaction modeled by the one-pion exchange (OPE) potential, while $v_R$ describes the interaction at both intermediate and short range. 
Let us consider these two terms in more details,
\begin{figure}[htbp]
\begin{center}
\includegraphics[scale=0.7]{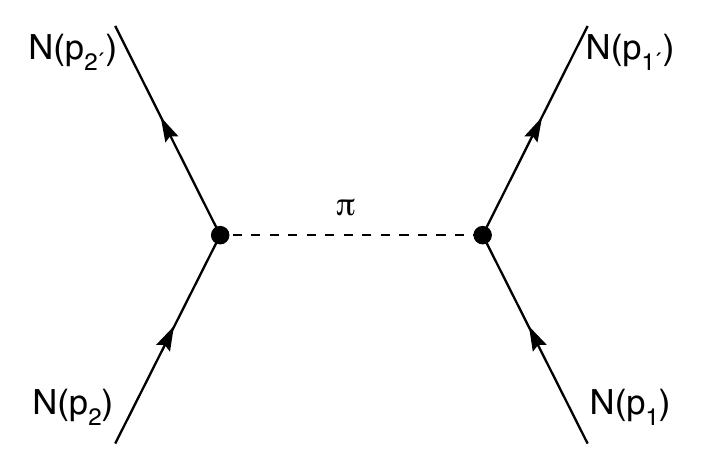}
\includegraphics[scale=0.7]{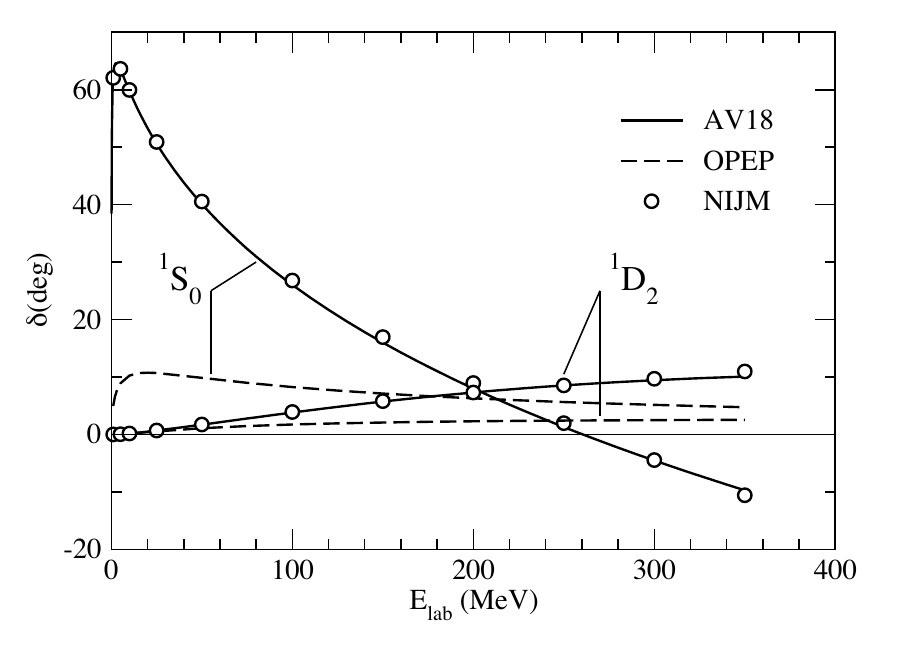}
\caption{\emph{Left panel}: Feynman diagram depicting the one-pion-exchange process, driving the $NN$ interaction at large distance.\emph{Right panel} Comparison between $^1$S$_0$ and $^1$D$_2$ phase shifts resulting from the Nijmegen analysis \cite{Nij} (open circles) and the predictions of the Argonne $v_{18}$ and OPE potentials.}
\label{OPE}
\end{center}
\end{figure}

\begin{description}
\item[${\tilde v}_{\pi}$ ] 
In the literature, the one-pion exchange potential has been derived using different choices of the $\pi NN$ coupling. 
For example one can write the interaction lagrangian in the form
\[
\mathcal{L}_{\pi NN} = 
\iu g_A\overline{N}\gamma_5\bm{\tau}\cdot\frac{\not{\!\partial\bm{\pi}}}{F_{\pi}}N \ ,
\]
where $g_A$ is the axial coupling constant, that can be extracted from neutron beta-decay, and $F_{\pi}$ is the pion decay constant. In the above equation, the $\gamma_5$ accounts for the fact that the pion field, represented by the $T=1$ isospin triplet  $\bm{\pi}=(\pi^+,\pi^-,\pi^0)$ with spin-parity $0^-$,  transform like pseudoscalar under Lorentz transformations. 
This interaction term, involving a pseudovector nucleon current, is usually employed in the modern versions of $\chi PT$. However, it totally is equivalent to the 
widely used  pseudoscalar $\pi$ NN interaction term, that can be obtained through an integration by parts
\[
\mathcal{L}_{\pi NN} = -\iu\underbrace{ \frac{2 m_N g_A}{F_{\pi}} }_{\mbox{$g_{\pi NN}$}}\overline{N}\gamma_5\bm{\tau}\cdot\bm{\pi}\,N.
\]
Note that this equivalence hold true because the interacting nucleons are on the mass shell. The $\pi NN$ coupling constant, resulting from the  
Goldberger-Treiman\cite{GolTre} relation is 
\[
g_{\pi NN}=2 m_{N}g_A/F_{\pi}\sim 14 \ . 
\]
The amplitude of the OPE process, described by the Feynman diagram of fig. (\ref{OPE}) reads
\[
M_{fi}=-\iu g_{\pi NN}\frac{\bar{u}_{s_2'}(p_2')\gamma_5 u_{s_2}(p_2')\bar{u}_{s_1'}(p_1')\gamma_5 u_{s_1}(p_1)}{k^2-m_{\pi}^2}\lan \bm{\tau}_1\cdot\bm{\tau}_2\ran,
\]
where $k=p_1'-p_1=p_2-p_2'$ , $k^2=k_{\mu}k^{\mu}=k_0^2-|{\bf k}|^2$ , $u_s(p)$ is the Dirac spinor associated with an on-shell nucleon, and
\[
\lan \bm{\tau}\cdot\bm{\tau}\ran=\eta_{2'}^{\dag}  \bm{\tau}\eta_2\eta_{1'}^{\dag} \bm{\tau}\eta_1,
\]
where $\eta_i$ is the Pauli spinor describing the isospin state of the $i$-th nucleon. In non-relativistic limit, the above amplitude provides the expression of  the 
$NN$ interaction, the expression of which in coordinate space is
\begin{align}
v_{\pi}&=\frac{ g^2_{\pi NN} }{4m_N^2}(\bm{\tau}_1\cdot\bm{\tau}_2)(\bm{\s}_1\cdot\bm{\nabla})(\bm{\s}_2\cdot\bm{\nabla})\frac{\eu\rp{m_{\pi}r}}{r}\nonumber\\
&=\frac{ g_{\pi NN}^2 }{(4\pi)^2}\frac{m_{\pi}^3}{4m_N^2}\frac{1}{3}(\bm{\tau}_1\cdot\bm{\tau}_2)\left\{\Bigg[ \st+S_{12}\Bigg(1+\frac{3}{x}+\frac{3}{x^2}\Bigg)\Bigg]\frac{\eu^{x}}{x}-\frac{4\pi}{m_{\pi}^3}\st\delta^3({\bf r})\right\},\nonumber
\end{align}
where $x=m_{\pi} r$. Note that the operator
\[
S_{12}=\sab,
\]
responsible of the non centrality of the OPE potential, is 
reminiscent of the tensor operator describing the non-central interaction between two magnetic dipoles. 

The OPE potential provides an accurate description of the long range part ($|{\bf r}|> 1.5$ fm) of the $NN$ interaction, as shown by the very good fit of the $NN$ scattering phase shifts
 in states of high angular momentum. In these states, due to the strong centrifugal barrier, the probability of finding the two nucleons at small relative distances becomes in fact negligibly small. 
 The notation $\tilde{v}_{\pi}$ refers to OPE potential  $v_{\pi}$ stripped of its zero-range component.

\item[$v_R$] At medium and short range many other complicated mechanisms, involving the exchange of two or more pions or heavier mesons (like the $\rho$ and $\omega$, 
 with masses $m_{\rho}=770$ MeV and $m_{\omega}=782 $ MeV, respectively) contribute to the $NN$ interaction processes. 
 Moreover, when the relative distance becomes very small ($|{\bf r}|< 0.5$ fm) the composite nature of nucleons is expected to play a crucial role.  
\end{description}

The potential obtained from the sum of ${\tilde v}_\pi$ and $v_R$ is usually written in the form
\[
v_{ij}=\sum_{ST}[v_{TS}(r_{ij})+\delta_{S1}v_{tT}(r_{ij})S_{12}]P_{2S+1}\Pi_{2T+1},
\] 
where the indices $S$ and $T$  denote the total spin and isospin of the interacting pair, respectively, while $P$ and $\Pi$ are the spin and isopsin projection operators, whose definition and properties are given in Appendix \ref{On}. The functions $v_{TS}(r_{ij})$ and $v_{tT}(r_{ij})$ describe the radial dependence of the interaction in the different spin-isopsin channels, and 
reduce to the corresponding components of the one-pion-exchange potential at large $r_{ij}$. Their shapes are adjusted in such a way as to reproduce the available $NN$ data (deuteron binding energy, charge radius and quadrupole moment and the NN scattering data). 
An altrernative representation of NN potential, based on the set of six operators,
\[
O^{n\leq 6}_{ij}=[\mathcal{I},\st,S_{12}]\otimes[\mathcal{I},(\bm{\tau}_1\cdot\bm{\tau}_2)] \ ,
\]
is given by,
\begin{equation}\label{stat}
v_{ij}=\sum_{n=1}^6 v^{n}(r_{ij})O^{n}_{ij}.
\end{equation}
The above potential provide a fairly good description of deuteron properties and the $S-wave$ scattering phase shift. However, to achieve a good description of the  $P$ wave
one has to include two additional components, associated with the momentum dependent operators defined as
\[
O^{n=7,8}_{12}={\bf L}\cdot{\bf S}\otimes[\mathcal{I},(\bm{\tau}_1\cdot\bm{\tau}_2)] \ ,
\]  
where $\bm{L}$ is the orbital angular momentum. The full expression of the potential,  yielding the best available fits of NN scattering data, with $\chi^2/$datum$\sim 1$,  
 includes ten additional operator more, bringing their total number to eighteen 
\begin{align}
O^{n=9,\dots,14}_{12}&=[{\bf L}^2,{\bf L}^2\st,({\bf L}\cdot\bm{S})^2]\otimes[\mathcal{I},(\bm{\tau}_1\cdot\bm{\tau}_2)] \ ,\nonumber\\
O^{n=15,\dots,18}_{12}&=[\mathcal{I},\st,\bm{S}_{12}]\otimes[T_{12},(\tau_{z1}+\tau_{z2})] \ ,\nonumber
\end{align}
where
\[
T_{12}=3(\bm{\tau}_1\cdot {\bf r})(\bm{\tau}_2\cdot {\bf r})-(\bm{\tau}_1\cdot\bm{\tau}_2) \ .
\]
The $O^{n=15,\dots,18}_{12}$ take care of small charge symmetry breaking effects, due to the different masses and coupling constants of the charged and neutral pions. 

The calculations discussed in this Thesis are based on a widely employed potential model, obtained within the phenomenological approach outlined in this Section, referred to as 
Argonne $v_{18}$  \cite{Argonne}. As an example of the quality of the phase shifts obtained from the Argonne $v_{18}$ potential, in fig. (\ref{OPE}) we show the 
results for the $^1S_0$ and $^1D_2$ partial waves, compared with the predictions of the OPE model (OPEP). We will adopt a static version of the full Argonne $v_{18}$, 
including only the first six operators. Use of this simplified potential appears to be justified in the context of our study of the dynamic form factors of neutron matter 
in low-momentum transfer regime. 

\section{Three-nucleon interactions}
It has been long realized that three-nucleon interactions play a critical role in determining the properties of both finite nuclei and infinite nuclear matter. 
The {\em exact} solution of the Scr\"odinger equation for the ground state of the three-nnucleon system with the Argonne $v_{18}$ potential yields a binding
energy per nucleon $E_0=7.6$ MeV, to be compared to the experimental value $E\rb{exp}=8.48$ MeV, while accurate calculations of the density dependence 
of the energy per particle of symmetric nuclear matter carried out with any phenomenological $NN$ potential fail to reproduce the empirical saturation 
properties. In order to bring theoretical results into agreement with the data  one has to add to the nuclear hamiltonian a contribution involving the potential 
$V_{ijk}$, describing three-nucleon interactions.

\begin{figure}[htbp]
\begin{center}
\includegraphics[scale=1.5]{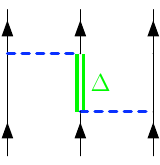}
\caption{Diagrammatic representation of the main process characterizing the attractive part of the three-nucleon interaction. 
The double line represents the excitation of a $\Delta$ resonsance ($M_\Delta = 1232$ MeV) in the intermediate state.}
\label{Delta}
\end{center}
\end{figure}

The theoretical description of three-nucleon potential $V_{ijk}$ has been first discussed in the pioneering work of Fujita and Miyazawa \cite{FujMiy}. 
They argued that the main contribution comes from a two-pion exchange process in which a nucleon-nucleon interaction leads to the 
excitation of one of the participating particles to a $\Delta$ ($M_\Delta = 1232$ MeV) resonance, which then decays due to the interaction with a third 
nucleon (see fig. (\ref{Delta}). 

The recent  models of the three-nucleon potentials are usually written in the form written in the form
\[
V_{ijk}=V_{ijk}^{2\pi}+V_{ijk}^{N} \ ,
\]
where the first contribution is the attractive Fujita-Miyazawa term of fig. (\ref{Delta}, while $V_{ijk}^{N}$ a purely phenomenological repulsive term. 
The parameters entering the definition of three-body potential are adjusted in such a way to reproduce ground state energies of $^3$H and $^3$He
and the saturation properties of symmetric nuclear matter  \cite{He}.

\begin{figure}[htbp]
\begin{center}
\includegraphics[scale=0.6]{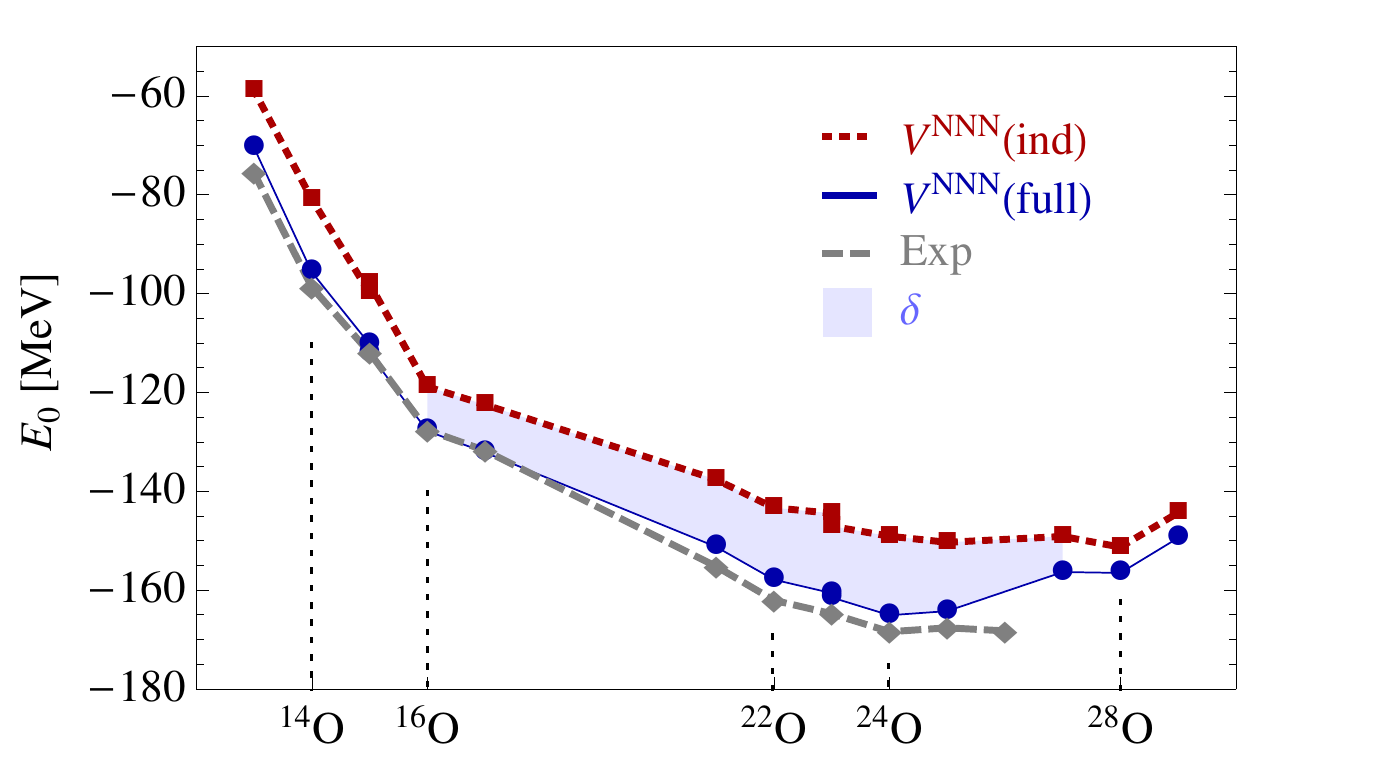}
\includegraphics[scale=1.5]{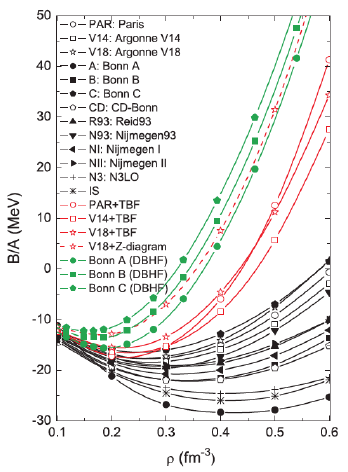}
\caption{\emph{Left panel:} Ground state energy of Oxygen isotopes. The red  line (ind) has been obtained using a chiral $N^3LO$ two-nucleon potential evolved through the
similarity renormalization group technique. The results represented by the blue line (full) also include the contributions of a $N^2LO$ chiral three-nucleon potential. 
All calculations have been carried out within the SCGF formalism \cite{3NFus}. Close to the neutron drip line the effect of three-nucleon forces (the shaded light blue 
region labelled $\delta$), turn out to be quite important. \emph{Right panel}: Equation of state of isospin symmetric nuclear matter, obtained with and without 
inclusion of three-nucleon interactions \cite{Li}. }
\label{3NF}
\end{center}
\end{figure}

Three-nucleon interactions deeply affect both the binding energy and several spectroscopic features of nuclei, mainly close to the 
neutron drip line. In this case, the last occupied nuclear orbital are in fact very close to the continuum spectrum, and the probability of 
many-body interactions is larger. As an example, in the left panel of figure (\ref{3NF}) we show the binding energy of several isotopes of Oxygen, obtained using the 
Self Consistent Green Function (SCGF) formalism \cite{Andrea}.
The shaded region emphasizes the increasing importance of the three-nucleon contribution as the neutron drip line ($^{24}O$) is approached. 
Typically, in this case its size is such that $\langle V_{ijk}\rangle/\lan v_{ij}\ran \sim 10\%$. 
For isospin symmetric nuclear matter, the role of the three-nucleon force in determinig the saturation properties is illustrated in the right panel of fig. (\ref{3NF}). 
The density dependence predicted by several different equation of states are plotted both with (green and red lines) and without (black line) inclusion of three-nucleon 
interactions.

%% file: Chap_CBF.tex
In non-relativistic nuclear many-body theory (NMBT), a nuclear system is seen as a collection of point-like protons and neutrons whose dynamics are described by the hamiltonian:
\[
H=\sum_{i}t(i)+\sum_{j>i}v(ij)+\dots
\]
where $t(i)$ and $v(ij)$ denote the kinetic energy operator and the \emph{bare} $NN$ potential respectively, while the ellipses refers to the presence of additional many-body interactions. Due to the presence of the repulsive core, carrying out perturbation theory in the basis provided by the eigenstates of the non interacting system requires a renormalization of the NN potential.  This is the foundation of many widely employed approaches developed to 
describe both finite nuclei and in infinite nuclear matter. For example, the schemes based on the no-core Shell Model \cite{Nav}, the Similarity Renormalization Group  \cite{SRG} and  G-matrix perturbation theory \cite{Baldo}, aim at obtaining the {\em in medium} $NN$ scattering amplitude 
from a \emph{bare} potential.  Alternatively, the many-body Schr\"oedinger equation can be solved using different approaches: self-consistent Green function theory\cite{Dick}, coupled cluster  method \cite{Martino}, as well as stochastic or variational techniques. Typically, each one of these methods 
have a preferred environment, finite nuclei or infinite nuclear matter, where they have been applied successfully.
However, stochastic and variational techniques have proven capable of providing accurate results both in light nuclei and 
uniform neutron and nuclear matter \cite{Bisconti,Carlson,Gandolfi}.

It has to be emphasized that within NMBT the interaction is completely determined by the analysis of \emph{exactly solvable} two and three-nucleon 
systems. As a consequence, the uncertainties associated with any many-body calculations are decoupled from the the determination of the dynamical 
model. In principle, given the hamiltonian, the properties of nuclear systems ranging from deuteron to neutron stars can be obtained in a fully consistent fashion, without including any additional adjustable parameters parameters. 

In this work we use a scheme formally similar to standard perturbation theory, in which non perturbative effects arising from the short range 
$NN$ repulsion are embodied in the basis functions, to be determined through a variational procedure. 
In the case of nuclear systems, the choice of the trial wave is particularly critical. To see this, consider, as an example, the fermion hard sphere system, i.e. 
a collection of particles interacting through a potential which is vanishing at all values of the interparticle distance $r$, except for the 
region $r<a$, in which it is infinite and positive.

\begin{minipage}[c]{.5\textwidth}
\hspace{-10mm}
\includegraphics[scale=0.5]{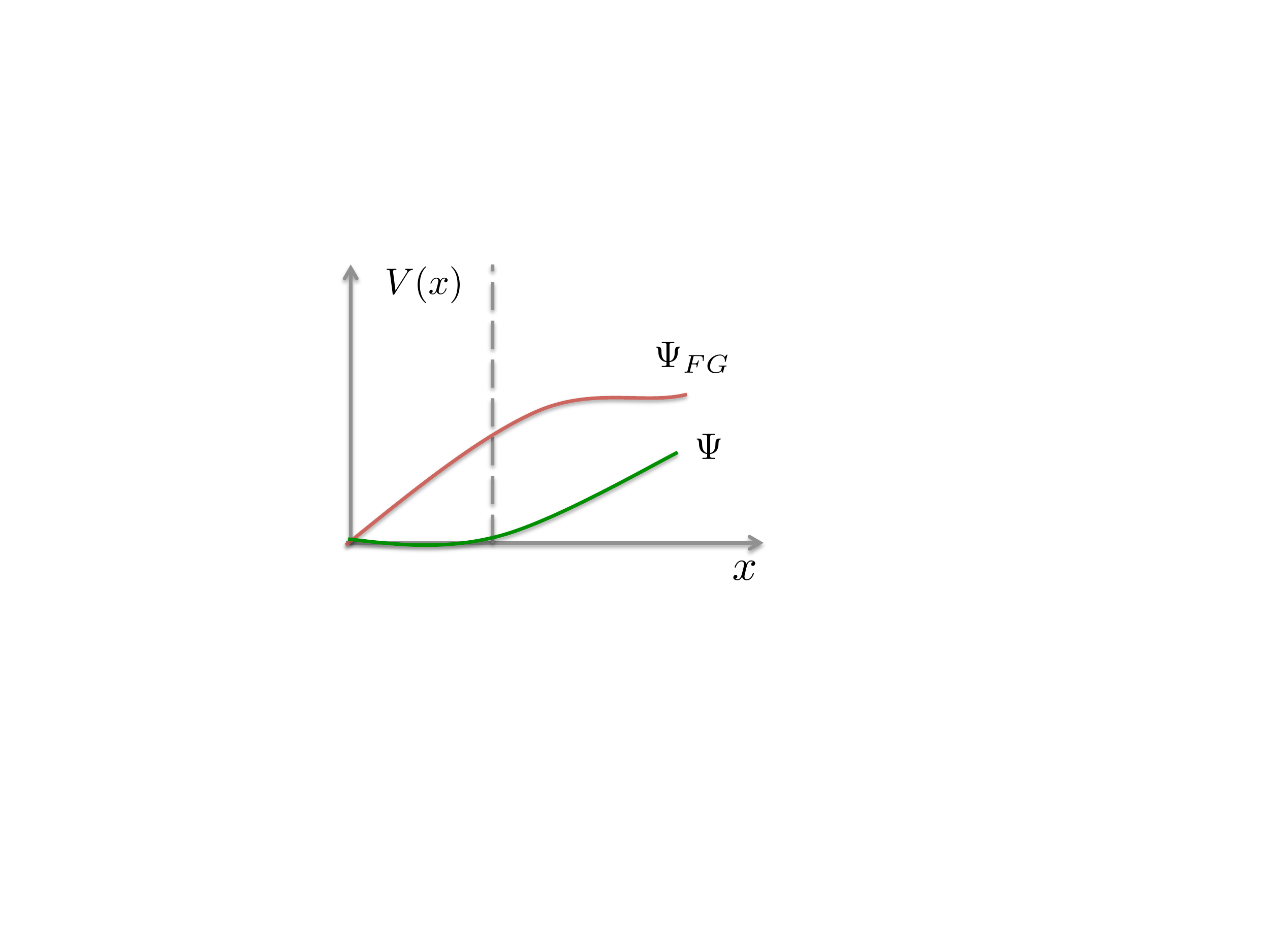}
\end{minipage}
\hspace{-26mm}
\vspace{3mm}
\begin{minipage}[l]{.64\textwidth}
Using the eigenstate of the non-interacting Fermi gas, $\Psi_{FG}$, there is no way to handle the infinity, as the behavior of the two-particle 
wave function is linear near a nodal point (see picture on the left). As a consequence, a finite region in which  $V(x)$ and $\Psi_{FG}$ overlap 
is always present, and the expectation value  $\langle V\rangle \to \infty$. On the other hand,  a good trial wave function 
must reflect the features of the actual ground state,  $\Psi$.
\end{minipage}
\\
Hence, the two-particle wave function associated with  $\Psi$ must be vanishing at $r<a$, and different from zero at $r>a$. 
The better we choose $\Psi$ we, the closer the variational energy is to the eigenvalue of the hamiltonian.

\section{Correlated Basis Function Theory}

In the Correlated Basis Function (CBF) approach, the basis states are {\em correlated} wave functions, obtained from the 
corresponding wave functions of the non interacting Fermi gas through the transformation \cite{feenberg,clark79}:
\[
| n \rangle =\frac{F| n\rb{FG}\rangle}{\langle n\rb{FG}|F^{\dag}F| n\rb{FG}\rangle^{1/2}} \  ,
\]
where $| n\rb{FG}\rangle$ is the determinant of single particle states describing $N$ non-interacting nucleons. The operator $F$ embodies 
the correlations among the particles induced by the $NN$ potentaial. It is usually written in the form
\[
F(1,\ldots,N)=\mathcal{S}\prod_{j>1=1}^{N}f_{ij} \ ,
\]
$\mathcal{S}$ being the symmetrization operator, needed to take into account the fact that, in general
\[
[f_{ij},f_{ik}]\neq 0 \ .
\]
The structure of the two-body \emph{correlation function}, $f_{ij}$, must reflect the complexity of $NN$ potential. 
As stated in the previous chapter, the calculations discussed in this Thesis have been performed using the truncated Argonne $v_{6}$ 
potential. As a consequence, we chose the same  operatorial structure for $f_{ij}$ 
\[
f_{ij}=\sum_{n=1}^{6}f^{n}(\br_{ij})O^n_{ij} \ .
\]
As already pointed out,  the $v_6$ model provides a fairly accurate description of the correlation structure of two-nucleon bound state. 
The shapes of the functions $f^{n}(r_{ij})$ are determined through the functional minimization of the expectation value of nuclear 
hamiltonian in the correlated ground state
\[
E_0= \langle 0|H|0 \rangle \ .
\] 
The evaluation of the above expectation value involves a degree of difficulty that rapidly increases with the number of particles in 
the system, $A$. It has been computed exactly using Monte Carlo techniques only for $A \leq 16$. For larger nuclei, as well as
for infinite nuclear matter, approximate calculations are carried out using the cluster expansion formalism, to be discussed in the 
next Section. 

\section{Cluster expansion formalism}
The correlation operator $F$ defined above, must exhibit a factorization, or  \emph{clustering}, property, dictated by the finite range 
of the $NN$ interaction. Suppose that we pick a subset of $p$ particles, labelled $i_1,\ldots,i_p$. These particles are then moved away 
 from the remaining $i_{p+1},\ldots,i_N$ particles. Under these conditions, the $A$-particle correlation operator $F$ must factorize into 
the product of two pieces, according to
\[
F(1,\ldots,N)\to F_{p}(i_{1},\ldots,i_{p})F_{N-p}(i_{p+q},\ldots,i_N) \ ,
\]
implying that the two subsystems become completely independent of one another. The above property provides the basis of the cluster 
expansion formalism, that allows one to express any matrix element of a many-body operator between correlated states as a 
sum of contributions associated with isolated subsystems (\emph{clusters}) consisting of an increasing number of particles. 
  
As an example, let us the consider the expectation value of the hamiltonian neglecting the three-nucleon potential. 
We will closely follow the derivation of the corresponding cluster expansion given in \cite{clark79}. The starting point is the definition of the generating functional:
\begin{equation}\label{GenFun}
I(\beta)=\langle 0 | \eu\rp{\beta(H-T_0)}| 0 \rangle \ ,
\end{equation}
where $|0\rangle=\hat{F}|(1,\ldots,N)\rb{FG}\rangle$ is the actual \emph{correlated} ground state of $N$ particles and
\[
T_0=\sum_{|{\bf p}_i|<p_F}t(i)\qquad,\qquad t(i)=\frac{|{\bf p}_i|^2}{2m} \ ,
\]
the kinetic energy operator. Starting from eq. (\ref{GenFun}) we can rewrite the expectation value of the hamiltonian in the form 
\begin{equation}\label{ener}
\langle H\rangle =(0|H|0)=T_0+\left.\frac{\partial}{\partial \beta}\ln I(\beta)\right|_{\beta=0} \ .
\end{equation}
Let us now expand the rhs of the above equation according to the cluster formalism. The functional in (\ref{GenFun}) involves the degrees of
freedom of all the $N$ particles. As the particles are indistinguishable, any subset of $p<N$ particles can be arranged in $N!/(N-p)!p!$ ways. 
Defining a generating functional for each subset, we get:
\begin{align}
\nonumber
I\rb{i}(\beta)  &= \langle i| \eu\rp{\beta(t(1)-\varepsilon_i^0)} |i\rangle \ , \\ 
\nonumber
I\rb{ij}(\beta) &=  \langle i j|F_2^{\dag}(12) \eu\rp{\beta(t(1)+t(2)+v(12)-\varepsilon_i^0-\varepsilon_j^0)} F_2(12)  |i j\rangle\rb{a}  \ , \\
\nonumber
& \  \vdots \\
\nonumber
I\rb{1,\dots,N}(\beta) &= I(\beta) \ ,
\end{align}
where the indices $i,j, \ldots$ label the non-interacting single-particle states of the "vacuum", or Fermi sea, $\varepsilon_i^0$ is the corresponding  kinetic energy, $v(ij)=v_{ij}$ and $ij \rangle = |ij \rangle - | ji \rangle$\footnote{As the particles are indistinguishable, the antisymmetrization
of two body matrix elements only requires  antisymmetrization of the state $|ij \rangle$.}. 

Let us now reexpress $I(\beta)$ in terms of $n$-body generating functionals ($n \leq N$), with $I_i(\beta) = 1,\quad i$. 
We start noting that $I_{ij}$ is close to the product of $I_i$ and $I_j$. It would in fact be exactly equal if we could neglect the interaction 
$v(12)$ and the associated correlation $F_2(12)$. This observation suggests to  rewrite $I_{ij}$ as,
\[
I\rb{ij}=I\rb{i} I\rb{j} Y_{ij} \ ,
\]
where the deviations of $Y_{ij}$ from unity arise from correlation effects. Relabeling $Y_i = I_{i}$ we can describe the above functions 
in terms of the $Y_{i,\ldots}$ according to
\begin{equation}\label{arr}
\begin{array}{rl}
I\rb{i}(\beta)\!&=Y_i \ ,\\ 
\vspace{-1mm}
I\rb{ij}(\beta)\!&=Y_iY_jY_{ij} \ , \\
&\,\vdots\\
I\rb{1,\dots,N}(\beta)\!&=I(\beta)=\Big[\prod_iY_i\Big]\Big[\prod_{j>i}Y_{ij}\Big]\cdot\cdot\cdot \Big[Y_{1,\ldots,N}\Big] \ ,
\end{array}
\end{equation}
implying
\begin{equation}
\ln I(\beta)=\sum_i\ln Y_i+\sum_{j>i}\ln Y_{ij}+\ldots+\ln Y_{1\ldots N} \ .
\end{equation}
Here the $p$-th term gathers all contributions involving all possible interaction among $p$ particles. In a diagrammatic language, this term can be represented by a $p$-vertex diagram representing the nucleons in the cluster, connected by lines corresponding to dynamical and statical correlations.  
Substituting the above equation in eq. (\ref{ener}) we can write the ground state expectation value of the hamiltonian in the form
\begin{equation}
\label{ene2}
\langle H\rangle=T_0+(\Delta E)_1+(\Delta E)_2+\ldots+(\Delta E)_N,
\end{equation}
where the contribution of the $p$-body cluster is given by the $p$-th term\footnote{Note that $(\Delta E)_1=0$, since $I_{i}=Y_i=1$.},
\[
(\Delta E)_p=\sum_{i_1<\ldots<i_p}\left.\frac{\partial}{\partial \beta}\ln Y_{i_1,\ldots,i_p}\right|_{\beta=0} \ .
\]
In order to express eq.(\ref{ene2}) in terms of the functions  $I_{i_1,\dots}$, we invert the relation in (\ref{arr}),
\[
\begin{array}{rl}
Y_{i}\!&=I\rb{i} \ , \\ 
Y_{ij}\!&=(I\rb{i}I\rb{j})^{-1}I\rb{ij} \  . \\
&\,\vdots\\
\end{array}
\]
The resulting two-body cluster term is 
\begin{equation}\label{Delta2}
(\Delta E)_2=\sum_{i<j}\left[\frac{1}{I\rb{ij}}\frac{\partial I\rb{ij}}{\partial \beta}-\bigg(\frac{\partial I\rb{i}}{\partial \beta}+\frac{\partial I\rb{j}}{\partial \beta}\bigg)\right]_{\beta=0} \ .
\end{equation}
As the number of particles in the cluster increases, the difficulties involved in the evaluation of the corresponding contribution become more and more 
severe, making numerical calculations impossible.
However, the diagrams associated with the different cluster contributions can be classified according to their topological structure,  and selected classes
 of diagrams can be summed up to all orders solving a set of integral equations, called Fermi Hyper-Netted Chain (FHNC) equation \cite{Fantoni1,Lovato}.
\section{Effective interaction} \label{Eff_int}
The in medium {\em effective} interaction is usually defined starting from eq. (\ref{ene2}) and setting
\begin{equation}
\label{defV}
\langle H \rangle= \langle 0_{FG} |T_0+V_{\rm eff}|0_{FG} \rangle \ . 
\end{equation}
The idea underlying the effective interaction approach is that the above equation yields
\emph{screened} potential,  that is well behaved, and can therefore be used to carry out perturbative calculations in the basis of 
eigenstates of the non interacting system. 

Here, we will follow a procedure developed in Refs. \cite{Pand,Ben}, whose authors derived the effective interaction from a cluster expansion 
of the left-hand side of eq. (\ref{defV}) truncated at the two-body level. 

Comparison between eqs. (\ref{defV}) and (\ref{Delta2}) shows that $V_{\rm eff}$ can be expressed in term of the 
effective two-nucleon interaction $w_{12}$, defined through 
\be
\label{eq:veff}
\langle 0_{FG} |V_{\rm eff}| 0_{FG}\rangle =\sum_{i<j}\langle i j| w_{12}|i j \rangle\rb{a} \ ,
\ee
with
\begin{align}\label{effective}
w_{12}&=\frac{1}{2}\Big[f_{12},[t(1)+t(2),f_{12}]\Big]+f_{12}v(12)f_{12}\nonumber\\
&=-\frac{1}{2m}\Big[f_{12},[\bm{\nabla}_{\bf r}^2,f_{12}]\Big]+f_{12}v(12)f_{12} \ .
\end{align}
In the second line of the above equation we have exploited the fact that the correlation functions only depend on the relative 
distance between the interacting particles, ${\bf r}$. Moreover, in the limit $N\to \infty$, the normalization $I\rb{ij}|_{\beta=0}^{-1}$ of eq. (\ref{Delta2})  differs from unity terms $O(1/N)$ at most, and can therefore be neglected. The details of the calculations leading to the final expression of $w_{12}$  
are given in Appendix \ref{EnFun}. The result is
\begin{align}
w_{12}&=f_{12}(-\frac{1}{m}(\bm{\nabla}^2 f_{12})-\frac{2}{m}({\bf \bm{\nabla}} f_{12})\cdot {\bf \bm{\nabla}}+v(12)f_{12})\nonumber\\
&\simeq f_{12}(-\frac{1}{m}(\bm{\nabla}^2)+v(12))f_{12}\nonumber=\sum_{n} v_{eff}^n(\br)O^n
\end{align}
Note that only the \emph{static} part of the effective interaction is retained. The results of numerical calculations show that the contribution of the 
term $(\bm{\nabla} f_{12})\cdot  \bm{\nabla}$, yielding an 
expcitely momentum-dependent contribution through the exchange part of the matrix elements, is in fact very small compared the one arising from 
the static term.

In order to include in $V_{\rm eff}$ the effects of three-nucleon forces, we follow the procedure adopted in Ref. \cite{Ben}, based on the approach originally proposed by Lagaris and Pandharipande \cite{Lag}. Within this scheme, interactions involving three or more nucleons are taken 
into account through a density dependent modification of the $NN$ potential at intermediate range, where two-pion exchange is believed to be the 
dominant mechanism. Neglecting, for simplicity, the charge-symmetry breaking components of the interaction, the resulting potential can be 
written in the form \cite{Lag}
\be
\label{TNR}
\widetilde{v}(ij) = \sum_{n=1,14} \left[ \widetilde{v}^n_\pi(\br_{ij})
+ v^n_I(\br_{ij}){\rm e}^{-\gamma_1 \rho} + v^n_S(\br_{ij}) \right] O^n_{ij} \ ,
\ee
where $\widetilde{v}^n_\pi$, $v^n_I$ and $v^n_S$ denote the long- (one-pion-exchange), intermediate-
and short-range part of the potential, respectively. The above modification results in a 
repulsive contribution to the binding energy of nuclear matter.
The authors of Ref.\cite{Lag} also include the additional attractive contribution 
\be
\label{TNA}
\Delta E_{TNA} = \gamma_2 \rho^2 (3-2\beta){\rm e}^{-\gamma_3 \rho} \ , 
\ee
with $\beta = (\rho_p - \rho_n)/(\rho_p + \rho_n)$, where $\rho_p$ and $\rho_n$ denote the 
proton and neutron density, respectively. The values of the parameters 
$\gamma_1$, $\gamma_2$ and $\gamma_3$ appearing in 
Eqs.(\ref{TNR}) and (\ref{TNA}) have been determined in such a way as to reproduce the empirical
binding energy and equilibrium density of nuclear matter \cite{Lag}.

Given the bare potential, the effective interaction is determined by the correlation operator $f_{12}$. 
Its shape can be obtained from the functional minimization of energy at two-body cluster level, yielding a set of coupled differential 
equations, to be solved with the boundary conditions,
\begin{align}
f_n(r\geq d)&=
\left\{
\begin{array}{lc}
 1 &  n=1  \\
 0 &  n=2,3,4    
\end{array}
\right.\nonumber\\
f_n(r\geq d_t)&=0 \qquad n=5,6\nonumber
\end{align}
and
\begin{align}
\left.\frac{d f_n}{dr}\right|_{r=d}&=0\qquad n=1,2,3,4\nonumber\\
\left.\frac{d f_n}{dr}\right|_{r=d_t}&=0\qquad n=5,6\nonumber
\end{align}
where $d$ and $d_t$ are variational parameters. The above condition simply express the requirements that i) for relative distances larger than the interaction range, the two-nucleon wave function reduces to the one describing non interacting particles and ii) tensor interaction have a longer range, implying 
$d_t > d$. 
We have solved the Euler-Lagrange equations for a wide range of nuclear matter density using the values for $d$ and $d_t$ obtained from the 
highly refined variational calculation of Ref. \cite{Ak}, carried out within the FHNC-SOC scheme. The derivation of the 
Euler Lagrange equations is discussed in Appendix \ref{EuLag} .

In figure (\ref{F-function}) the different components of the effective interaction at density $\rho=0.16$ fm$^{-3}$ are compared to the corresponding components of the Argonne $v_6$ potential. It clearly appears that the screening arising from correlations leads to a significant quenching 
of the \emph{bare} $NN$ interaction.

\begin{figure}[htbp]
\begin{center}
\includegraphics[scale=0.9]{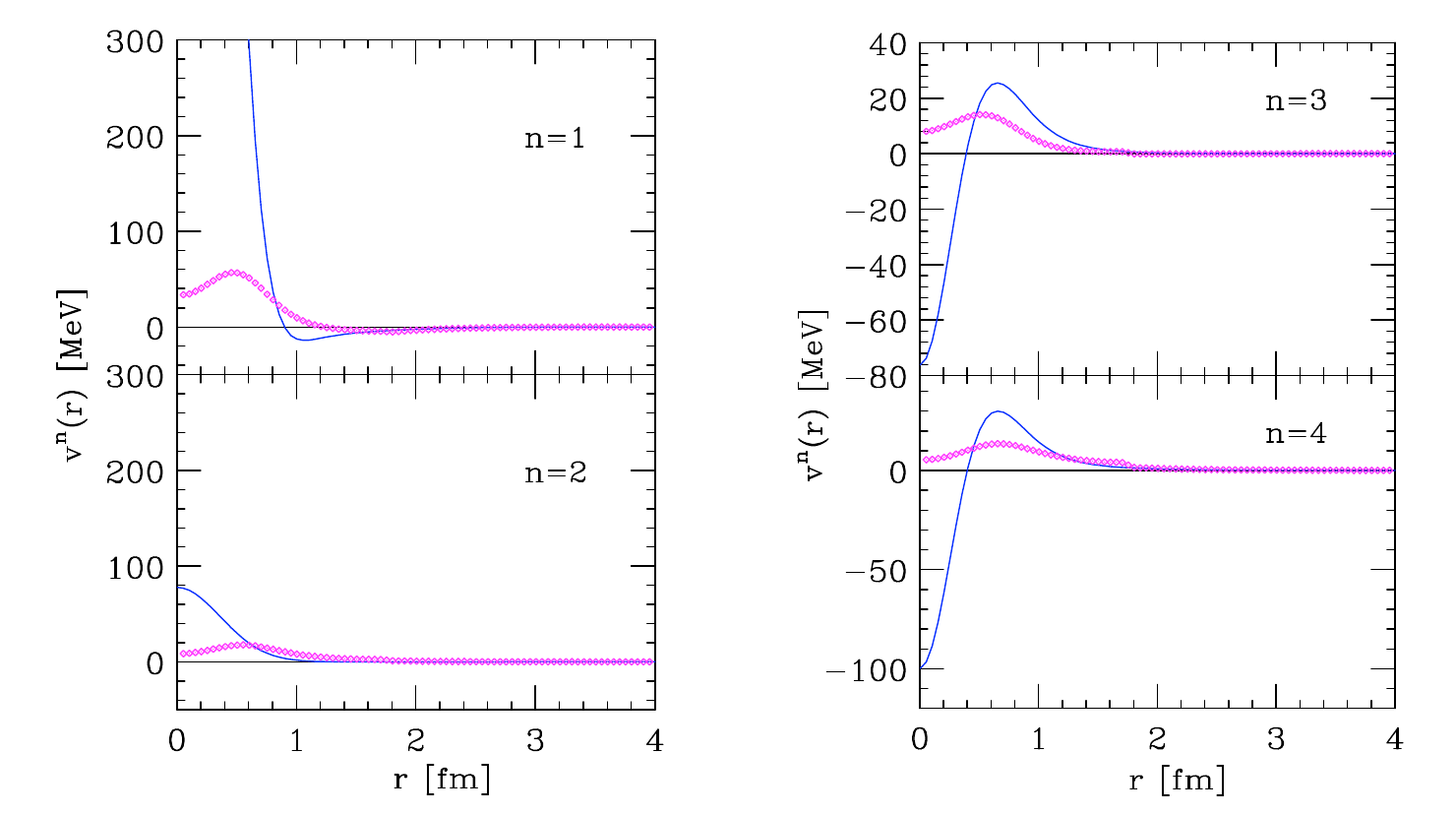}
\includegraphics[scale=0.9]{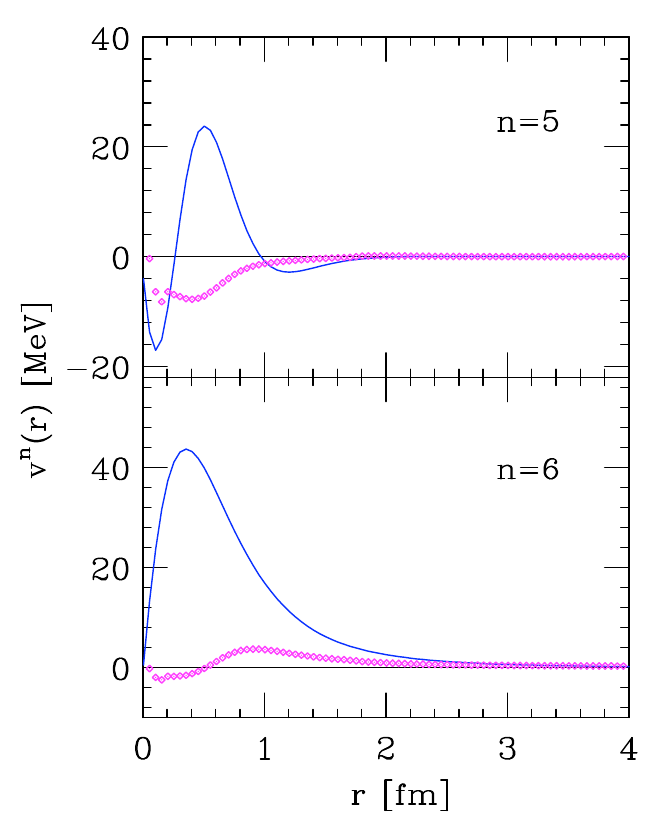}
\caption{Comparison between the components of the \emph{bare} Argonne $v_6$ potential (blue dashed line) and those  of the CBF effective potential 
 (pink solid line)  at  $\rho=0.16$ fm$^{-3}$. }
\label{F-function}
\end{center}
\end{figure}

\section{Ground state energy}
Within the scheme described in the previous Sections, 
the final expression of the ground state energy reads
\[
\frac{E}{N} = \langle T_0 \rangle  + \frac{1}{2}\sum_{ij}\Big\{\langle i j |w_{12}|i j\ran- \langle i j |w_{12}|j i\ran \Big\} , 
\]
where $\langle T_0 \rangle$ is the Fermi gas kinetic energy and we have explicitly written the direct and exchange 
contributions to the matrix elements of $w_{12}$. 

\begin{figure}[htbp]
\begin{center}
\includegraphics[scale=0.9]{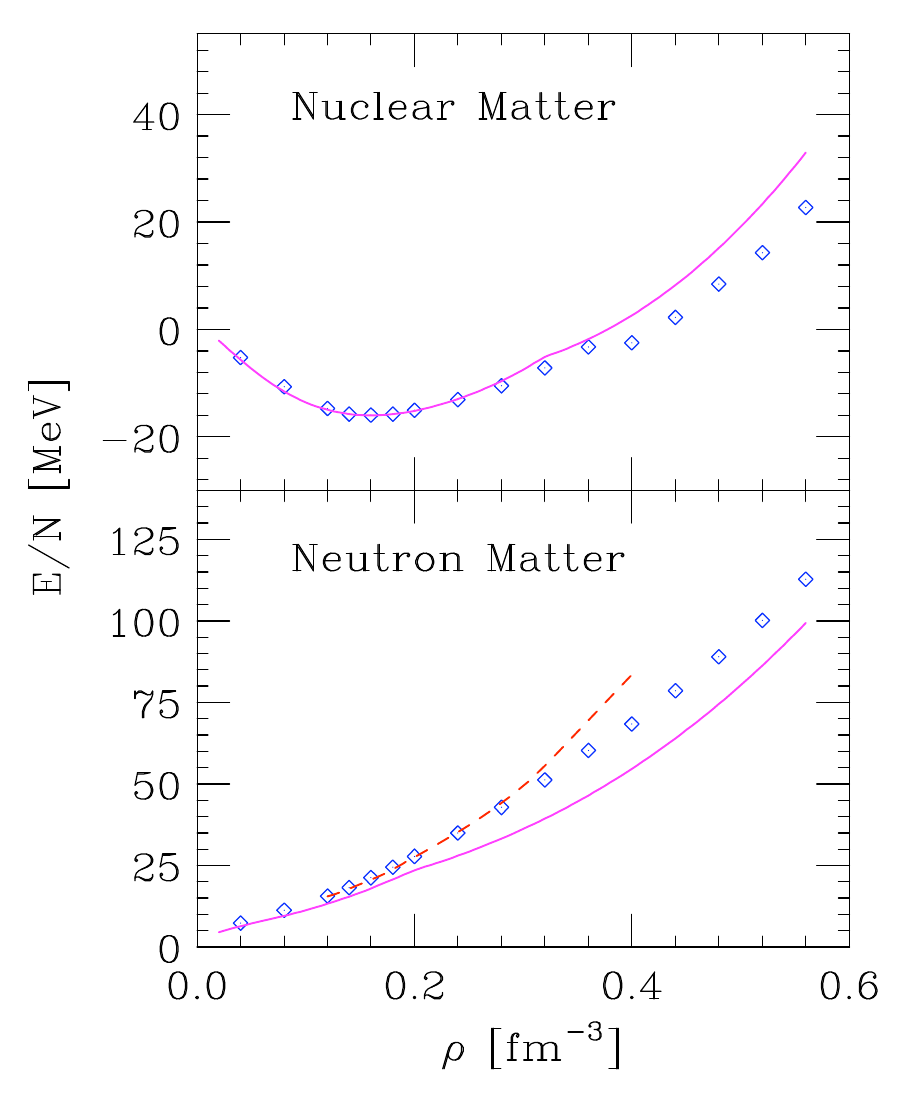}
\caption{\emph{Upper panel}:Energy per particle of symmetric matter as function of the density.\emph{Lower panel}: Same as in the upper panel, but 
for pure neutron matter. The diamonds and the solid lines represent the result obtained from eq. (\ref{En}) and those of  Ref.  \cite{Ak1}, respsctively. 
The red dashed line in the lower panel shows  the result of the AFDMC approach \cite{Fantoni}.}
\label{Eos}
\end{center}
\end{figure}

One of the attractive features of our approach is that it allows one to treat \emph{any} forms of isospin asymmetric matter at the same level 
of accuracy. Labeling with $x_\gamma$ the fraction of particles of type $\gamma$\footnote{$\gamma=1,2,3,4$ corresponds to spin-up proton ($p \up$), 
spin-down proton ($p \down$), spin-up neutron ($n \up$) and spin-down neutron ($n \down$), respectively.}, we can write
\[
\rho=\sum_{\gamma=1}^{4}\rho_{\gamma} \ \ , \ \  \rho_{\gamma}=x_{\gamma}\rho  \ \ , \ \  \sum_{\gamma=1}^4 x=1 \ .
\]
Each set of $x_\gamma$ uniquely defines a type of matter, for example, pure neutron matter corresponds to 
\[
x_1=x_2=0 \ \ \ \ \ , \ \ \ \ \  x_3=x_4=\frac{1}{2} \ .
\]
The expression of the ground state energy of any kind of matter is (see appendix \ref{EnFun}),
\be
\label{En}
\frac{E}{N}=\frac{3}{5}\sum_{\gamma}\frac{p_{F,\gamma}^2}{2m}+\frac{\rho}{2}\sum_{\gamma\mu}x_{\gamma}x_{\mu}\sum_{n=1}^6\int d^3\br\, w_{12}^n
\Big[ A_{\gamma\mu}^n-B_{\gamma\mu}^n \ell(p_{F,\gamma}r) \ell(p_{F,\mu}r)\Big]
\ee
where $w_{12}^n$ is the component of the effective $NN$ interaction associated with the operator $O^n$, $p_{F,\gamma}=(6\pi^2\rho_{\gamma})^{1/3}$ and the Slater function $\ell$ is defined as 
\[
\ell(p_{F,\mu}r)=\sum_{\bk} \eu\rp{\iu \bk \cdot \br}\Theta(p_{F,\gamma}-|\bk|) \ ,
\]
while
\[
A_{\gamma\mu}^n=\langle\gamma\mu|O^n|\gamma\mu \rangle \quad B_{\gamma\mu}^n=\langle\gamma\mu|O^n|\mu\gamma \rangle
\]
denote the matrix elements of the six operators in spin-isospin space, where. Their explicit expressions for the direct channel are 
\[ A=
\begin{array}{r}
\quad O_1 \quad O_2\quad O_3\quad O_4\qquad O_5\qquad O_6\qquad\\
\begin{array}{c}
  p\up \,\,  p\up \\
  p\up \,\, p\down \\
  p\up \,\,  n\up\\
  p\up \,\,  n\down\\
\end{array}
\left(\,
\begin{array}{cccccc}
1 & 1 & 1 & 1 & a(\theta) & a(\theta)\\
1 & 1 &-1 &-1 & -a(\theta) & -a(\theta)\\
1 & 1 & 1 &-1 & a(\theta) & -a(\theta)\\
1 &-1 &-1 & 1 & -a(\theta) & a(\theta)
\end{array}\right)
\end{array}
\]
with $a(\theta)=(3\cos^2\theta-1)$, $\theta$ being the angle between $\br$ and the $z$-axis. The corresponding expressions for the exchange channel 
are
\[ B=
\left(
\begin{array}{cccccc}
1 & 1 & 1 & 1 & a(\theta) & a(\theta)\\
0 & 0 & 2 & 2 &-a(\theta) &-a(\theta)\\
0 & 2 & 0 & 2 & 0 & 2a(\theta)\\
0 & 0 & 0 & 4 & 0 & -2a(\theta)
\end{array}\right).
\]
The matrices are obviously symmetric.  Equation (\ref{En}) is completely general, and allows one to calculate the ground state energy 
for any forms of nuclear matter, once the $x_\gamma$ are fixed. 
In fig. \ref{Eos}, taken from Ref. \cite{Valli}, the energies of pure neutron matter (lower panel) and isospin symmetric nuclear matter 
(upper panel) obtained from eq.  (\ref{En}) are compared to the results of Refs. \cite{Fantoni} and \cite{Ak1}. In \cite{Ak1} (solid lines) the calculation have been carried out using a variational approach based of the FHNC-SOC formalism, with a nuclear hamiltonian including the Argonne $v_{18}$ 
and Urbana IX potentials. The results of \cite{Fantoni} (dashed line of the lower panel) is have been obtained within the Auxiliary Field Diffusion Monte Carlo (AFDMC) approach using, the truncated Argonne $v_8$ interaction and the same three-body potential. It clearly appears that 
our effective interaction provides a description in excellent agreement with that resulting from state-of-the-art theoretical approaches over a 
broad density range. Note that the correct saturation properties of symmetric nuclear matter are only obtained thanks to the inclusion of the effects
of three- and many-body interactions. It should also be emphasized that, using eq. $v\rb{eff}$ and the model of Ref. \cite{Lag}, amounts to 
effectively including contributions of clusters involving more than two nucleons.

As a final remark, it is worth mentioning that our approach, that does not involve any adjustable parameters, also provides a 
very reasonable estimate of the compressibility module\footnote{ To be defined in eq.  (\ref{Kmodulus}).} of symmetric 
nuclear matter: $K=230$ MeV.

%

%% file: Chap_Landau.tex
Landau theory 
provides a conceptual framework for understanding the main features of neutron matter near $T\sim 0$. The theory describes fermionic systems that are assumed to be in \emph{normal} state, i.e. not showing any tendency to aggregate particles into clusters exhibiting bosonic features. In an infinite system, these clusters are the result of an attractive interactions, the occurrence of  which completely distorts the physics near the ground state. The best known example is liquid $^3$He\footnote{Although the theory can also be  extended to describe $^3$He-$^4$He mixtures.}, which is so far the only system in which quantum effects appear before solidification. Further exotic examples are electrons, both in the conductive band of metals and in white dwarfs, and, of course, nuclear matter. It is a remarkable fact that these systems, while being very different from one another, can actually be described by the same 
theory. This is due to the combined effects of the fermionic nature of the constituent particles and the zero temperature limit. In this context, statistical correlations largely dominate over dynamical correlations. The system is ``frozen'', almost degenerate, its state being very close to the 
corresponding state of the non-interacting system, where only statistical correlations induced by the Pauli principle matter. In the  ground state, 
the particles fill up the lowest energy levels, up to the Fermi energy $\epsilon\rb{F}$, while the states of higher energy are empty. As a consequence,  
the Fermi energy coincides with the chemical potential of the system, $\mu$. 

Interacting systems close to the ground state, can be equally well described in term of two set of quantities, the single-particle energy levels, $\epsilon_{s}$, and the corresponding  occupation numbers, as described by the distribution function by $n(\epsilon_{s})$. The latter provides the the probability 
density for state of energy $\epsilon_{s}$ to be occupied. In a cold, non interacting, Fermi gas this distribution function reduces to the step function
\begin{equation}\label{n}
n(\epsilon\rb{s})=\theta(\epsilon\rb{s}-\epsilon\rb{F}). \ .
\end{equation}
In principle, switching on the interaction, the system can no longer be described in term of single particle states. In order to evaluate any macroscopic observables, one should find the stationary states of the system as a whole. Nevertheless, low-lying excited states are strongly influenced by the ground state itself, since interactions can only affect the states near the Fermi level, where empty and occupied states are very close to each other. 
Particles with energies lying deep in the Fermi sea still behave as free particles, being largely unaffected by interactions. As a consequence, all scattering properties, giving rise to multiparticle excitations, are restricted to a narrow region near the Fermi surface. 

The features of the energy spectrum 
of the system in this region of energy can be inferred from very general considerations, which are valid regardless of the magnitude and 
the specific nature of the interaction. Each low-lying excited state of a macroscopic system can be seen as simple combination 
of a set of elementary excitations, dubbed ``quasiparticles'', which largely behave like non-interacting particles. Their dispersion 
relation ($\omega = \omega({\bf k})$), i.e. the equation linking energy end momentum,  is the fundamental brick to build up the excitaton 
spectrum of the system. 

The concept of \emph{elementary excitations} of a many-particle system can be best understood considering the vibrations of a lattice\cite{Abr}, 
discussed in the next Section.

\section{Quasiparticles}
Let us consider a perfect crystal, and focus on lattice excitations. As long as the amplitudes of its vibrations are small, the 
lattice be regarded as a set of $3N$ \emph{coupled}  harmonic oscillators ($N$ denotes the number of atoms). Introducing normal coordinates, 
the system is reduced to a collection of $3N$ oscillators with frequencies $\omega_i$, and its energy takes the form
\begin{equation}
\label{E}
E=\sum_{i=1}^{3N}\omega_i\left(n_i+\frac{1}{2}\right) \ ,
\end{equation}
where the $n_i$ are integer non-negative numbers. Lattice vibrations can be can be described as a superposition of monochromatic 
plane waves propagating through the crystal, characterized by a wave vector ${\bf k}$ and a frequency $\Omega(\bk)$ which, in general, 
is not a single-valued function of  ${\bf k}$. For small ${\bf k}$, we can have, for example, a linear dependence on the wave vector, i.e.  
$\Omega(\bk)=u(\theta,\phi) |{\bf k}|$. 

An alternate description is based on quantum mechanical wave-particle duality, stating that each plane wave can be associated with a particle 
carrying momentum ${\bf k}$ and energy $\Omega({\bf k})$. According to this picture, an excited state of the lattice can be seen as a collection 
of particles, called {\em phonons}, moving freely within the volume of the crystal, and its energy is determined  by the number of phonons in 
the state $i$. 

In phonon language an anharmonic term in the energy can be interpreted as a phonon-phonon interaction, giving rise to 
scattering processes or creation of additional phonons.

The description of lattice excitations in terms of phonons is closely  related the description of normal Fermi liquids in terms of quasiparticles, proposed 
by Landau. At near zero temperatures, quasiparticles are the relevant degrees of freedom. However, in the presence of interactions (the anharmonic term, in phonon language), quasiparticles become unstable, as they do not describe the exact eigenstates of the system,  and transitions via decay or scattering lead to a damping of the excitations. Decays occur for high energy excitations, while scattering becomes the dominant damping mechanism with growing number 
of excitations. Using simple phase-space considerations \cite{Landau}, it can be shown that the quasiparticle concept is meaningful only in the 
vicinity of the Fermi surface and at low temperature. The inverse quasiparticle lifetime can be written as
\begin{equation}
\label{lifetime}
\frac{1}{\tau}\approx a (\epsilon-\mu)^2+bT^2 \ ,
\end{equation}
where $\mu$ is the chemical potential and $a,b$ are positive constants. In the vicinity of the Fermi surface and a sufficiently low $T$, quasiparticles are long-lived states, and pretty much behave  like stable particles.  Equation (\ref{lifetime}) states the domain of applicability of Landau theory. 
Within this range \emph{all the ground state properties of the system can be described through the  distribution function associated with the ground state 
of non interacting quasiparticles}, which in turn corresponds to the corresponding state of  \emph{interacting} physical particles. 

Adding an interaction, or raising the temperature, leads to a change in the distribution function near the Fermi surface. As quasiparticles are elementary excitations of the system as a whole, Landau theory assumes $\varepsilon\rb{{\bf p}}$ to be a \emph{functional} of quasiparticle distribution function, i.e.
\[
\varepsilon\rb{{\bf p}}=\varepsilon\rb{{\bf p}}(n\rb{{\bf p}}) \ .
\]
As a consequence, the total energy of the system can be written as,
\[
E=E_0+\sum_{k \sigma}\varepsilon_{\sigma}(\bk)\delta n_{\sigma}(\bk) \ .
\]
Note that, as in the phonon description of lattice excitations, the information on the ground state energy $E_{0}$ is totally lost.
 The definition of quasiparticle energy can be obtained performing a functional derivative of $E$
\[
\varepsilon_{\sigma}(\bk)=\frac{\partial E}{\partial n_{\sigma}(\bk)} \ .
\]
On the other hand, the distribution function $n_\sigma(\bk)$ at equilibrium is determined by minimizing the total energy
\[
\delta E(n)=T\delta S(n)+\mu\delta N(n) \ .
\]
In the above equation, $S(n)$ denotes the entropy, which is the same as in the free gas,  its form being derived from purely combinatorial considerations\footnote{A strong assumption that has to be done is that the number of interacting states is the same as the number of states in the non-interacting system.}. As a consequence, one finds the ralations
\[
\delta E=\sum_{k \sigma}\epsilon_{\sigma}[n\rb{\sigma}({\bf k})]\delta n_{\sigma}(\bk) \ , \ 
\delta N=\sum_{k \sigma}\delta n_{\sigma}(\bk) \ , \   \delta S=-\sum_{k \sigma}\delta n_{\sigma}(\bk)\textrm{ln}\frac{n_{\sigma}(\bk)}{1-n_{\sigma}(\bk)},
\]  
leading to an implicit equation for the quasiparticle energy, as the Fermi-Dirac distribution 
\begin{equation}\label{n_eq}
	n\rb{\sigma}({\bf k})=\left\{ 1+ \eu\rp{ \beta  \{ \varepsilon\rb{\sigma}[n\rb{\sigma}(\bf{k})] - \mu  \}  }  \right\}^{-1} \ ,
\end{equation}
in turn depends on $\varepsilon\rb{\sigma}[n\rb{\sigma}(\bf{k})]$.

\section{Landau Parameters}
The quasiparticle energy is an additive quantity, as long as the number of quasiparticles is very small with respect to $N$. Under these conditions, their interaction probability is negligible and quasiparticles can be assumed to be free, the quasiparticle energy is defined as the first functional derivative 
with respect to  $\delta n$ and terms $O(\delta n^2)$ can be neglected. 

Switching on quasiparticle interactions, or warming up the system, we need to also include the terms $O(\delta n^2)$. As a result, we obtain 
\begin{align}
\label{LL1}
& \delta E = \sum_{k \sigma}\varepsilon^0_{\sigma}(\bk)\delta n_{\sigma}(\bk)+\frac{1}{2}
\sum_{\sigma\sigma' k k'}f_{\sigma\sigma'}({\bf k},{\bf k}')\delta n_{\sigma}(\bk)\delta_{\sigma'}(\bk')  \ , \\
\label{LL2}
& f_{\sigma\sigma'}({\bf k},{\bf k}') =\frac{\delta^2 E}{\delta n_{\sigma}(\bk)\delta n_{\sigma'}(\bk') } \ ,
\end{align}
where $\varepsilon^0_\bk$ is the energy of non interacting quasiparticle and $f_{\s\s'}(\bk,\bk')$ describes the interaction between two quasiparticles carrying momenta $\bk$ and $\bk'$. Since $\delta n\rb{s}(\bk)=n\rb{s}(\bk) -n^0(\bk)$ is appreciably different from zero only near the Fermi surface, all the momenta involved in $f_{\s\s'}(\bk,\bk')$ are restricted to the Fermi sphere $\bk\sim\bk'\sim \bk\rb{F}$. As a consequence, at fixed density, i.e. fixed
$\bk\rb{F}$,  $f_{\s\s'}(\cos \theta)$ only depends now on the relative orientation of the two momenta.

The function $f_{\sigma\sigma'}({\bf k},{\bf k}')$ must reflect the main properties of the interaction between bare particles. In the case of electrons, 
for example, it must be spherically symmetric and involve both the direct and exchange terms. In the case of nuclear matter, on the other hand, the composite nature of nucleons results in the appearance of  a non central component, non-diagonal in spin space, that is usually taken into account through a tensor term
according to 
\[
f_{\sigma\sigma'}(\cos\theta)=f(q) I+g(q) ( {\boldsymbol \sigma}_1\cdot {\boldsymbol \sigma}_2 ) + h(q)S_{12}(\hat{\bq}) \ ,
\]  
with ${\bf q} = ({\bf k} - {\bf k}^\prime)/2$ and
\[
S_{12}(\hat{\bq}) = 3 \frac{ ( {\boldsymbol \sigma}_1 \cdot {\bf q} )( {\boldsymbol \sigma}_2 \cdot {\bf q} ) }{ |{\bf q}|^2 } - 
({\boldsymbol \sigma}_1 \cdot {\boldsymbol \sigma}_2 ) \ .
\]
The three functions $f(q),g(q)$ and $h(q)$ embody all features of $NN$ interaction,  within the operatorial representation employed in 
our work, i.e  the Argonne $v^6$. Note that here the tensor term in $v^6$ has its counterpart in momentum space. 

Landau parameters are generally defined as the projections of $f_{\sigma\sigma'}$ on the basis of states of definite angular momentum. 
Expanding $f_{\sigma\sigma'}(\cos \theta)$ in Legendre polynomials, and multiplying by the density of states at the Fermi surface, $N(0)$, 
one obtains the dimensionless quantities
\[
F_{\sigma\sigma'}(\cos\theta)=N(0)f_{\s\s'}(\cos\theta)=\sum_{\ell=0}^{\infty}\underbrace{F^{\sigma\sigma'}_\ell}_{} P_{\ell}(\cos\theta).
\]
The set of parameters $F^{\s\s'}_\ell={F_\ell,G_\ell,H_\ell}$ are known as \emph{Landau} parameters. They provide a link between the microscopic 
dynamics and macroscopic observables, and require either phenomenological information or a local microscopic hamiltonian to be determined. 
While not being directly obtainable from Landau theory itself, the Landau parameters have very important implications. Depending on their values, 
the system exhibits different phases,  such as giant resonances or pion condensation.

In the case of a purely central interaction, the derivation of macroscopic observables at equilibrium is well known \cite{Landau}. As an example, let us  consider a scalar probe inducing density fluctuations of the system. In coordinate space this amounts to compression and rarefaction of matter density, 
similar to the propagation of a sound wave. In momentum space, on the other hand, it corresponds to an oscillation of the Fermi surface, that "inhales and exhales". This radial perturbation can be written in the form $\delta \bk_F=\lambda\bk_F$, independent of spin. In this case, the compressibility of the system is found to be
\be\label{Klandau}
\mathcal{K}\equiv-\frac{1}{V}\left(\frac{\partial V}{\partial P}\right)_T=\frac{N(0)}{\rho^2(1+F_{0})} \ .
\ee
Now, suppose to add an external magnetic field ${\bf B}$, oriented along the $z$-axis,  which couples with the particle spin through a term  
$\sim \mu \ (\bm{\s}\cdot {\bf B})$, where $\mu$ denotes the magnetic moment. A measure of the strength of the response is provided by the magnetic susceptibility, defined as
\be\label{chilandau}
\chi_{\alpha\beta}=\left.\frac{\partial M_{\alpha}}{\partial B_{\beta}}\right|_{B=0},
\ee
where $M$ is the total magnetization of the system and the indices $\alpha$ and $\beta$ are refer to the euclidean space. Within Landau theory $\chi$ is 
determined by the different behavior of  spin-up and spin-down particles, which oscillates with opposite of phase $\delta \bk_F^{\uparrow/\downarrow}=\pm\bk_F^{\uparrow/\downarrow}$. Note, however, that in the presence of non central forces, the Fermi surface is no longer spherical. In addition, the 
quasiparticle effective charges may be modified, implying in turn that the effective magnetic moment may have components which are not scalar under rotation, i.e. it may have different values in different directions. The general form of  $\mu$, taking into account the effects of tensor forces 
is \cite{Olss1} 
\[
\mu_{\alpha \beta}=\mu_n\delta_{\alpha \beta}+\frac{3}{2}\mu_T\left(\frac{\bk_\alpha\bk_\beta}{\bk^2}-\frac{\delta_{\alpha \beta}}{3}\right),
\]
where $\mu$ is the usual magnetic moment, while $\mu_T$ represents the magnitude of non-diagonal term. The experimental value of $\mu_T$ has 
not been determined yet. In Ref.  \cite{Arima}, Arima and collaborators proposed the value $|\mu_T|\sim 0.049 |\mu|$.

The calculation of the susceptibility in the presence of non central interactions was first discussed by Haensel \cite{Haensel}, and then carried out in  
a very general fashion by the authors of Ref.  \cite{Olss1}, who used different forms of both the tensor effective interaction and the magnetic moment. 
These results are typically expanded in powers of $\mu_T$. However, this quantity is still largely unknown, and seems to be very small. 
In a normal system its effect seems to be completely negligible, and has been ignored in our work. Within this approximation, the diagonal element of susceptibility reduces to 
\[
\chi=N(0)\mu^2\frac{1}{1+G_0}\left(1+\frac{1}{8}\frac{1}{1+G_0}\frac{(H_0-H_1)^2}{1+G_2/5}\right) \ .
\]
Note that in the limit $H_l \to 0$, we retrieve the usual expression, suitable for a purely central potential. 

Corrections to the mass of the particles arise from a different kind of oscillation. Consider a displacement of the Fermi sphere by a 
small momentum ${\bf q}$, without changing its size and shape. The energy-density of the system is increased by 
\[
\delta E = \rho \frac{|{\bf q}|^2}{2m} \ .
\]
Assuming that near the Fermi surface the quasiparticle energies can be written using an effective mass expansion, i.e. that  
$\varepsilon^0_{\sigma}(\bk) \approx |{\bf k}|^2/2m$ for $|{\bf k}| \approx k_F$, one can also obtain $\delta E$ from eq. \eqref{LL1}.
Equating the two results, and substituting $N(0) = k_F m^\star/ \pi^2$, leads to
\[
\rho \frac{|{\bf q}|^2}{2m} = \rho \frac{|{\bf q}|^2}{2m^\star} + \frac{1}{6 m^\star} \rho F_1 |{\bf q}|^2 
\]
implying  
\[
m^*=\frac{m}{1+\frac{1}{3}F_1}.
\]

%% file: Chap_Value.tex
\section{Quasiparticle interaction}
As pointed out in the previous Chapter, the quasiparticle interaction must embody all the relevant features of the bare interaction, which 
in this work is written using the six operators needed to take into account the spin-ispspin dependence and the presence of a non central 
component. Obviously, as our analysis is restricted to pure neutron matter, $({\boldsymbol \tau} \cdot {\boldsymbol \tau})=1$, and 
the number of operators can be reduced to three: $\hat{I},({\boldsymbol \sigma}_1 \cdot {\boldsymbol \sigma}_2)$ and $S_{12}(\hat{\bq})$. 
As a consequence, one has to introduce three different sets of Landau parameters, associated with the direct, spin-exchange and tensor terms
of the interaction, respectively. 

WIthin Landau theory, the contribution of second order in $\delta n$ to the energy shift $\delta \mathcal{E}= \mathcal{E}-\mathcal{E}_0$ 
can be written in the form
\bea
\label{E1}
\delta\mathcal{E}_{\rm L} & =& \frac{1}{2V}\sum_{i,j}\sum_{\bk_i,\bk_j}\delta n_{i}(\bk_i)\delta n_j(\bk_j) \\
& \times & \langle ij |\Big\{ \Big[f(q)\hat{I} \Big]+\Big[g(q)  ({\boldsymbol \sigma}_1 \cdot {\boldsymbol \sigma}_2) \Big] +\Big[h(q)S_{12}(\hat{\bq}) \Big] \Big\}|ij\rangle \ ,
\eea
where the sum is extended to quasiparticle pairs in states $ij$ living on the Fermi surface. 

In order to derive the Landau parameters from the effective interaction described in Chapter \ref{CB}, we have to combine the right hand side of the above equation with the corresponding result obtained using $V_{\rm eff}$ and the Hartree-Fock approximation, eq. \eqref{En}. 
Variation of the energy with respect to $\delta n=n-n\rb{FG}$ yields the second order contribution
\bea
\nonumber
\delta\mathcal{E}_{\rm CBF}=&&\!\!\!\!\!\!\!\!\!\!\!\!\frac{1}{2V}\sum_{\bk_i,\bk_j}\sum_{i,j}\delta n_{i}(\bk_i)\delta n_j(\bk_j) \\
\times \langle ij |\Big\{ 
+&\!\!\!\!\!\! \Big[&\!\!\!\!\!\!\int d^{3}r(w_1+w_2)(1-\hat{P}_{\s}\eu\rp{-\iu{\bf q \cdot r}})\hat{I} \Big]+\nonumber\\
+&\!\!\!\!\!\! \Big[&\!\!\!\!\!\!\int d^{3}r(w_3+w_4)(1-\hat{P}_{\s}\eu\rp{-\iu{\bf q \cdot r}})
({\boldsymbol \sigma}_1 \cdot {\boldsymbol \sigma}_2) \Big]+\nonumber\\
+&\!\!\!\!\!\! \Big[&\!\!\!\!\!\!\int d^{3}r(w_5+w_6)(1-\hat{P}_{\s}\eu\rp{-\iu{\bf q \cdot r}})\,S_{12}(\hat{{\bf r}}) \Big]\Big\} | i j  \rangle \ .
\label{E2}
\eea
Note that, owing to to the presence of the spin-exchange operator 
$\hat{P}_{\s} = [1-({\boldsymbol \sigma}_1~\cdot~{\boldsymbol \sigma}_2)]/4$, the operators involved in the above equation,  
$\hat{P}_{\s\tau}\hat{I},\hat{P}_{\s}({\boldsymbol \sigma}_1 \cdot {\boldsymbol \sigma}_2), \ldots$, are different from those appearing 
in eq. \eqref{E1}. Therefore, it turns out to be more convenient considering the matrix elements, rather than the operators themselves.  
For example, in the case of a pair of particle with spins $\uparrow\uparrow$ we require that
\[
\frac{\delta^2 \mathcal{E}_L}{\delta n_{\uparrow}\delta n_{\uparrow}}=\frac{\delta^2 \mathcal{E}_{CBF}}{\delta n_{\uparrow}\delta n_{\uparrow}},
\]
where:
\[
\frac{\delta^2 \mathcal{E}_L}{\delta n_{\uparrow}\delta n_{\uparrow}}=f(q)\underbrace{   \langle \uparrow\uparrow|\hat{I}|\uparrow\uparrow\rangle }_{=1}+g(q)\underbrace{ \langle \uparrow\uparrow|\hat{\vec{\s}}_1\cdot\hat{\vec{\s}}_2|\uparrow\uparrow\rangle}_{=1}+h(q)\underbrace{\langle \uparrow\uparrow|S_{12}(\hat{q})|\uparrow\uparrow\rangle}_{2 P_2(cos\xi)}.
\]
On the other hand, in the $\uparrow\downarrow$ sector we find
\[
\frac{\delta^2 \mathcal{E}_L}{\delta n_{\uparrow}\delta n_{\downarrow}}=f(q)\underbrace{   \langle \uparrow\downarrow|\hat{I}|\uparrow\downarrow\rangle }_{=1}+g(q)\underbrace{ \langle \uparrow\downarrow|\hat{\vec{\s}}_1\cdot\hat{\vec{\s}}_2|\uparrow\downarrow\rangle}_{=-1}+h(q)\underbrace{\langle \uparrow\downarrow|S_{12}(\hat{q})|\uparrow\downarrow\rangle}_{-2 P_2(cos\xi)},
\]
where $P_2(cos\xi)$ is the second Legendre polynomial and $\cos\xi$ is the angle between $\hat{q}$ and the $z$-axis. Using the above results one 
can construct the symmetric and antisymmetric branches according to the standard definition
\bea
&&f^{s}(q)=f(q) \ ,\nonumber\\
&&f^{a}(q)=g(q)+2 h(q)P_2(cos\xi) \ ,
\nonumber
\eea
showing that inclusion of the tensor term results in a change of the interaction strength depending on the direction of the momentum transfer.
\begin{figure}[htbp]
\includegraphics[scale=0.4]{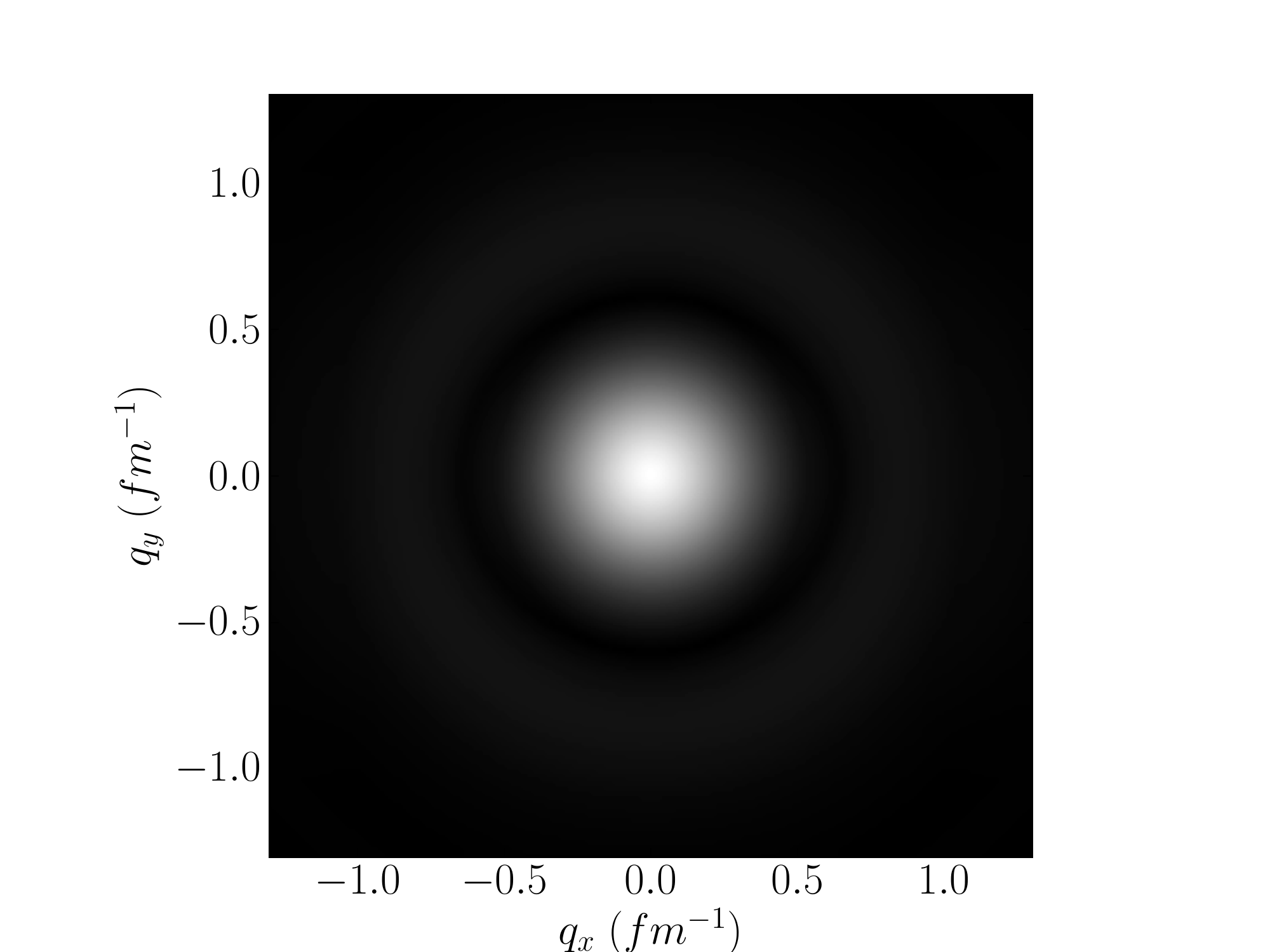}\hspace{-15mm}
\includegraphics[scale=0.4]{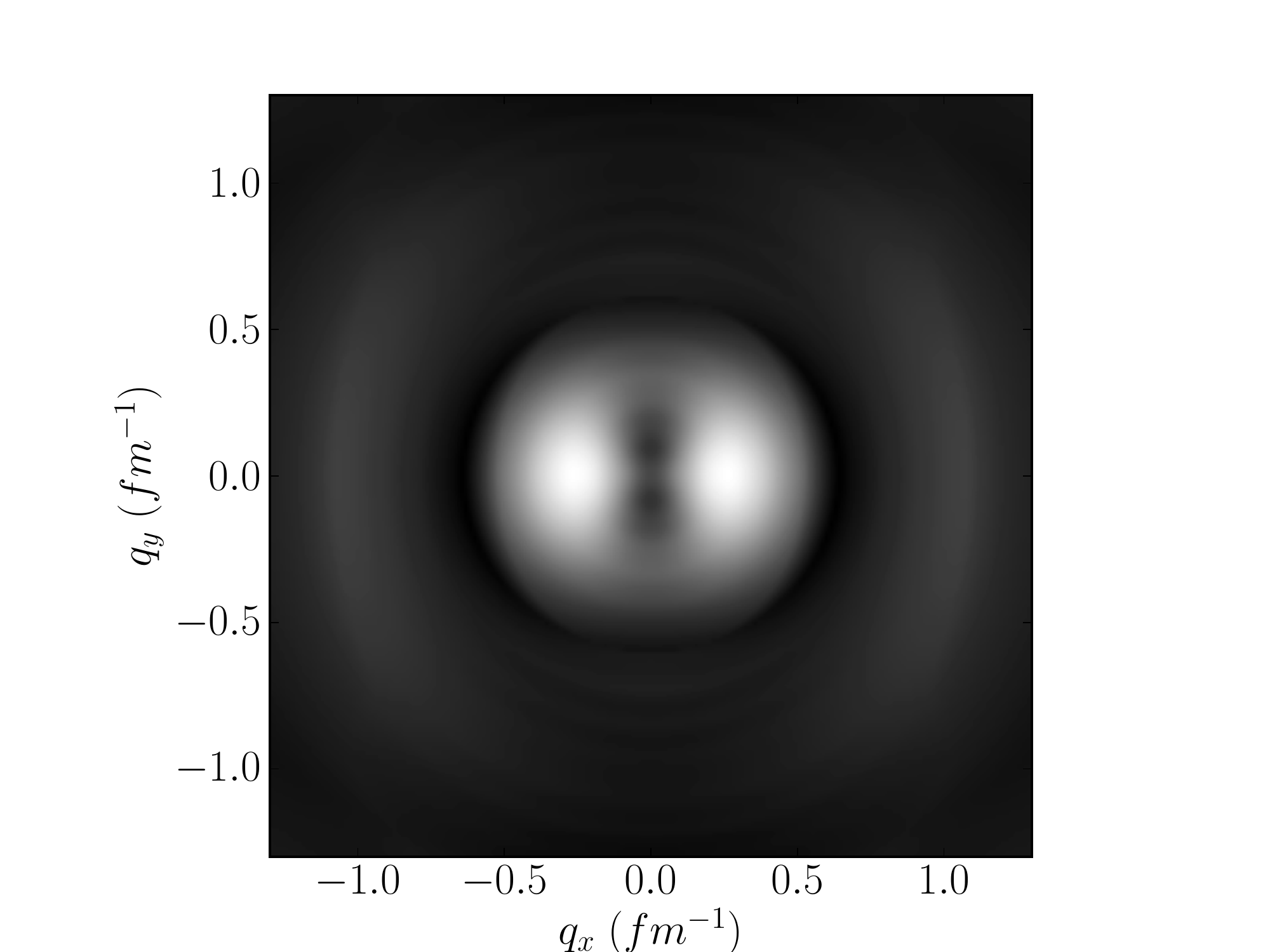}
\caption{Density plots of the functions $f^s$ (left) and $f^a$ (right)  obtained from the matrix elements
of the effective interaction of Chapter \ref{CB}, as a function of the $x$ and $y$ components of the momentum transfer ${\bf q}$. Note that 
the functions are azimuthal symmetric, and shown for  $q_z=0$ plane. The non spherically symmetric dependence arising from the tensor term is clearly visible.}
\label{FTdensity}
\end{figure}
This feature is illustrated in fig. (\ref{FTdensity}), displaying the functions $f^s$ (left) and $f^a$ (right) obtained from the matrix elements
of the effective interaction of Chapter \ref{CB}. 

Carrying out the second functional derivatives of $\mathcal{E}_{CBF}$\footnote{The last term in eq. (\ref{E2}) is the Fourier transform of the tensor operator $S_{12}(\hat{\bq})$. A simple derivation is given in Ref. \cite{FT}, while a more formal discussion can be found in  Appendix \ref{FourierTr}.}, we can readily identify the Landau parameters from
\bea
&&f(q)=\frac{2\pi}{V}\int d|{\bf r}|\, {\bf r}^2\Big[2(w_1+w_2)-g_0(3w_3+3w_4+w_1+w_2)\Big],\nonumber\\
&&g(q)=\frac{2\pi}{V}\int d|{\bf r}|\, {\bf r}^2\Big[2(w_3+w_4)-g_0(-w_1-w_2+w_3+w_4)\Big],\nonumber\\
&&h(q)=-\frac{4\pi}{5}\int d|{\bf r}|\, {\bf r}^2 g_2\Big[w_5+w_6\Big],\nonumber
\eea
with
\[
g_\ell(r,q)=\frac{2l+1}{2}\int_{-1}^{1}\eu\rp{-\iu q r y}P_\ell(y)dy \ \ \ , \ \ \  \eu\rp{-\iu q r y}=\sum_\ell g_\ell P_\ell(y) \ .
\]

\section{Landau parameters from the CBF effective interaction}

\begin{figure}[htbp]
 \hspace{-8mm}
\includegraphics[scale=0.55]{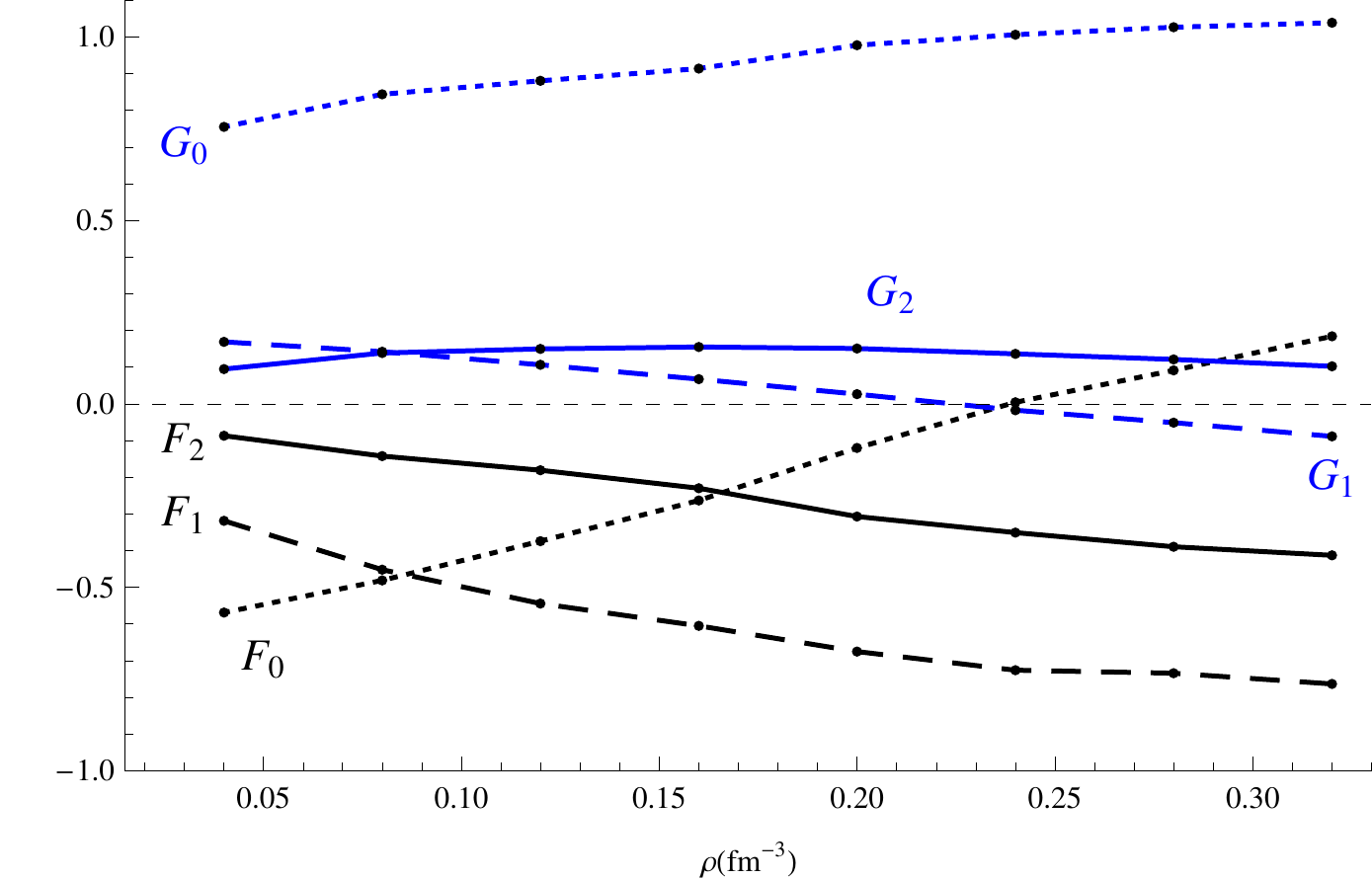}
\includegraphics[scale=0.55]{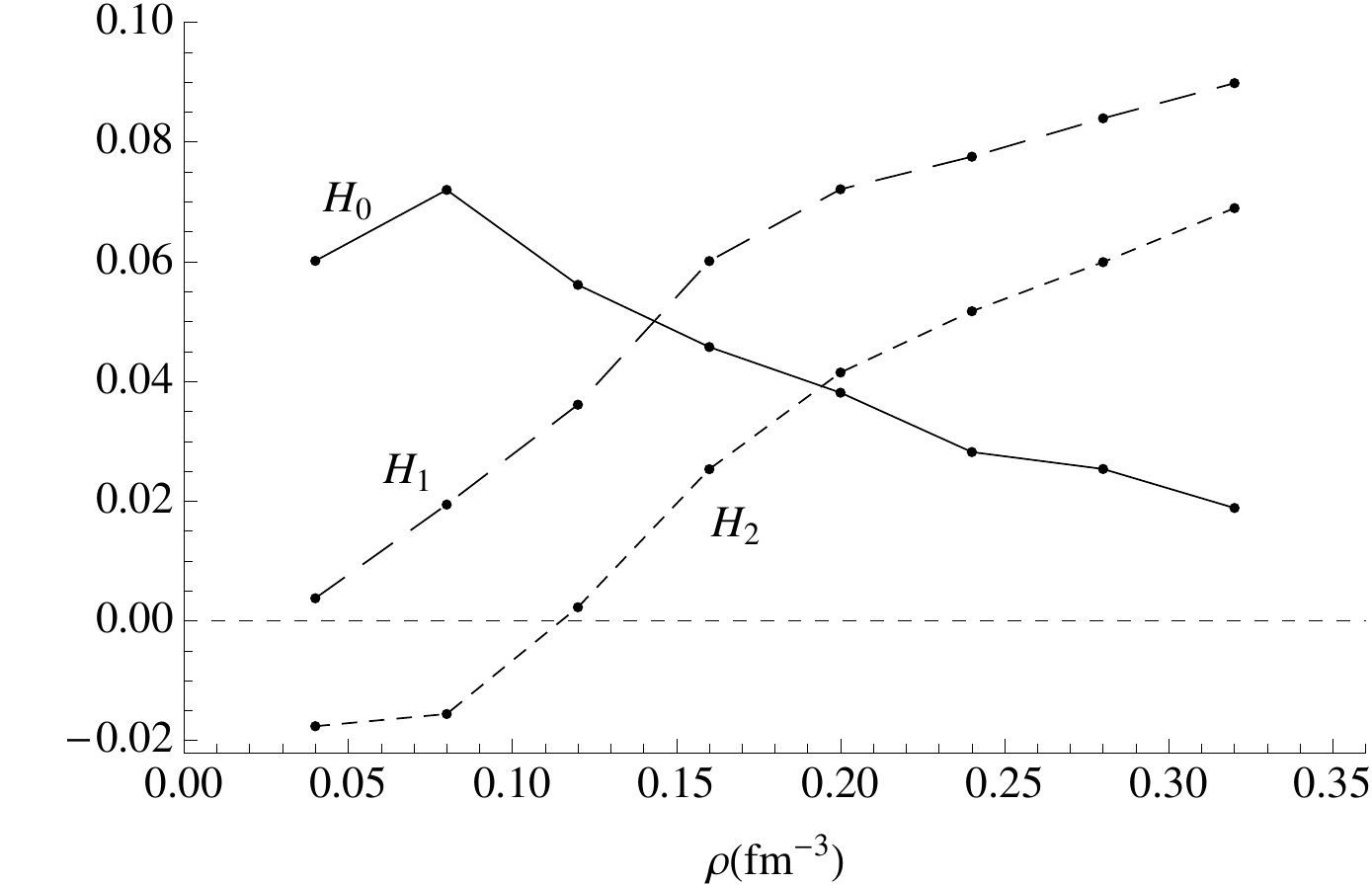}
\caption{ Left panel: density-dependence of the Landau parameters $F_\ell$ and $G_\ell$ of
pure neutron matter obtained from the matrix elements of the
CBF effective interaction. Right panel: same as in the upper panel, ut for the Landau parameters $H_\ell$.}
\label{Values}
\end{figure}

The values of the dimensionless Landau parameters $F_\ell$, $G_\ell$ and $H_\ell$ obtained from the matrix elements of the CBF 
effective interaction are listed in Table \ref{LT1} and \ref{LT2}, for $\ell=0,1,2$. Their density-dependence is displayed in fig. \ref{Values}.

\begin{table}[htbp]
\[
\begin{array}{ccccccc}
\toprule
 \rho(fm^{-3}) & F_0&  F_1  & F_2 &  G_0  &  G_1 & G_2\\
\midrule
0.04 & -0.569 & -0.319 &-0.087      & 0.755 & 0.169& 0.094\\
0.08 &-0.481 & -0.452 & -0.143      & 0.844 & 0.142& 0.138\\
0.12 &-0.374 & -0.544  & -0.180     & 0.881 &  0.107&0.150 \\
0.16 &-0.263 & -0.605 & -0.230  & 0.914 & 0.067&0.155 \\
0.20 &-0.120 & -0.675 &  -0.307  & 0.978 & 0.027& 0.150\\
0.24 &0.004 & -0.726  & -0.351   & 1.006 & -0.017&0.136 \\
0.28 &0.092 & -0.734  &  -0.390  & 1.027 & -0.052&0.121 \\
0.32 &0.184 & -0.763  &  -0.412    & 1.039 & -0.088&0.102 \\
\bottomrule
\end{array}\]
\caption{Landau parameters $F_\ell$ and $G_\ell$ of pure neutron matter obtained from the matrix elements of the
CBF effective interaction (see left panel of fig. \ref{Values}). }
\label{LT1}
\end{table}
 \begin{table}[t]
\[
\begin{array}{cccc}
\toprule
 \rho(fm^{-3})&  H_0 & H_1 & H_2\\
\midrule

 0.04 & 0.060 & 0.004 & -0.018\\
 0.08 & 0.072 & 0.019 & -0.016\\
 0.12 & 0.056 & 0.036 & 0.002\\
 0.16 & 0.046 & 0.060 & 0.025\\
 0.20 & 0.038 & 0.072 & 0.041\\
 0.24 & 0.028 & 0.078 & 0.052\\
 0.28 & 0.025 & 0.084 & 0.060\\
 0.32 & 0.019 & 0.090 & 0.069\\
\bottomrule
\end{array}
\]
\caption{Same as in Table \ref{LT1}, but for the Landau parameters $H_\ell$ (see right panel of fig. \ref{Values}). }
\label{LT2}
\end{table}

Within Landau theory, the parameters of Tables \ref{LT1} and \ref{LT2} provide a description of both the static and dynamical properties 
of pure neutron matter. In the literature, one can find limited sets of Landau parameters, obtained from phenomenology and semi-quantitative 
treatment  \cite{par1,par2,par3}. Formally $F_\ell,G_\ell,H_\ell$ can be related to the {\em in medium} forward scattering amplitude. 
Nevertheless, as pointed out in the previous Chapter, their physical interpretation is deeply related to the static properties of matter. 
One of the major achievements of Landau theory is that \emph{once the Landau parameters has been fixed, all dynamical 
properties can be naturally derived from the static ones}.


In principle, the description of short and long range correlations should be carried out in a consistent fashion, using perturbation theory in 
the basis of correlated states. While static properties can be derived directly from the equation of state (EOS)\footnote{In general, the EOS is a non trivial 
relation linking the thermodynamic variables specifying the state of a physical system.}, the calculation of the dynamic responses involves non 
trivial difficulties. A prominent issue is the treatment of collective modes, whose contribution becomes more and more important as the momentum transfer decreases. In particular, the calculation of the response function at $|\bq| \to  0$ within the CBF formalism requires the description of the propagation of correlated one particle-one hole states  through the medium. 
A consistent description of the response of isospin symmetric nuclear matter, performed using the CBF effective interaction, has been reported 
in ref. \cite{Farina}. However the analysis is restricted to the density channel. 

In this Thesis, we adopt a different approach. The dynamic response is no longer obtained microscopically, summing the contribution of all 
particle-hole pairs.  We take instead the point of view Landau theory, in which the degrees of freedom are the quasiparticles, representing 
the elementary excitations of the system. This approach is both simpler and more general than that discussed in Ref.  \cite{Farina}, as it is
capable to treat the coherent and incoherent contributions to the response on equal footing.
All the main features of CBF interaction, reflecting both the nature of the bare interaction and correlation effects, are summarized by few parameters,  
the values of which must satisfy a set of stability conditions.

The main effect taken into account by the CBF effective interaction is the screening of the nuclear medium, arising from short range correlations, that 
leads to a strong suppression of the repulsive core of the bare $NN$ potential.
This pattern is exactly what we observe in the Landau parameters.
The large value of $G_0$, compared to $F_0$, reflects the negative value of $f_{\up\down}< 0$. This means that the effective interaction between 
particles with opposite spin is attractive, thanks to Pauli principle that keeps particles with parallel spin apart. In a different system, such as $^3$He, 
this lead to the opposite relation, $F_0 \gg G_0$, since the interaction is not screened, and particles with opposite spin experience the repulsive 
short-range interaction. As a consequence $f_{\up\down}$ becomes more repulsive than $f_{\up\up}$. 
Within the CBF scheme, the repulsive core is screened, and the attractive interaction emerges. Also note that at low density the $\up\down$ channel 
is the most important, and greatly influence the values of Landau parameters.

Increasing the density (or the pressure) $F_0$ increases more rapidly than the other parameters, as it is strongly related to the compressibility of the system. Its growth is not as fast as in  $^3$He, where the rapid variation is again attributed to the repulsive core of the potential 
($|f_{\up\down}^{^3\textrm{He}}|\gg |f_{\up\down}^{NM}|$). At  $\rho \approx 0.16 {\rm fm}^{-3}$, the equilibrium density of 
symmetric nuclear matter, fig. \ref{Values} shows that $G_0 \sim 0$, or $|f_{\up\up}|\sim|f_{\up\down}|$, implying that the repulsive interaction between particles with parallel spin has the same magnitude as the attractive interaction. Increasing the density $|f_{\up\up}|$ becomes larger,  in spite of screening.

Let us consider the harmonic $\ell=1$. The values of the Landau parameters are almost constant in the range shown in Fig. \ref{Values}, since the centrifugal barrier prevents the particles from coming close to one another. In particular, $F_1<0$  has a direct physical interpretation, since it is related to the effective mass. 

For incoherent excitations the stability conditions $F_\ell,G_\ell>-(2\ell+1)$ are satisfied. Nevertheless, in the low density limit the attraction in 
the  $\up\down$ channel appears to be dominant, and may lead to the appearance of an instability region,
 in which $F_0< -1$ for $\rho<0.04\, \textrm{fm}^{-3}$. 

We now turn to the discussion of collective modes. The occurrence of these excitations depends on several relations between Landau parameters
that were extended to the case of nuclear matter, with the inclusion of non central interactions at the end of 70s \cite{friman}. Comparing the time scale of collective modes with the quasiparticlep mean free path, ($\lambda$), we can distinguish al least two different classes. When the length scale is longer 
than $\lambda$, hydrodynamics may be used. In this limit the only relevant variables are density fluctuations and the average fluid velocity, that 
correspond to $\ell=0$ and $\ell=1$ distortions of the Fermi surface. The distortions arising from all higher harmonics are removed by collisions. 
In the opposite limit, i.e. short length scale, the higher harmonics persist, because of the absence of collisions, and a detailed description in term of 
quasiparticle distribution function is required \cite{Abr}. The range $-1<F_0<0$ corresponds to strong Landau damping in the collisionless limit, 
since $v_{C}<v_F$, where $v_C$ denotes the velocity of the collective mode. The collective mode are destroyed by incoherent excitation. Nevertheless,  
we have undamped oscillation in the hydrodynamic limit. For small positive value of $F_0\geq0$ the occurrence of the collective mode depends on 
the  magnitude of $F_0$. A \emph{sufficient} condition for stability was first found by Haensel, who studied the positive solutions of the the 
zero sound dispersion relation \cite{Haensel1}. Considering the harmonics $\ell=0,1$ he obtained 
\be\label{0}
F_0>\frac{-F_1}{1+\frac{1}{3}F_1}.
\ee
Another \emph{sufficient} condition was derived by Mermin \cite{Mer}:
\be\label{aq}
\sum_\ell\frac{F_\ell}{1+\frac{F_\ell}{2\ell+1}}>0 \ \ , \ \  \textrm{for}\,\ell=0,1 \ \ , \ \ \frac{F_0}{1+F_0}>\frac{-F_1}{1+\frac{1}{3}F_1}.
\ee
However, the above relation is more restrictive than the forst one for $F_0>0$\footnote{We remind the reader that for $F_0<0$ solutions can exist, but are damped, as  $v_{C}<v_F$}. Note that with $F$ we denote either $F$ or $G$. 
In the same paper  \cite{Mer}, Mermin demonstrates one more relation between Landau parameters\footnote{While this is true for neutron matter, for nuclear matter there are different isospin states and a general condition involves all channels \cite{Liu,Gogny}.}:
\be\label{b}
\sum_l\left(\frac{F_\ell}{1+\frac{F_\ell}{2\ell+1}}+\frac{G_\ell}{1+\frac{G_\ell}{2\ell+1}}\right)=0.
\ee
In general, eq. (\ref{b}) implies that at least one set parameters $F,G$ satisfies condition (\ref{aq}). Hence, at $T=0$ at least one of the two channels must 
exhibit a zero sound wave. The values reported in fig. (\ref{Values}) satisfy the condition (\ref{aq}): zero sound always occur in the spin 
channel. On the other hand, in the density channel the condition of eq. (\ref{0}) is fulfilled for $\rho>0.32$ only.

\section{Static properties of neutron matter}

Before discussing the application of Landau theory to the calculation of the dynamic structure functions, in this Section we report the results of the
calculations of a variety of equilibrium properties of neutron matter at $T=0$.

We will focus on effective mass, $m^*$, isothermal compressibility, $\mathcal{\chi}^\rho$, and magnetic susceptibility, $\chi^\sigma$, which, in the static limit, can be related to the energy-density, density and spin-density responses, respectively. As these quantities can be obtained both  from matrix elements of the CBF effective interaction and from the Landau parameters
listed in Tables \ref{LT1} and \ref{LT2}, the analysis discussed in this Section provides a valuable consistency test of our approach.

\subsubsection{Single-Particle Spectrum and effective mass}
Within the CBF approach, starting from the expression of the energy
\begin{equation}\label{energy}
\mathcal{E}=\sum_{{\bf p},\sigma}\frac{{\bf p}^2}{2m}n_{\sigma}({\bf p})+\underbrace{\frac{1}{2 }\sum_{{\bf p},{\bf p}'}\sum_{\sigma\sigma'}\int d^3\br\Big[A(r)-B(r)\eu\rp{\iu {\bf p\cdot r} }\eu\rp{-\iu {\bf p}'{\bf \cdot r}}\Big]n_{\s}({\bf p})n_{\s'}({\bf p}')}_{\mbox{$\sum_{j>i}\langle ij|w^{eff}_{12}(\br)|ij\rangle_a$}} .
\end{equation}
we can define, the single-particle spectrum in Hartree-Fock approximation as
\begin{eqnarray}\label{a}
e_{\s}(p)&=&T(p)+U(p)=\frac{p^2}{2m}+2\sum_{j>i}\delta(|{\bf p}_i|-p)\delta_{\s_i,\s}\langle ij|w^{eff}_{12}(\br)|ij\rangle_a \nonumber\\
&=&\frac{p^2}{2m}+\int \frac{d\Omega_p}{4\pi}\sum_{{\bf p}'}\sum_{\sigma'}\int d^3\br\Big[A(r)-B(r)\eu\rp{\iu {\bf p\cdot r} }\eu\rp{-\iu {\bf p}'{\bf \cdot r}}\Big]n_{\s'}({\bf p}') \ .
\end{eqnarray}
 The effective mass is then defined as
\[
m^*=p\left[\frac{d e_{\s}(p) }{d p}\right]^{-1} \ .
\]
Within the CBF effective interaction approach, the potential is perfectly known and well behaved, and we can apply the above expression without any 
problems. Hence, the effective mass is defined for all values of the momentum, $p$\footnote{
The derivative with respsct to $p$ is performed as follows
\begin{eqnarray}
\bm{\nabla}_{\bf p} U(p)\!\!\!&=&\!\!\!\bm{\nabla}_{\bf p}\sum_{{\bf p}'} \Big(V(0)-V({\bf p-p}')\Big)n({\bf p}')\nonumber\\
&=&\!\!\!-\sum_{{\bf p}'}\bm{\nabla}_{\bf p}V({\bf p-p}')n({\bf p}')=-\int d^3 {\bf r} B(r) \left(\bm{\nabla}_{\bf p} \eu\rp{\iu {\bf p\cdot r}}\right)\sum_{{\bf p}'} \eu\rp{\iu {\bf p'\cdot r}}n({\bf p}')\nonumber\\
&=&\!\!\!-\bm{\nabla}_{\bf p}\int dr r^2B(r) \left(\frac{\partial J_0(p r)}{\partial p} \right)\frac{N}{\nu}\ell(p_F r) ] \ .\nonumber
\end{eqnarray}
}. 
On the other hand, starting from eq. (\ref{energy}) we can vary the distribution function to obtain an estimate of the energy in the Landau scheme. 
Note that this is a general procedure, in which the interaction is unknown, and can be different in different systems. 
The most convenient procedure consists in changing variable, so that the derivative is applied to $\delta n$,  which is different from zero 
only near $p=p\rb{F}$\footnote{In this case, we perform a change of variable and then use the Landau ansatz $\partial n/\partial \epsilon_p=-\delta(\epsilon^0_p-\epsilon_F)$:
\begin{eqnarray}
\bm{\nabla}_{\bf p}U(p)&=&\bm{\nabla}_{\bf p}\sum_{{\bf p}'} \Big(V(0)-V({\bf p-p}')\Big)n({\bf p}')=\overbrace{....}^{{\bf q}={\bf p-p}'}=\nonumber\\
&=&-\sum_{{\bf q}}V({\bf q})\bm{\nabla}_{\bf p} n({\bf p-q})=-\int d^3 {\bf r} B(r)\sum_{{\bf q}}V({\bf q})\frac{\partial n({\bf p-q})}{\partial \epsilon_{{\bf p-q}}}\underbrace{\bm{\nabla}_{\bf p}\epsilon_{{\bf p-q}}}_{\mbox{$\vec{v}_{{\bf p-q}}$}}\nonumber\\
&=&\overbrace{....}^{{\bf p}'={\bf p-q}}=\int d^3 {\bf r} B(r)\sum_{{\bf p}'}V({\bf p-p}')\delta(\epsilon^0_{{\bf p}'}-\epsilon_F)\vec{v}^0_{{\bf p}'}\nonumber\\
&=&\frac{F_1^S}{3}\vec{v}_{{\bf p}}.\nonumber
\end{eqnarray} 
}
The inset of fig (\ref{Static}) the ratio between effective and bare neutron mass, $(m^*/m)$, evaluated at $p=p_F$, as a function of matter density. 
It clearly appears that the results obtained from the CBF single particle energies (dots) and from Landau theory (solid line) are the same 
within the numerical accuracy. 

\subsubsection{Compressibility}
Compressibility provides a measure of the ``elasticity'' of a system, determining its response to a density compression. 
Classically, it is related to the speed of sound $v_s\approx\sqrt{K/\rho}$ at equilibrium. From the definition 
\[
\mathcal{K}=-\frac{1}{V}\left(\frac{\pd V}{\pd P}\right)_T  \ ,
\]
through standard thermodynamic relations one obtains
\[
\frac{1}{\mathcal{K}}=\rho\Big[\rho\frac{\partial^2}{\partial \rho^2}+2\frac{\partial}{\partial\rho} \Big]\mathcal{E} \ ,
\]
where $\mathcal{E}=E/N$ denotes the energy per particle\footnote{In studies of isospin symmetric nuclear matter it is useful to introduce 
the so called ``compressibility modulus'', defined as 
\be\label{Kmodulus}
K_{\infty}=\left.9 \rho^2\frac{\partial^2 \mathcal{E}}{\partial \rho^2}\right|_{\rho=\rho_0} \ .
\ee
From the above equation, it follows that $K_{\infty}\approx K$ at saturation density ($\rho_{0}=0.16$ fm$^{-3}$),  where the first derivative 
of $\mathcal{E}$ vanishes. Note that interpretation of $K_{\infty}$ is strictly associated with symmetric matter,  
since the energy of pure neutron matter does not exhibit a minimum. }.

\begin{figure}[htbp]
\hspace{-8mm}
\includegraphics[scale=0.45]{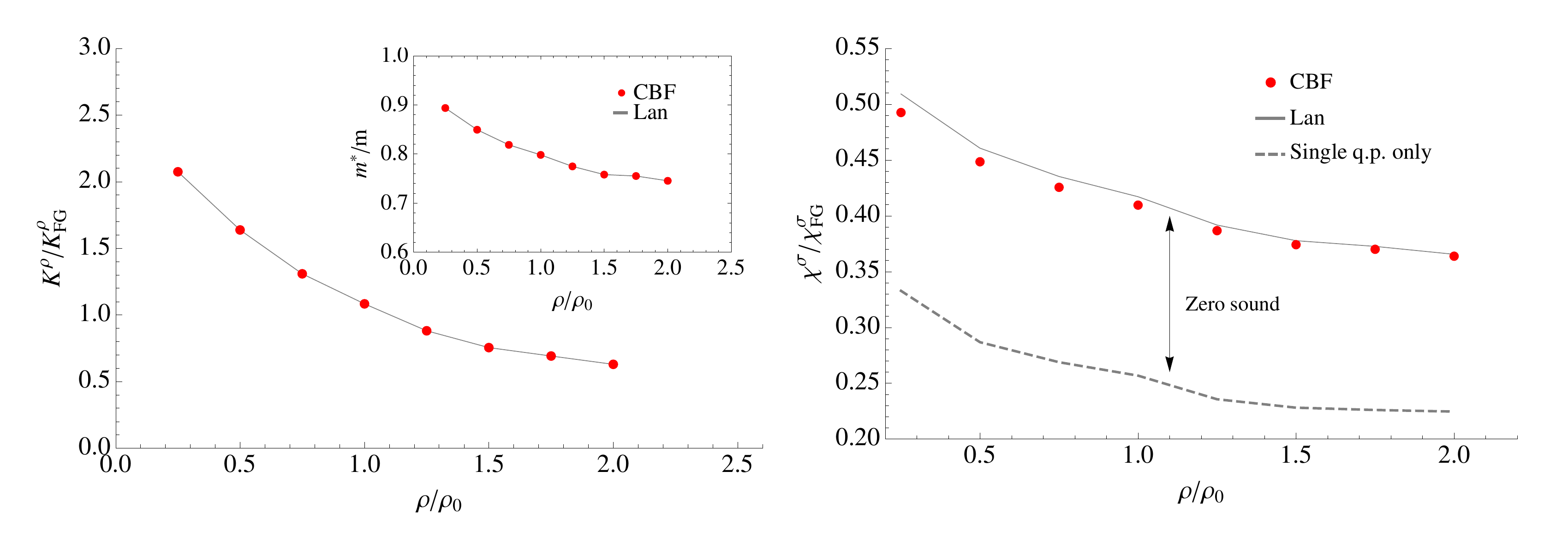}
\caption{Left panel: Compressibility of neutron matter, normalized to its Fermi gas value, as function of density in units of $\rho_0=0.16 $ fm$^{-3}$.
The dots and the solid line correspond to the results obtained from Landau's theory, eq.(\ref{Klandau}), and the equation of state computed using the CBF effective interaction, eq.(\ref{energy}), respectively. Right panel: same as in the left panel, but for for the spin susceptibility, $\chi^\sigma$. The dashed line shows the
susceptibility obtained from the dynamic spin structure function including only the incoherent contribution.  The inset of the left panel shows the density dependence of the ratio between effective
and bare neutron mass.}
\label{Static}
\end{figure}

%
Using the above definition we can easily evaluate $K$ from eq. (\ref{energy}) and compare it with the result obtained from 
the corresponding definition of Landau theory, eq. (\ref{Klandau}). 
The comparison shown in the left panel of fig. (\ref{Static}), indicates that the results of the two approaches are in perfect agreement 
with one another over a broad density range.

\subsubsection{Spin susceptibility}
Another important thermodynamical property is the spin susceptibility, that measures the response to an external magnetic field under
the assumption that orbital effects can be safely ignored (see eq. \ref{chilandau}). Here, we outline the standard procedure to evaluate  
the spin susceptibility  within the CBF framework.
The energy per particle of asymmetric nuclear matter can be accurately approximated using the expression 
\be
\frac{1}{N} \ E(\alpha,\beta,\gamma) = E_0 + E_\sigma \alpha^2 + E_\tau \beta^2 + 
 E_{\sigma\tau} \gamma^2 \ ,
\ee
with
\bea
\nonumber
\alpha & = & (x_3-x_4) + (x_1-x_2) \\ 
\beta & = & (x_3+x_4) - (x_1+x_2)  \\
\nonumber
\gamma & = & (x_3-x_4) - (x_1-x_2) \ .
\eea
The above coefficients define the spin-isospin content of matter: for example, $x_\lambda = 1/4$ for all values of $\lambda$ yields 
$E/N=E_{{\rm SNM}}=E_0$, while pure neutron matter corresponds to $x_1=x_2=0$ and $x_3=x_4=1/2$,  
$E/N=E_{{\rm PNM}}=E_0 + E_\tau$. Note that this obviously implies that $E_\tau$ can be identified with the symmetry energy.

Let us consider fully spin-polarized neutron matter. The two degenerate ground states, corresponding to $x_3=1$ 
and $x_4=0$ ($\alpha=1$, spin-up) and $x_3=0$ and $x_4=1$ ($\alpha= -1$, spin-down), respectively,  have energy 
\be
E^\uparrow = E^\downarrow = E_{{\rm PNM}} + \widetilde{E}_\sigma \ ,
\label{Epol}
\ee
with $\widetilde{E}_\sigma = E_\sigma + E_{\sigma\tau}$. For arbitrary polarization, $\alpha$, the 
energy can be obtained from the expansion 
\be
E(\alpha) = E(0) + \left. \frac{\partial E}{\partial \alpha} \right|_{\alpha=0} \alpha 
 + \frac{1}{2}  \left. \frac{\partial^2 E}{\partial \alpha^2} \right|_{\alpha=0} \alpha^2 
 + \ldots  \ .
\label{alpha:exp} 
\ee
As $E$ must be an even function of $\alpha$ (see eq.(\ref{Epol})), the linear term in the above expansion must vanish. 
Hence, neglecting terms of order $\alpha^3$, we can write
\be
\Delta E = E(\alpha) - E(0) = \frac{1}{2}  
\left. \frac{\partial^2 E}{\partial \alpha^2} \right|_{\alpha=0} \alpha^2 \ .
\ee
In the presence of a uniform magnetic field ${\bf B}$ the energy of the system becomes
\be
E_B(\alpha) = E(\alpha) - \alpha \mu B , 
\ee
where $B$ denotes the magnitude of the external field, the direction of which is chosen along the 
spin quantization axis,  and $\mu$ is the neutron magnetic moment.

Assuming that equilibrium corresponds to $\alpha=\alpha_0$, i.e. that 
\be
\left. \frac{\partial E}{\partial \alpha} \right|_{\alpha=\alpha_0} - \mu B = 0 \ ,
\ee
we obtain
\be
\alpha_0 = \mu B \left( \frac{\partial^2 E}{\partial \alpha^2} \right)^{-1}_{\alpha=0} \ .
\ee
From the definitions of the total magnetization 
\be
M = \mu (\rho_3 - \rho_4) = \mu \alpha_0 \rho = \mu^2 
\left( \frac{\partial^2 E}{\partial \alpha^2} \right)^{-1}_{\alpha=0} B \rho \ ,
\ee
and of the spin susceptibility $\chi$
\be
M = \chi B \ ,
\ee
we finally obtain
\be
\chi = \mu^2 \left( \frac{\partial^2 E}{\partial \alpha^2} \right)^{-1}_{\alpha=0} \rho = 
\mu^2 \frac{1}{2(E^\uparrow - E_{{\rm PNM}})} \ \rho \ .
\label{def:chi}
\ee
In the right panel of fig. \ref{Static} we compare the susceptibility calculated with eqs. (\ref{chilandau}) and (\ref{def:chi}), normalized to 
the Fermi gas result.  As for the compressibility, the two approaches are in very close agreement. It has to be pointed out that in the static limit the contribution of the Landau parameters associated with the tensor interaction, $H_\ell$ (listed in Table \ref{LT2}) turns out to be less than 0.1\%, 
and their effects are hardly visible. Therefore, we can further simplify the calculations, neglecting non-central contributions in spin channel
altogether. This approximation does not spoil the level of accuracy of our result. The dashed line in 
fig. (\ref{Static}) shows the contribution of single quasiparticle excitation (see section \ref{sumrule} below). This means that in the static 
limit a large fraction of the response comes from the excitation of a collective mode. 

\begin{figure}[htbp]
\hspace{-12mm}
\includegraphics[scale=0.55]{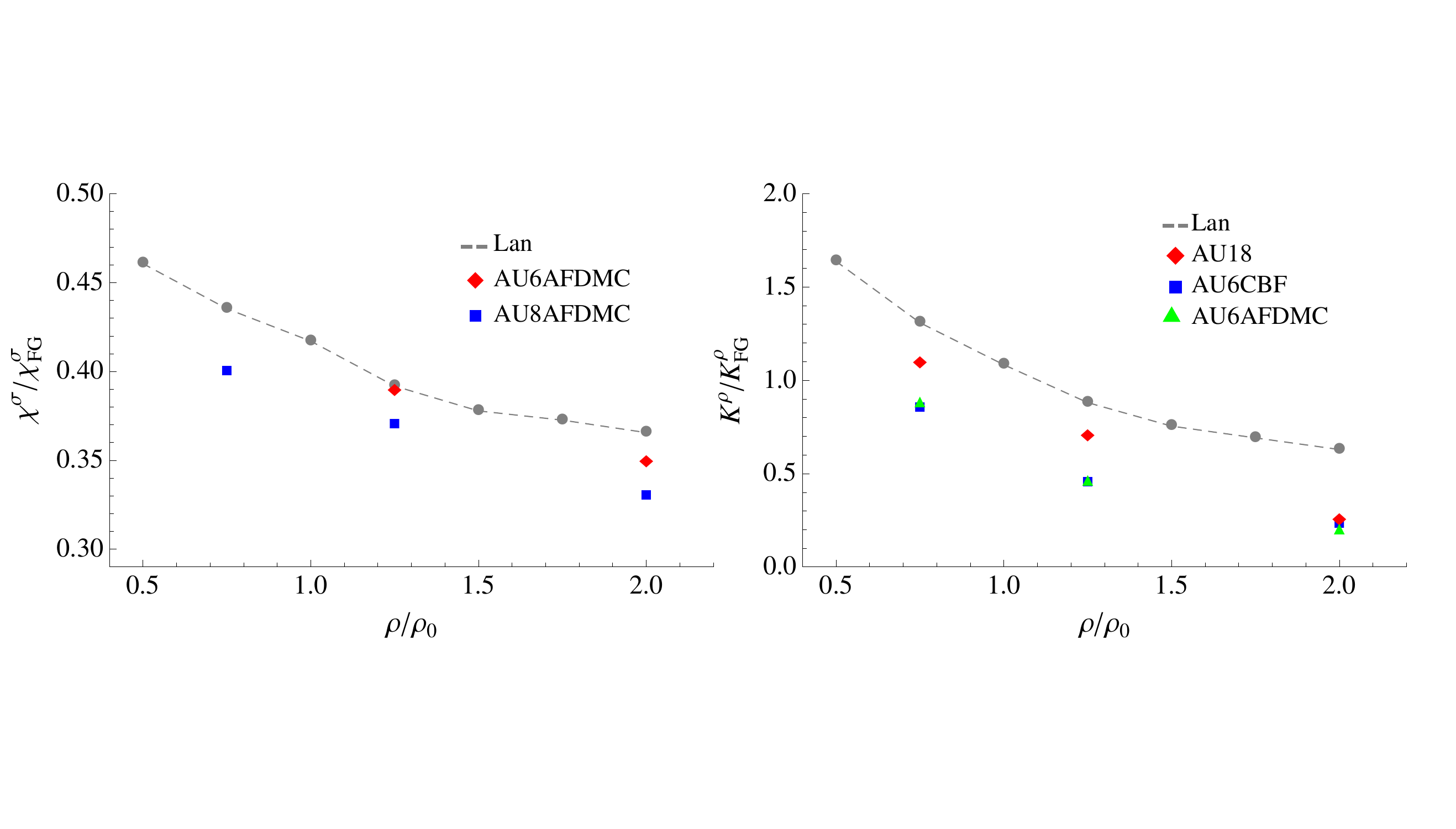}
\caption{Left panel: comparison  between the compressibility
 computed within Landau's theory and the corresponding results obtained using the variational FHNC-
SOC approach. Right panel: comparison
 between the spin susceptibility computed within Landau's theory and the corresponding
results obtained using the Auxiliary Field Diffusion Monte Carlo approach.}
\label{Stat_com}
\end{figure}

In the left panel of Fig. \ref{Stat_com}, the magnetic susceptibility of neutron matter computed within Landau theory
 is compared to the results
 of Ref. \cite{Fantoni}, obtained using the Auxiliary Field Diffusion Monte Carlo approach and nuclear hamiltonians 
 including the truncated $v_6^\prime$ and $v_8^\prime$ forms of
the Argonne $v_{18}$ potential, supplemented with the Urbana IX three-nucleon potential. In the right panel, we compare the density dependence of
the compressibility resulting from our calculations to the results of a variational calculations ,
  carried out using the full Argonne  $v_{18}$ $NN$ potential and the Urbana IX three-nucleon potential. 
  Our results appear to be in reasonable agreement with those obtained from highly refined theoretical approaches,
the differences at large density being likely to be ascribable to the different treatment of three-nucleon forces, which
 are known to play a critical role at $\rho > \rho_0$.

%% file: Chap_Weak.tex

This Chapter is devoted to the analysis of the Boltzmann-Landau (BL) equation, the solution of which can be interpreted as the dynamical 
form factor within Landau theory. Our discussion will closely follow the treatment developed in classic textbooks of quantum many-body 
theory \cite{Noz,Abr}.

\section{Response in the low-momentum transfer regime}

In the second quantization formalism, linear-response theory \cite{Bay} provides a microscopic description of the dynamic form factor
of interacting many-body systems. Consider an external scalar probe described by the potential
\[
U(\br ,t)=\eu^{\iu [{\bf q\cdot r}-(\omega+\iu \eta)t]} U({\bf q},\omega) \ ,
\] 
with $\eta=0^+$, coupled to the density of the system, assumed to be in its ground state, via 
\[
\int d^3r \rho({\bf r})U({\bf r},t) \ . 
\]
The induced oscillation in the local density is proportional to the initial perturbation according to
\[
\delta \rho({\bf q},\omega)=K({\bf q},\omega) U({\bf q},\omega),
\]
where $K$ is the response function, that can be written in the form
\[
\chi({\bf q},\omega)=\sum_{n\neq 0}\!\phantom{}\,|(\rho_{{\bf q}}^{\dag})_{n0}|^2\frac{2 \omega_{n0}}{(\omega+i\eta)^2-\omega_{n0}^2} \ .
\]
In the above equation,  $\rho_{{\bf q}}$ is the spatial Fourier transform of particle density operator,  
$(\rho_{{\bf q}})_{n0}$ denotes its matrix element between the ground state, $|0\rangle$, and an excited state $|n\rangle$, 
and $\omega \rb{n0}=E\rb{n}-E_0$ is the excitation energy. The operator $\rho_{{\bf q}}^{\dag}$ acting on the ground state generates one or more 
particle-hole (ph) excitations, or a collective mode (zero sound). 

It can be shown that in the limit of low-momentum transfer, ${\bf q}\to 0$, the matrix elements $(\rho_{{\bf q}}^{\dag})_{n0}$ involving a single ph excitation do not depend on $|{\bf q}|$, while the contributions of the collective 
mode and multipair excitations exhibit a $\propto\sqrt{|{\bf q}|}$ and $\propto|{\bf q}|$ behavior, respectively. 
As a consequence, the $|{\bf q}|$-dependence of the corresponding contributions to the response is 
$\propto|{\bf q}|^0, \ \propto |{\bf q}|^1$ and $ \propto |{\bf q}|^2$.

The formalism based on the BL equation does not take into account multipair excitation, but allows one to treat coherent and incoherent ph 
excitations on equal footing. Therefore, it turns out to be useful writing the response in the form
\[
\chi({\bf q},w)=\overbrace{\sum_{n\neq 0}\!\phantom{}' \,|(\rho_{{\bf q}}^+)_{n0}|^2\frac{2 w_{n0}}{(w+i\eta)^2-w_{n0}^2}}^{\mbox{$\chi_{Landau}$}}+\chi_{multipair} \ , 
\]
where the sum includes ph excitations only.
In view of its $|{\bf q}|$-dependence, neglecting $\chi_{multipair}$ is expected to be a reasonable approximation in the $|{\bf q}| \to 0$ limit.

\begin{figure}[htbp]
\begin{center}
\includegraphics[scale=0.5]{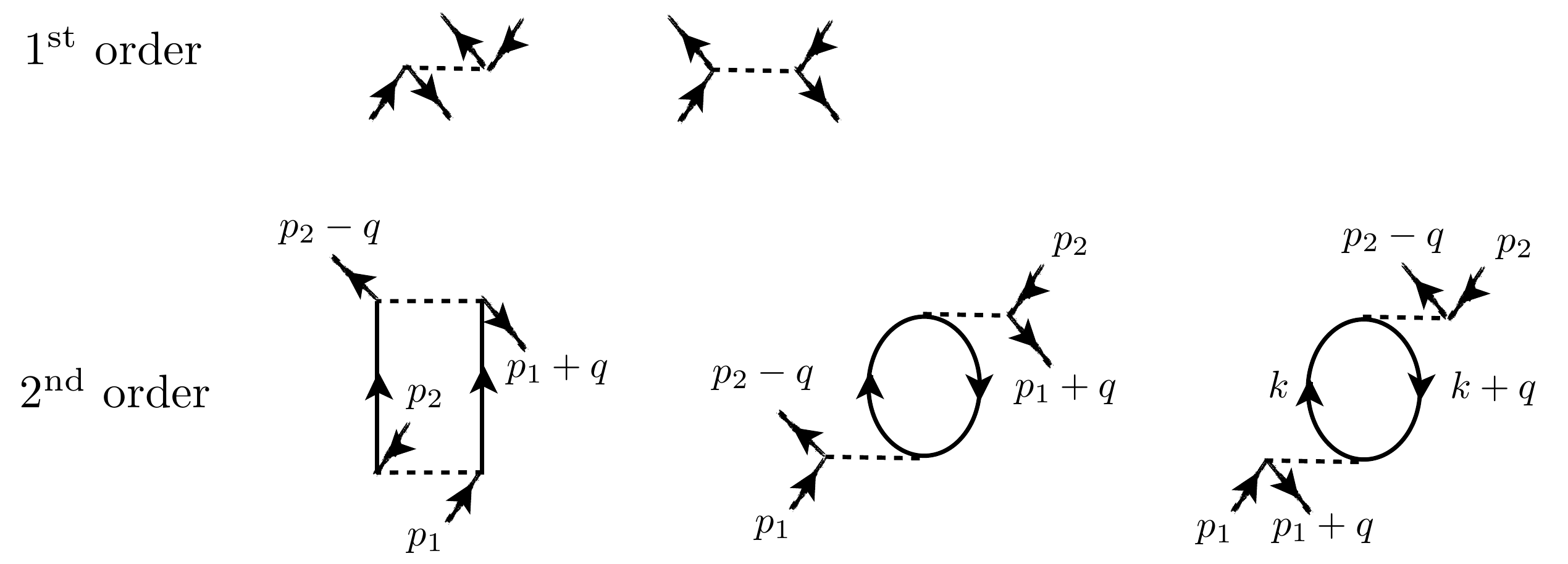}
\caption{Feynman diagrams describing one particle-one hole interactions up to second order in perturbation theory. 
The second order diagrams in the second line are labelled by the Mandelstam variables $s, \ u, \ t$. In the limit $|{\bf p_1-p_2}|=q\to 0$ 
only the loop in the last diagram is divergent  \cite{Abr,Dic}. }
\label{diagrams}
\end{center}
\end{figure}

As $|{\bf q}|\to 0$,  the main contribution associated with a single ph excitation exhibits a specific form, that can 
be easily understood.  Consider the Feynman diagrams of fig. \ref{diagrams}, showing all possible processes involving 
a particle ($\uparrow$) with momentum $p_1$ and a hole ($\downarrow$) with momentum $p_2$,  up to second order in the 
interaction, represented by the dashed line. 
The second order process can be can be classified in three topologically different sets, usually labelled according to the Mandelstam variables $s,\ u, \ t$, 
which reflect the different momenta associated with the loop.
The fermionic ph loop, that can be readily evaluated for the non interacting system, is referred to as  Lindhart function. It turns out to be exactly 
the same as the function $\Omega_{00}$ obtained from the kinetic equation (see below). In the low momentum transfer limit,
the first two second order diagrams can be disregarded with respect to the third, $t$-channel, one. In this last term, the integral of the fermionic loop 
is in fact divergent as $q\to 0$. Summing up the $t$-channel diagrams to all order, one finds the expression of the response in 
the so called ring approximation 
\be\label{respRing}
\chi\rp{ring\, appr}\sim\frac{\Omega_{00}(\lambda)}{1+V\Omega_{00}(\lambda)} \ .
\ee
The same expression, with an interaction potential $V$ expressed in terms of Landau paraeters is obtained from the solution of BL 
equation. 

\section{Response near equilibrium}
One of the main features of  Landau theory is that, under certain assumptions, the dynamic form factors of the system are fully constrained 
by the Landau parameters. This is true if the excited state of the system, produced by the interaction with the external probe, is such that 
the quasiparticle distribution in not far from that describing the system in equilibrium. 
In this case the non equilibrium, non homogeneous state can be specified by the quasiparticle distribution  $n_{k\sigma}({\bf r},t)$ depending on 
position and time. Note that we can fix ${\bf k}$ and ${\bf r}$ at the same time, since the characteristic length of the spatial 
inhomogeneity, $\lambda$, is much larger than the localization 
length of the quasiparticle $\lambda_f$. Typically the change in the distribution function occurs on the Fermi surface and is proportional to 
the temperature, implying  $\Delta p \sim T/v_f$ and 
due to the uncertainty principle,  $\lambda_f \sim v_f/T$. In the region $\lambda \gg \lambda_f$ the system is locally homogenous (globally quasi homogenous) and we can use a classic description based on a local distribution function to specify the momentum density at position 
{\bf r} and time $t$. 

The expressions of all observables of the system must also exhibit an explicit ${\bf r}$ and $t$ dependence. For example, 
\begin{eqnarray}
\delta \mathcal{E}(t)&=&\int d^3\br \delta E(\br,t)=\int d^3\br\sum_{\bk\sigma}\varepsilon_{\bk\sigma}(\br,t)\delta n_{\bk\sigma}(\br,t) \ ,\nonumber\\
\varepsilon_{\bk\sigma}(\br,t)&=&\varepsilon^0_{\bk,\sigma}+\int d^3 \br'\sum_{\bk',\sigma'}f_{\sigma\sigma'}(\bk,\bk',\br,\br')\delta n_{\bk'\sigma'}(\br',t)+... \ ,\nonumber
\end{eqnarray}
where $E(\br,t)$ is the total energy density and $f_{\sigma\sigma'}(\bk,\bk',\br,\br')$ describes the effective interactions between quasiparticles at 
a relative distance $\br-\br'$. At zero-th order the system is homogenous, implying $\varepsilon^0_{\bk,\sigma}(\br,t)=\varepsilon^0_{\bk,\sigma}$.
The effective interaction is different from zero only if the quasiparticles are very close, typically at distances comparable with the scattering 
length, i.e. $|\br-\br'|\sim a \sim 1/v_f$. Since $n_{\bk'\sigma' }(\br,t)$ is almost constant over a distance $\lambda \gg a$, we can 
expand $\delta n_{\bk'\sigma' }(\br',t)=\delta n_{\bk'\sigma'}(\br,t)+\ldots$ and disregard terms $O(a/\lambda)$. 
It follows that 
\begin{equation}
\delta E(\br,t)=\sum_{\bk\sigma}\varepsilon^0_{\bk\sigma}\delta n_{\bk\sigma}(\br,t)+\frac{1}{2}\sum_{\bk \bk'\sigma\sigma'}f_{\sigma\sigma'}(\bk,\bk')\delta n_{\bk\sigma}(\br,t)\delta n_{\bk'\sigma'}(\br,t)+...\nonumber
\end{equation}
with
\begin{equation}
f_{\sigma\sigma'}(\bk,\bk')=\int d^3\br' f_{\sigma\sigma'}(\bk,\bk',\br-\br') \ \ \ , \ \ \ {\bf r} \sim {\bf r}' \ . \nonumber
\end{equation}

\section{Kinetic Equation}

Within Landau theory \cite{Landau}, all the relevant macroscopic quantities are linked to the miscroscopic dynamics through the quasiparticle distribution function. The evolution of the system is then determined by the kinetic equation for $\delta n$. Let us start from the Boltzmann equation for 
quasiparticle distribution function
\begin{equation}\label{KE}
\frac{d n_{\bk,\sigma}(\br,t)}{dt}=I[n] \ ,
\end{equation}
where $d/dt$ is a total time derivative and the quantity $I[n]$, called collision integral, describes the fluctuation of the number of quasiparticles in 
a phase space volume arising from their mutual interactions. The quasiparticle energy density is treated as the classic hamiltonian of the 
single quasiparticle, and the external potential $U(\br,t)$ is added to account for the interaction between the quasiparticle and the probe. 

Eq (\ref{KE}) can be expanded according to 
\begin{equation}
\label{eqa}
\frac{\partial n_{\bk \sigma}(\br,t)}{\partial t} + 
\frac{1}{\hbar}\left\{ \frac{\partial n_{\bk\sigma}(\br,t)}{\partial \br_{\alpha}}\frac{\partial\varepsilon_{\bk\sigma}}{\partial \bk_{\alpha}}-\frac{\partial n_{\bk\sigma}(\br,t)}{\partial \bk_{\alpha}}\frac{\partial(\varepsilon_{\bk\sigma}(\br,t)+U(\br,t))}{\partial \br_{\alpha}}\right\}=I[ n ] \ ,\nonumber
\end{equation}
with
\[
\dot{\bk}=-\frac{\partial(\varepsilon+U)}{\partial\br} \ \ \ , \ \ \ \dot{\br}= \frac{\partial\varepsilon}{\partial\bk} \ ,
\]
and linearized, since the total distribution function is well defined only for $|{\bf k}| \sim k_F$. For small deviations from 
equilibrium
\[
\delta n_{\bk\sigma }(\br,t)=n_{\bk \sigma}(\br,t)-n^0_{\bk \sigma} \ ,
\]
where the $n^0_{\bk \sigma}$ is the distribution at equilibrium, is in fact non vanishing only in the vicinity of the Fermi surface. 
Therefore, the final expression is
\begin{equation}
\frac{\partial\delta n_{\bk\sigma}(\br,t)}{\partial t} + \frac{\partial\delta n_{\bk\sigma}(\br,t)}{\partial \br_{\alpha}}\frac{\partial\varepsilon^0_{\bk\sigma}}{\partial \bk_{\alpha}}
-\frac{\partial n^0_{\bk\sigma}(\br,t)}{\partial \bk_{\alpha}}\frac{\partial(\varepsilon_{\bk\sigma}(\br,t)+U(\br,t)}{\partial \br_{\alpha}} =I[ n ] \ ,
\end{equation}
with
\begin{equation}
\frac{\partial n^0_{\bk\sigma}}{\partial \bk_{\alpha}}=\frac{\partial n^0_{\bk\sigma}}{\partial \varepsilon_{\bk \sigma}^0}\frac{\partial \varepsilon^0_{\bk\sigma}}{\partial \bk_{\alpha}}\qquad \frac{\partial \varepsilon^0_{\sigma \bk}}{\partial \bk}= {\bf v}_{k} \ ,
\end{equation}
leading to
\begin{align}
\nonumber
\label{eqq}
\frac{\partial n_{\bk\sigma}(\br,t)}{\partial t}&+{\bf v}_k\cdot\bm{\nabla}_r\delta n_{\bk\sigma}(\br,t) \\
& +\frac{\partial n^0_{\bk\sigma}}{\partial \varepsilon^0_{\bk\sigma}}{\bf v}_k\cdot \left({\bf F}(\br,t)-\sum_{\sigma' \bk'}f_{\sigma\sigma'}(\bk,\bk')\bm{\nabla}_{\br}\delta n_{\bk'\sigma'}(\br,t)  \right)=I[n] \ ,
\end{align}
where $\bm{F}(\br,t)=-\bm{\nabla}_{\br} U(\br,t)$ and 
\begin{equation}
\frac{\partial n^0_{\bk\sigma}}{\partial \varepsilon^0_{\bk\sigma}}=-\delta(\varepsilon_F-\varepsilon^0_{\sigma \bk}) \ .
\end{equation}

\subsection{The collision integral}

The evaluation of the collision integral is essential for deriving the dynamic response. The behavior of the system depends mainly on the value of 
$\omega \tau$ where $\omega$ is the frequence of the external perturbation and $\tau$ is the quasiparticle collision time. According to Landau theory, 
the probability of collision between two excitations (quasiparticles) decreases according to the square of the spread of Fermi distribution 
function, implying in turn hence $\tau \sim T^{-2}$. At large $T$, $\omega \tau << 1$ and the perturbation propagates according to ordinary 
hydrodynamics, since locally the system is in thermodynamic equilibrium thanks to the many quasiparticle collisions. For example,  
the speed of sound is related to compressibility through the usual thermodynamic relation $u=\sqrt{\partial P/\partial \rho}$, and its damping 
depends linearly on $\tau$ \footnote{The attenuation coefficient is $\gamma\sim \omega^2\eta /\rho u^3$ where $\eta$ is the shear viscosity, $\rho$ is the density and $u \sim v_f$ the quasiparticle velocity. 
Note that $\eta/\rho \sim v_f^2 \tau$ where $v_f$ does not depend on $T$} \cite{Abr}. 
On the other hand, if $\omega \tau \sim 1$ diffusion is extremely damped, and the perturbation does not propagate. 

In the region of $T\sim0$, i.e. $\omega\tau\gg1$, eq. (\ref{eqq}) turns out to allow a new solution. Compared to the classical sound wave, this solution, driven by quantum effects, describes a somewhat different physics: a collective mode propagating in the liquid without experiencing any damping. 
This mode was first predicted by Landau, who dubbed it ``zero sound''. Let us assume that 
\begin{equation}
\delta n\sim \eu\rp{-\iu \omega t} ,
\end{equation}
where $1/\omega$  is the time scale of local dynamic, which implies $\omega \tau \gg 1$. It follows that the integral $I[n]$,  
being of order of $\delta n/\tau$, can be neglected compared to $\partial n/\partial t$.

\subsection{Spin-dependent solution in collisionless limit}

Consider an external magnetic field coupled to the spin according to $U(\br,t)=-g\mu_B {\bf S}\cdot{\bf B}(\br,t)$, where ${\bf B}$ is aligned in 
the $z$-direction. The linear spin-density response is defined trough the magnetization of the matter
\begin{equation}
{\bf M}=\bar{\chi}{\bf B} \ \ \ , \ \ \ \bar{\chi}_{\alpha\beta} = \left. \frac{\partial {\bf M}_{\alpha}}{\partial {\bf B}_{\beta}} \right|_{B=0},
\end{equation}
where $B = |{\bf B}|$, the magnetization ${\bf M}$ is given by
\begin{equation}
{\bf M}_z=-\frac{g \mu_B}{2}(\delta \rho_{\uparrow}-\delta \rho_{\downarrow})= \frac{g\mu_B}{2V}\sum_{\bk}(\delta n_{\bk\uparrow}-\delta n_{\bk\downarrow})=\frac{g\mu_B}{2V}\sum_{\bk}\delta n^a_{\bk} \ ,
\end{equation}
and the spin susceptibility is\footnote{In the literature one can also find the slightly different definition
\begin{equation}
\chi=\left(\frac{2}{g \mu_B}\right)^2\bar{\chi}=\frac{2}{g \mu_B}\left(\sum_{\bk}\delta n^a_{\bk}\right)\frac{1}{B}.
\end{equation}
} 
\begin{equation}
\bar{\chi}=\frac{g \mu_B}{2V}\left(\sum_{\bk}\delta n^a_{\bk}\right)\frac{1}{B} \ .
\end{equation}
The kinetic equation will help us to determine the change in the Fermi distribution function $\delta n^a$ in the direction of momentum ${\bf k}$ 
due to the external field $H({\bf r},t)$. Performing a Fourier Transform of eq. (\ref{eqq}) we find
\begin{align}
\nonumber
\label{eq11}
&(w-{\bf v}_{k}\cdot\bq)\delta n_{\bk\sigma}(w,\bq) \\
&+{\bf v}_{k}\cdot\bq \,\frac{\partial n^0_{\bk\sigma}}{\partial \varepsilon^0_{\bk\sigma}}\left[-\left(\frac{g\mu_B}{2}\right)\bm{\sigma}_z H(w,\bq)+\sum_{\bk'\sigma'}f_{\sigma\sigma'}(\bk,\bk')\delta n_{\bk'\sigma'}(\bq,w)\right]=0 \ .
\end{align}
This above equation is similar to an integral equation, and can be solved by iteration. Substitution of the zero-th order expression in the last term yields
\begin{eqnarray}
\delta n_{\bk\uparrow}&=&\frac{{\bf v}_{k}\cdot\bq}{(w-{\bf v}_{k}\cdot\bq)}\frac{\partial n^0_{\bk\uparrow}}{\partial \varepsilon^0_{\bk\uparrow}}\left[-\frac{g\mu_B}{2}B+...\right]\nonumber\\
\delta n_{\bk\downarrow}&=&\frac{{\bf v}_{k}\cdot\bq}{(w-{\bf v}_{k}\cdot\bq)}\frac{\partial n^0_{\bk\downarrow}}{\partial \varepsilon^0_{\bk\downarrow}}\left[\frac{g\mu_B}{2}B+...\right],
\end{eqnarray}
implying $\delta n_{\bk\uparrow}=-\delta n_{\bk\downarrow}$. It follows that in each point of the Fermi surface,  corresponding to a momentum $\bk$, 
the oscillation is $\sim ({\bf v}_{k}\cdot\bq)/(w-{\bf v}_{k}\cdot\bq)$.

The equation for $\delta n^a$ is \footnote{We have used the the relations
\begin{eqnarray}
f_{\uparrow\uparrow}\delta n_{\uparrow}+f_{\uparrow\downarrow}\delta n _{\downarrow}&=&(f_{\uparrow\uparrow}-f_{\uparrow\downarrow})\delta n_{\uparrow}=\frac{(f_{\uparrow\uparrow}-f_{\uparrow\downarrow})}{2}(\delta n_{\uparrow}-\delta n_{\downarrow})=f^a(\bk,\bk')\delta n^a_{\bk'}\nonumber\\
f_{\downarrow\uparrow}\delta n_{\uparrow}+f_{\uparrow\uparrow}\delta n _{\downarrow}&=&-f^a(\bk,\bk')\delta n^a_{\bk'}.\nonumber
\end{eqnarray}
}
\begin{align}\label{eq}
&(w-\bq\cdot{\bf v}_k)\delta n^a_{\bk}(\bq,w) \\
&+2(\bq\cdot{\bf v}_k)\frac{\partial n^0_{\bk}}{\partial \varepsilon^0_{\bk}}\left(-\frac{g\mu_B}{2}B(\bq,w)+\sum_{\bk'}f^a(\bk,\bk')\delta n^a_{\bk'}(\bq,w) \right)=0.
\end{align}
We now make fro the solution the ansatz 
\begin{equation}
\delta n_k^a(\bq,w)=-\frac{\partial n^0_{\bk}}{\partial \varepsilon^0_{\bk}}\nu_{\bk}(\bq,w) \ .
\end{equation}
Note that the above expression is in general quite reasonable: the first factor constrains the change in the distribution function to the Fermi surface, 
while $\nu_{\bk}$ provides a quantitative description of this variation in the direction of $\bk$. In the case of  a $\Theta$-function distribution, 
one can rewrite the above expression using the angles $\theta',\phi'$ between $\bq$ and $\bk'$ 
\begin{equation}
\delta n_{\bk'}=\Theta(k_f+\nu_{\bk'}(\theta',\phi')-k')-\Theta(k_f-k')\simeq \hbar\,v_f \nu_{\bk'}(\theta',\phi')\delta (\varepsilon^0_{\bk'}-\varepsilon_f)  \ .
\end{equation}
Here, we only consider azimuthally symmetric solutions,  corresponding to $\nu_{\bk'}(\theta)$\footnote{In principle, eq.  (\ref{eq}) can describe any 
kind of collective modes, that differ in the angular dependence $\nu_{\bk}(\theta,\phi)$ as well as in velocity. The arbitrary dependence of $\nu_{\bk'}$ on $(\theta', \phi')$ determines different solution. For example, a $\eu\rp{\iu m \phi}$ dependence corresponds to collective modes leaving the total volume of the Fermi sphere unchanged.}.

Expanding  both $\nu_{\bk'}(\cos\theta')$ and $f^a(\cos\theta,\cos\theta')$ in Legendre polynomials
\begin{equation}
\nu_{\bk'}=\sum_\ell P_\ell(cos \theta' )\nu_\ell\qquad f^a(\cos\theta,\cos\theta')=\sum_\ell P_\ell(cos \theta' )\hat{f}^a_\ell(\cos\theta),
\end{equation}
eq (\ref{eq}) becomes
\begin{equation}\label{eq2}
\nu_k+2 \frac{\cos\theta}{s-\cos\theta}\sum_{k'}f^a(\cos\xi)\frac{\partial n^0_{\bk'}}{\partial \varepsilon^0_{\bk'}}\nu_{\bk'}(\cos\theta')=2\frac{\cos\theta}{s-\cos\theta}\left(-\frac{g\mu_B}{2}B(\bq,\omega)\right),
\end{equation}
where the sum yields
\begin{eqnarray}
2\sum_{\bk'}f^a(\cos\xi)\frac{\partial n^0_{k'}}{\partial \varepsilon^0_{k'}}\nu_{\bk'}(\cos\theta')\!\!\!\!&=&\!\!\!\!-2V\int \frac{d^3\bk'}{(2\pi)^3} f^a(\cos\xi)\delta(\varepsilon_f-\varepsilon^0_{\bk'})\nu_{\bk'}(\cos\theta)\nonumber\\
&=&\!\!\!\!\!-V\frac{N(0)}{4\pi}\int d(\cos\theta')d\phi' f(\cos\xi)\nu_{\bk}(\cos\theta') \ .\nonumber
\end{eqnarray}
We can simplify the expression using the addition theorem of Legendre polynomials and expanding again in Legendre polynomials in $\theta$ space:
\begin{equation}
2\sum_{\bk'}f^a(\cos\xi)\frac{\partial n^0_{\bk'}}{\partial \varepsilon^0_{\bk'}}\nu_{\bk'}(\cos\theta')=-\sum_l \frac{1}{2\ell+1}G_\ell P_\ell(\cos\theta)\nu_\ell,
\end{equation} 
where $s=\omega/(|{\bf q}| v_f)$ and $G_\ell=VN(0)f^a_\ell$. The equation in (\ref{eq2}) become:
\begin{equation}\label{eq3}
\frac{\nu_\ell}{2\ell+1}+\sum_{\ell'}\Omega_{\ell \ell'}(s)G_{\ell'}\frac{\nu_{\ell'}}{2\ell'+1}=-\Omega_{\ell0}(s)\left(-g\mu_B B\right)
\end{equation}
with
\begin{equation}
\Omega_{\ell \ell'}(s)=\int \frac{d\mu}{2}P_\ell(\mu)\frac{\mu}{\mu-s}P_{\ell'}(\mu),
\end{equation}

The explicit expressions for $\ell=0,1,2$ are the following:
\begin{eqnarray}
&&\Omega_{00}=1+\frac{s}{2}\ln\frac{s-1}{s+1}\qquad \Omega_{\ell1}=s\Omega_{\ell0}+\frac{1}{3}\delta_{\ell 1}\nonumber\\
&&\Omega_{20}=\frac{1}{2}+\frac{3 s^2-1}{2}\Omega_{00}\qquad \Omega_{22}=\frac{1}{5}+\frac{3 s^2-1}{2}\Omega_{20} \ .
\end{eqnarray}

Considering only terms with $\ell=0,1,2$ in (\ref{eq3}) we obtain three coupled linear equations for $\nu_0$, $\nu_1$  and $\nu_2$, that can be 
cast in the matrix form
\[
\left(
\begin{array}{lll}
1+\Omega_{00}G_0 & \frac{1}{3}\Omega_{01}G_1    & \frac{1}{5}\Omega_{02}G_2\\
\Omega_{10}G_0     & \frac{1}{3}(1+\Omega_{11}G_1) & \frac{1}{5}\Omega_{12}G_2\\
\Omega_{20}G_0     & \frac{1}{3}\Omega_{21}G_1     & \frac{1}{5}(1+\Omega_{22}G_2)
\end{array}\right)
\cdot
\left(\begin{array}{l}
\nu_0\\
\nu_1\\
\nu_2
\end{array}\right)
=-(g\mu_BB(\bq,\omega))
\left(\begin{array}{l}
\Omega_{00}\\
\Omega_{10}\\
\Omega_{20}
\end{array}\right).
\]
Finally, the spin-density response defined by 
\begin{equation}
\delta\rho^a=\frac{1}{V}\sum_{\bk}\delta n^a_{\bk} \ \ \ , \ \ \  \delta n^a_{\bk}=-\frac{\partial n^0_{\bk}}{\partial \varepsilon^0_{\bk}}\nu_{\bk},
\end{equation}
reads
\begin{equation}
\delta \rho^a=\frac{1}{V}\sum_{\bk}\delta(\varepsilon_{\bk}^0-\varepsilon_f)[\nu_0+P_1(\cos\theta)\nu_1+P_2(\cos\theta)\nu_2]=
\frac{N(0)}{2V}\nu_0 \ .
\end{equation}
Note that for $\ell \neq 0$ the integral of Legendre polynomials vanishes, since they are orthogonal. The value of $\nu_0$ can be obtained from the matrix equation above. Using only $\ell=0,1$ the matrix reduces to dimension $2 \times 2$, and 
\begin{equation}\label{finalCHI}
\chi=\frac{N(0)}{V}\frac{\Omega_{00}(s)}{1+[G_0+s^2\frac{G_1}{1+G_1/3}]\Omega_{00}(s)}.
\end{equation}
\subsection{Coherent and incoherent contributions}
\begin{figure}[htbp]
\hspace{-12mm}
\vspace{-10mm}
\includegraphics[scale=0.6]{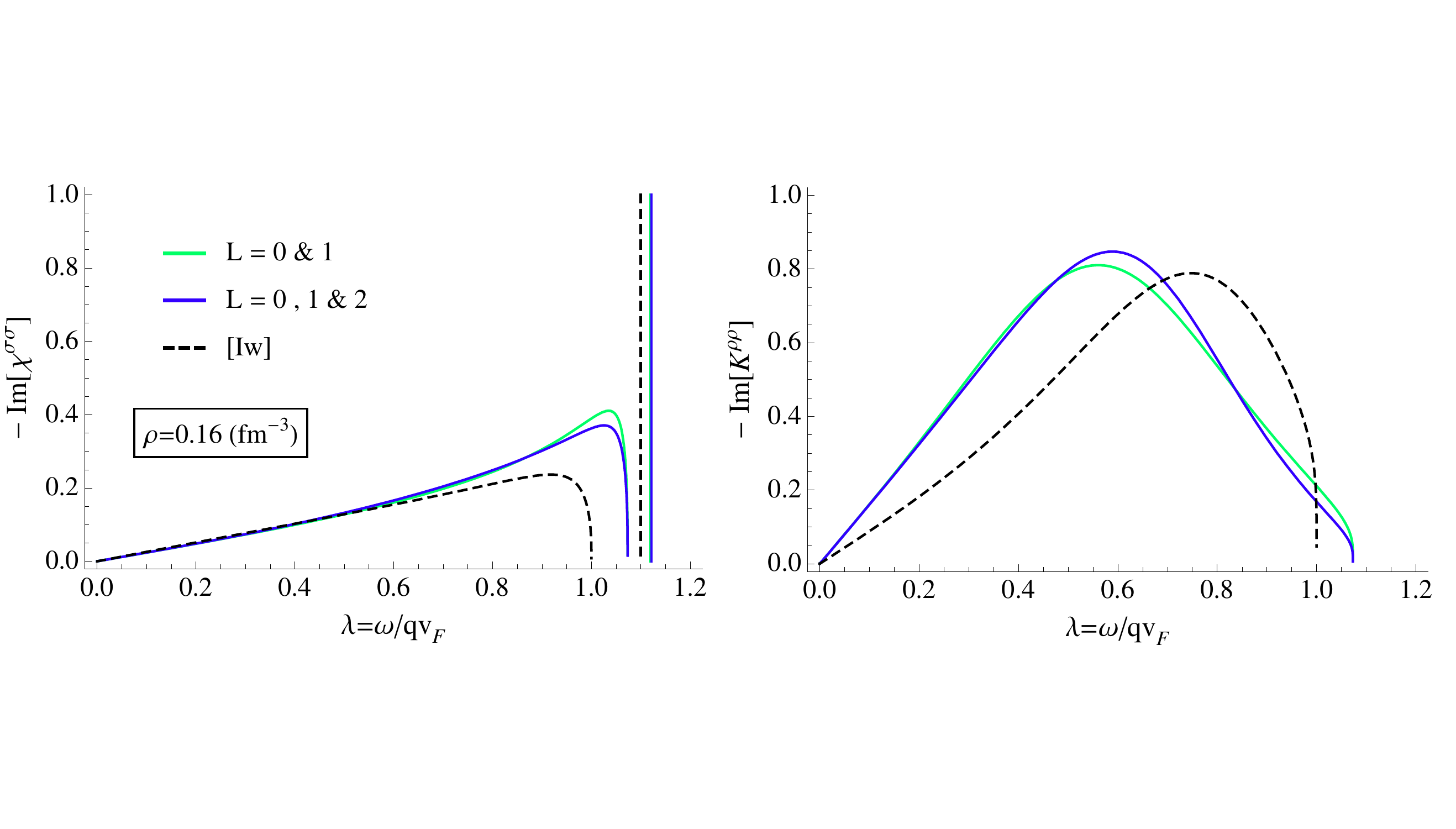}
\vspace{-10mm}
\caption{Left panel: spin-density structure function of neutron matter at $\rho = \rho_0 = 0.16 \ {\rm fm}^{-3}$, as function of 
$\lambda= \omega/(|{\bq}|v_F$. The curves have been
obtained including only the Landau parameters with $\ell =0, \ 1$ or taking into account the contribution associated with $\ell =2$.
 For comparison, the thick dashed line shows the results of Ref. \cite{Iwamoto}. The curves are normalized as discussed in the text. Right panel: same as in the left panel, but for the density structure function.}
\label{f2s}       
\end{figure}

\begin{figure}[htbp]
\hspace{-12mm}
\includegraphics[scale=0.6]{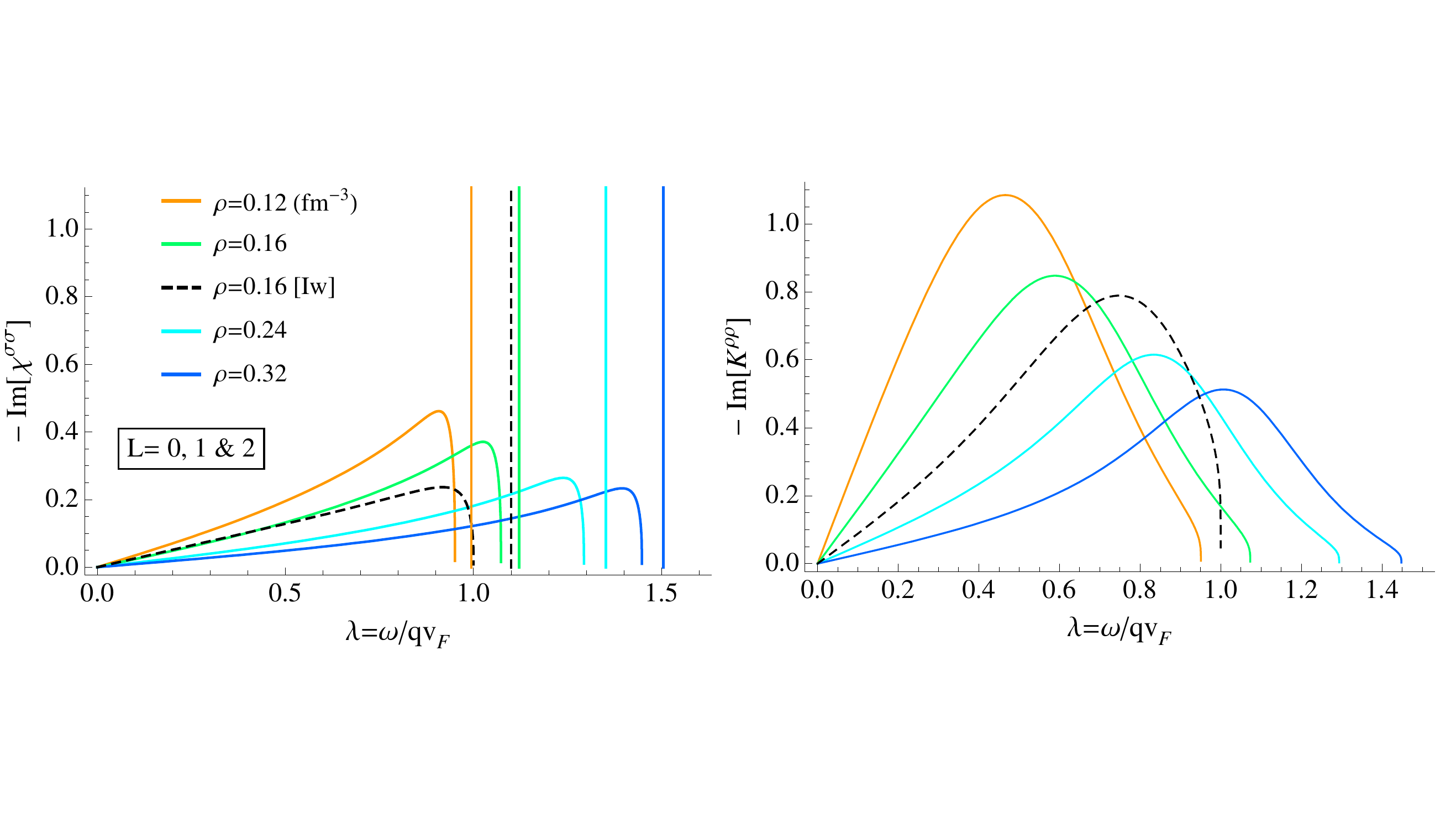}
\vspace{-15mm}
\caption{Left panel: spin-density structure function of neutron matter, as a function of $\lambda=\omega/(|{\bq}|v_F$, at different densities.
The calculations have been carried out including the Landau parameters with $\ell =$ 0, 1 and 2.
For comparison, the thick dashed line shows the results of Ref. \cite{Iwamoto}, corresponding to $\rho = \rho_0 = 0.16 \ {\rm fm}^{-3}$.
The curves are normalized as discussed in the text. Right panel: same as in the left panel, but for the density structure function.}
\label{f2rho}       
\end{figure}

One of the striking properties of Landau theory is that it is able to manage coherent and incoherent response on equal footing. In eq (\ref{eq2}) we can distinguish at least two cases depending on the value of the parameter $s$. 
\begin{description}
\item[ $s<1$] The speed of the perturbation,  $u_s=\omega/|{\bf q}|$, is smaller then $v_f$\footnote{Note that the condition 
$\cos\theta_s=u_s/v_f$  in eq. (\ref{eq2}) is the same as the condition for emission of Cherenkov radiation. However, here it refers to zero sound waves,  
triggered by a single quasiparticle excitation.}. 
Note that for $\omega/|{\bf q}| <v_f$ the denominator of eq.(\ref{eq2}) can vanish, thus indicating the presence of a resonance between the external perturbation and the single quasi particle excitation. In this case the perturbation can lead to the excitation of incoherent single ph pairs, and is 
consequently strongly damped. This phenomenon is reminiscent of the propagation of light in a medium,  in which the imaginary part of the 
refraction index is proportional to the magnitude of the damping. In the case under consideration, the same role is played by the imaginary part of 
the response,  that can be obtained from
\begin{equation}\label{Ima}
\Omega_{00}=1-\frac{s}{2}\ln\frac{s+1}{s-1}=1-\frac{s}{2}\ln\left|\frac{s+1}{s-1}\right|+i\frac{\pi}{2}s \ \ \ , \ \ \  s<1 \ .
\end{equation}
Typically, at $T=0$  an incoherent qp excitation from the ground state consists of the disappearance of a particle with $|{\bf k}| \leq k_F$ and 
the appearance of one with $|{\bf k}+{\bf q}| > k_F$. The corresponding excitation energy is
\begin{equation}
\Delta E=\varepsilon_{\bk+\bq}-\varepsilon_{\bk}=\frac{(\bq+\bk)^2}{2m^*}-\frac{\bk^2}{2m^*}=\frac{1}{2m^*}(\bq^2+2\bk\cdot\bq) \ .
\end{equation}
For foxed $|\bk|$ and $|\bq|$, the maximum and minimum values of the above energy are $\Delta E_{max}=(\bq^2+ 2 |\bk| |\bq| )/2m^*$ and 
$\Delta E_{min}=(\bq^2- 2 | \bk | | \bq | )/2m^*$. These relations correspond to two hyperboles, delimiting the spectrum of incoherent excitation 
(fig. \ref{f}). For $s<1$ the dispersion relation of external perturbation corresponds to the region between the two hyperboles, 
and the attenuation arising from the interaction with single ph excitations would made it vanish.
\begin{figure}[ht]
\includegraphics[scale=0.65]{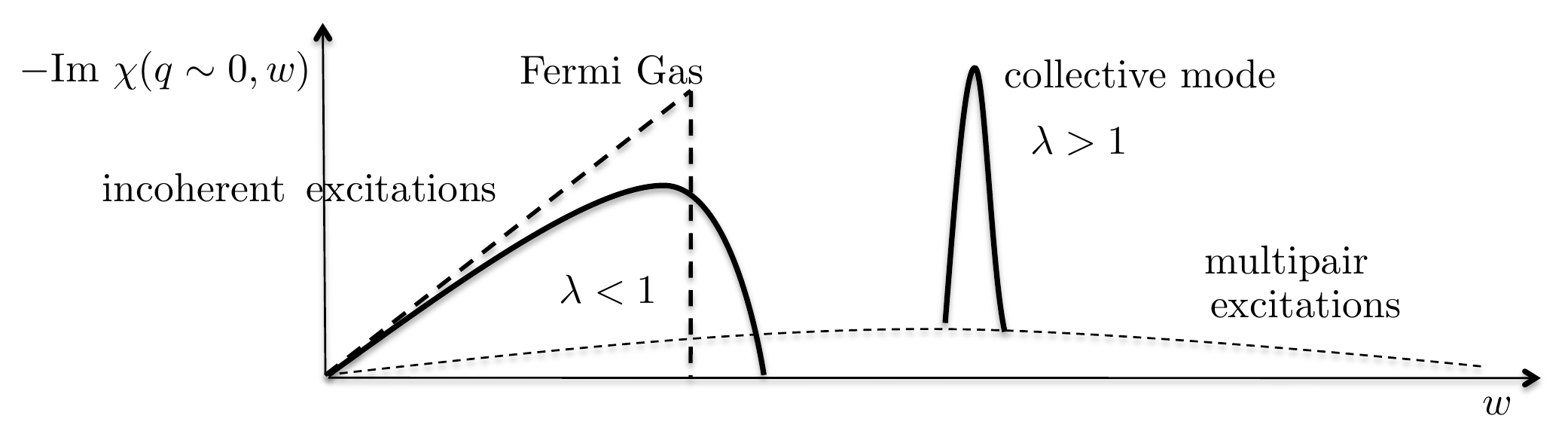}
\caption{Response in the $|{\bq}| \to 0$ limit, as a function of the energy transfer $\omega$. Incoherent, coherent and multipairs contributions are 
clearly visible \cite{Noz} .}
\label{f}       
\end{figure}
\item[ $s>1$] The velocity of the wave associated with the external perturbation is larger than the Fermi velocity. Hence, it can only couple to 
the collective mode only. 
Generally, in this case the response is obtained adding an infinitesimal imaginary part to the frequency, i.e. replacing  
$\omega \to \omega-\iu\eta$. From the formal point of view, this is the rigorous way to perform this calculation, letting $\eta \to 0$ at the 
end.  This replacement is irrelevant when dealing with incoherent excitations,  since in this case the response has a non vanishing 
imaginary part, as shown by eq. (\ref{Ima}) above. However, it turns out to be important in the region corresponding to $s>1$,   
where the response can be written in the form
\begin{align}
\chi&=\lim_{\eta\to 0}\frac{N(0)}{V}\frac{\Omega_{00}(s)}{1+[G_0+s^2\frac{G_1}{1+G_1/3}]\Omega_{00}(s)-\iu\eta}\nonumber\\
&=\mathcal{P}[...]+\iu\pi \,\frac{N(0)}{V}\Omega_{00}(s)\, \delta(1+[G_0+s^2\underbrace{\frac{G_1}{1+G_1/3}}_{\mbox{$A$}}]\Omega_{00}(s)) .\nonumber
\end{align}
To obtain the above equaiotn, we have used the well-known expression to distinguish the real part, associated with the principal part integration,  
and the imaginary part yielding a Dirac $\delta$-function. Let us now focus on the imaginary part alone. After some manipulations we find
\begin{align}
& \Omega_{00}(s_0)\delta(1+[G_0+s^2A]\Omega_{00}(s))&\\
 &  \ \ \ \ \ \ \ \ \ \ \ \ \ \ = \Omega_{00}(s_0)\frac{\delta(s-s_0)}{|2s_0 A\Omega_{00}(s_0)+(G_0+s_0^2A)\Omega_{00}'(s_0)|} \ ,\nonumber\\
& \Omega_{00}(s_0)=\frac{-1}{G_0+s_0^2A} \ , \nonumber
\end{align}
where $s_0$ is the root if the equation obtained requiring the argument of the $\delta$-function to vanish. 

In this case the energy of the coherent wave is above the incoherent excitation spectrum. Therefore,  the collective mode is undamped. We note that this excitation is propagating in Fermi liquids at $T=0$. 
\end{description}
In fig. (\ref{f2s}) we compare the imaginary part of the response function in both the density and spin channels. According to standard linear response theory the correlation function is proportional to the imaginary part  of the response
\[
\lan \s(\omega,-\bq)\s(0,\bq)\ran\sim {\rm Im} \chi \ . 
\]
The calculation has been carried out using the set of Landau parameters listed in the tables \ref{LT1} and \ref{LT2},  including 
contributions associated with angular momentum components up to $L=2$. For comparison, we also show the results of Ref. \cite{Iwamoto} (dashed line). 
It clearly appears that including the  $\ell=2$ parameters hardly affects the response in both the spin and density channel. Moreover, 
a zero sound mode appears in the spin channel, consistently with the results of Ref.  \cite{Iwamoto}. Comparing the right and left panels of  
fig. \ref{f2s}, it is apparent that the collective mode strongly depletes the incoherent response, and largely contributes to the response in the 
$|\bq|\to 0$ limit. 
Figure \ref{f2rho} shows the response for different values of the density. Note that all curves are normalized so that the maximum of the corresponding Fermi gas structure functions be unity.

\section{Sum Rules in the $|\bq|\to 0$ limit}\label{sumrule}
In the extreme low-momentum limit, the dynamic response function is heavily affected by the static properties of matter. 
Linear response theory provides the qualitatively behavior of the the matrix elements of both the density and spin-density operator, as they 
must reproduce the static observables. Moreover, some conservation laws can be rearranged in such a way as to constrain the 
response function. The number of non-trivial sum rules grows with to the complexity of the hamiltonian, but they are usually limited by the degree of accuracy of the approximations employed in the calculations. Typically, these constraints are expressed by moments of the correlation functions, defined as
\[
m^n=\lim_{|\bq|\to 0}\int \omega^{n} \chi(\bq,\omega) d \omega \ ,
\]
and commutator relations. 

Consider a hamiltonian including central interactions only. The corresponding response function turns out to be diagonal in both 
the spin and density channels. The value of the moments with $n$  \emph{odd} can be simply expressed, e.g. in the spin channel, as
\begin{equation}\label{sumrules}
m^n= C\lan \Big[\bm{\s}(q),\underbrace{\Big[ H,\ldots\Big[ H}_{\mbox{n times}}, \bm{\s}^{\dag}(q)\Big]\ldots\Big]\ran
\end{equation}
where the value of $C$, given by $C=2\pi/\rho$ in our case, depends on the pre-factor in the definition of the correlation function \cite{Safier}. The first quantization form of $\bm{\s}$ is
\[
\bm{\s}(q)=\sum_{j}^{N}\bm{\s}\rb{j} \eu\rp{\iu {\bf q \cdot r_j}} \ .
\] 
As stated in Ref. \cite{Safier}, Landau theory does not exactly satisfy the spin-density sum rule for moments with $n\geq 3$. Moreover, in our case we the  hamiltonian includes tensor interactions, although the corresponding Landau parameters are neglected.

An alternative consistency test directly relates both the compressibility and the susceptibility to the corresponding  dynamic form factors
\begin{align}
\label{boh1}
m^{-1}_{\rho}&=\lim_{|\bq|\to 0}\int \frac{K^{\rho\rho}(\bq,\omega)}{\omega} d \omega=\frac{\pi K^{\rho}}{\rho} \ ,\\
m^{-1}_{\s}&=\lim_{|\bq|\to 0}\int \frac{\chi^{\s\s}(\bq,\omega)}{\omega} d \omega=\frac{\pi \chi^{\s}}{\rho} \ .\label{boh2}
\end{align}
Our strategy is based on evaluating the left hand side of the above equations, separating the contributions of the collective mode from the one
arising from single ph excitations, and then comparing it to the right hand side.
 \begin{figure}[htbp]
 \hspace{-16mm}
\includegraphics[scale=0.6]{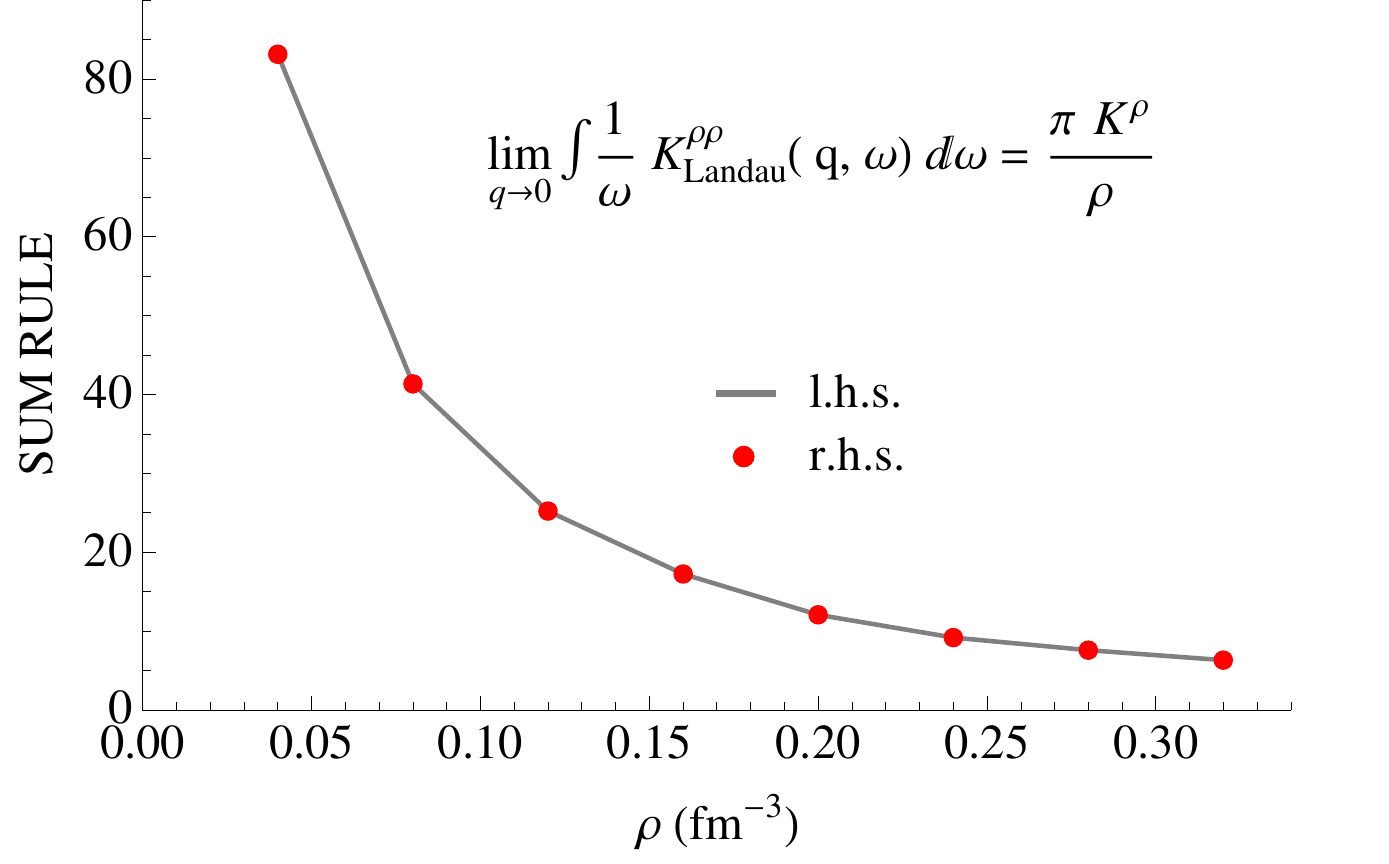}
 \hspace{-8mm}
\includegraphics[scale=0.6]{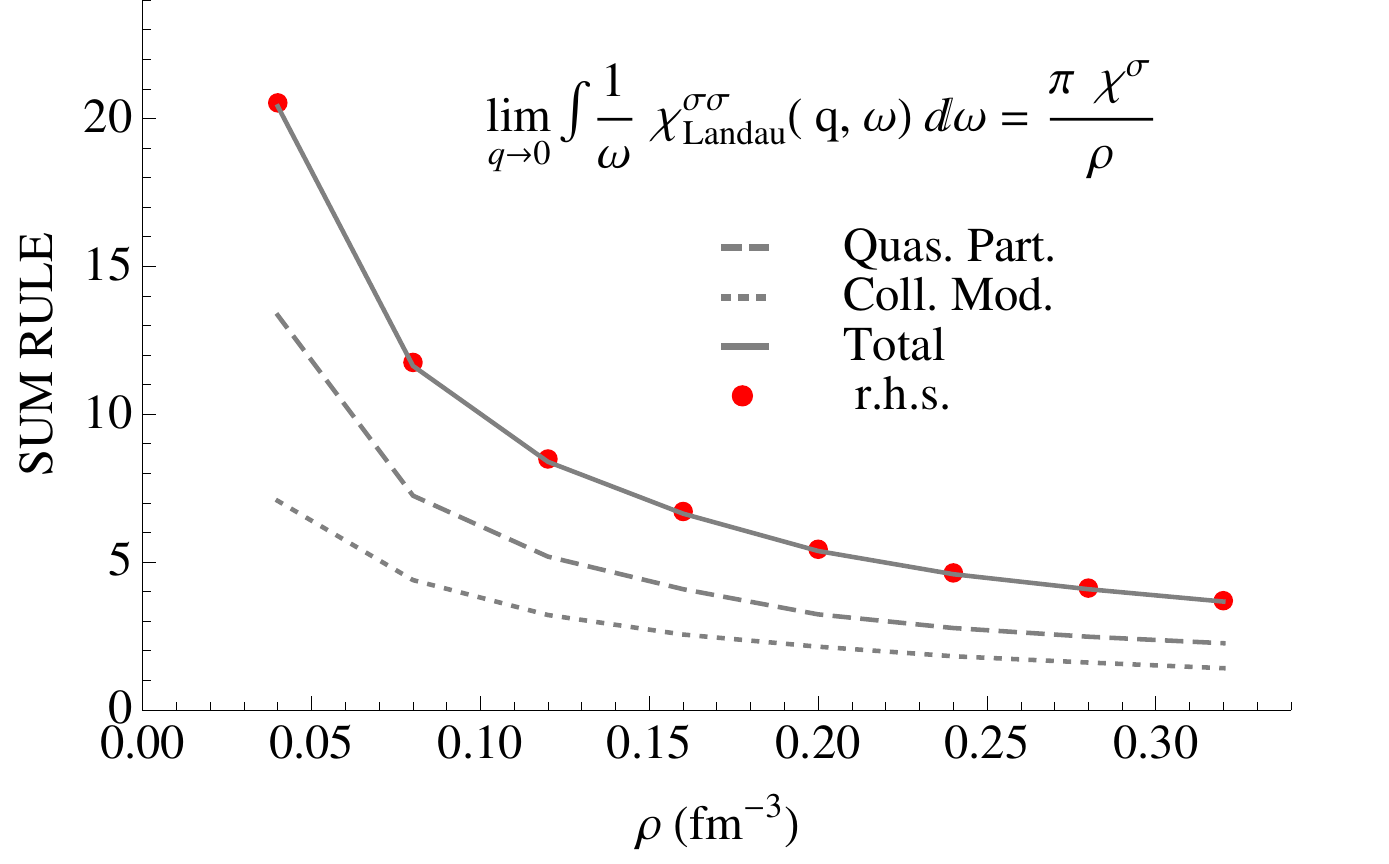}
\caption {Left panel: The $m^{-1}$ sum rule in the density channel. Right panel: Same as in the left panel, but for the spin channel. In this case, the contributions of the collective mode (dotted line) and incoherent excitations (dashed line) are displayed separately. The observed 
trend reflects the one shown in fig. \ref{Static},  showing that the collective mode plays a relevant role in the low momentum limit.}
\label{SUM1}       
\end{figure}

In fig. \ref{SUM1}, we show the behavior of $m^{-1}_{\rho/\s}$. The solid lines represents the right hand side of eq. \eqref{boh1} and \eqref{boh2},  
while the dots correspond to the left hand side. The results have been obtained for a neutron matter density $\rho=0.16$ fm$^{-3}$, for both the 
density and spin density channels. We can clearly see the two different contributions coming from single quasiparticle and collective mode
 excitation. As already pointed out, the latter gives a large contribution in the static limit. 

\begin{figure}[htbp]
\begin{center}
\includegraphics[scale=0.6]{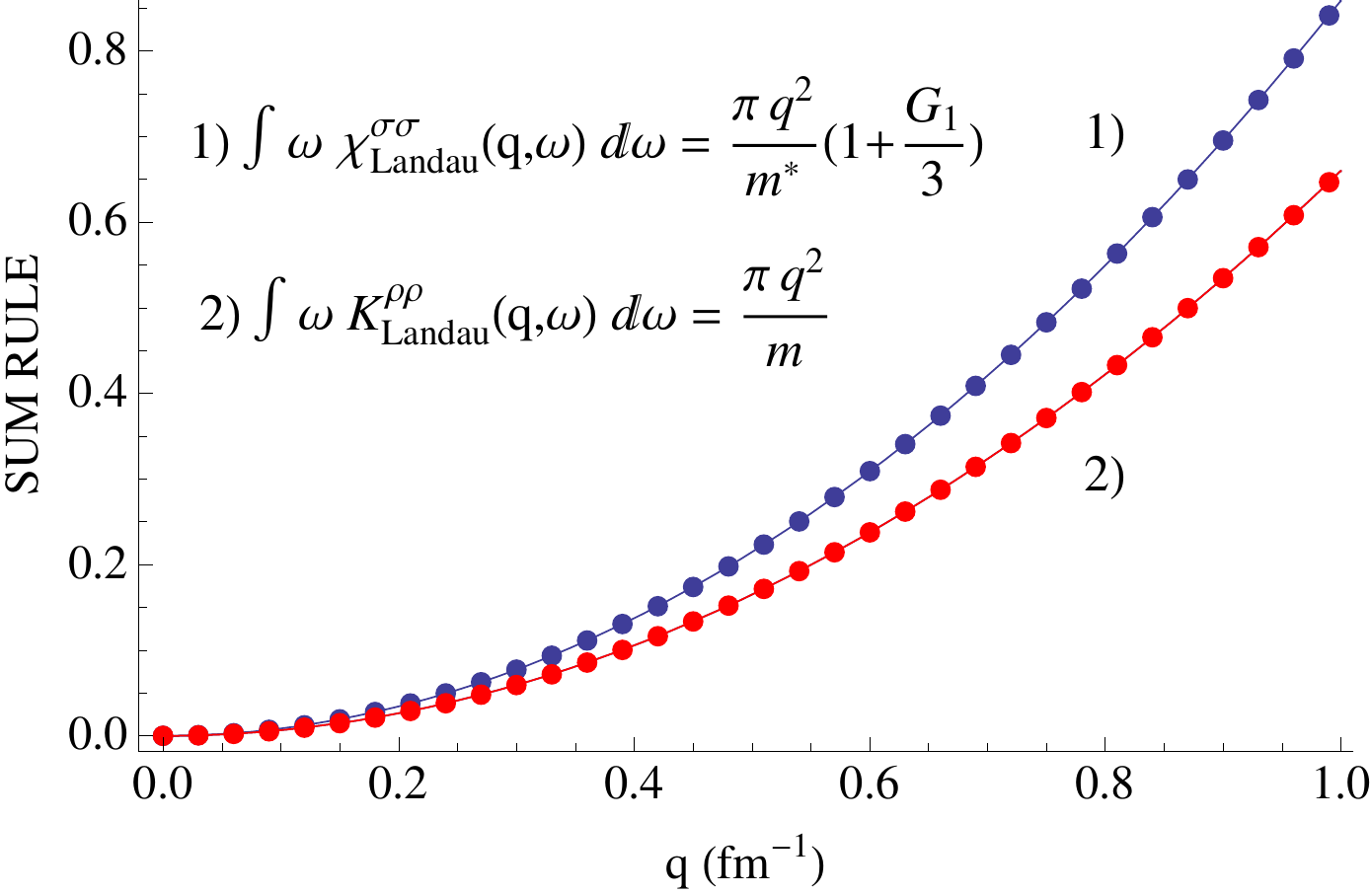}
\caption{The $m^{1}_{\rho/\s}$ sum rule at $\rho=0.16$ fm$^{-3}$, as a function of momentum transfer. The solid line represents the 
right hand side of eqs. \eqref{m11} and \eqref{m12}, while the dots correspond to the left hand side. The observed behavior do not change significantly with the density. }
\label{SUM}  
\end{center}     
\end{figure}

Let us now consider the $n=1$ sum rules. The integral of the Landau theory response in the left hand side can be evaluated explicitly 
both for the density and spin-density channels (see fig.  (\ref{SUM})) . Within this approach we get,
\begin{align}
\label{m11}
m^{1}_{\rho}&=\lim_{|\bq|\to 0}\int \omega\, K\rb{Landau}^{\rho\rho}(\bq,\omega) d \omega=\frac{\pi q^2}{m}, \\
\label{m12}
m^{1}_{\s}&=\lim_{|\bq|\to 0}\int \omega\, \chi\rb{Landau}^{\s\s}(\bq,\omega) d \omega=\frac{\pi q^2}{m^*}(1+\frac{1}{3}G^1) \ .
\end{align}
While in the density channel the above expression is exactly the same as the one obtained from eq. (\ref{sumrules}), the corresponding 
expression for the spin channel turns out to be slightly different. The right hand side is still depending on both the symmetric and antisymmetric 
Landau parameters, and is no longer a function of the bare mass and the momentum transfer only ($\pi \bq^2/m$). This is intrinsically related to the 
degree of approximation inherent in Landau theory itself. In \cite{Olss1}, the difference between eq. (\ref{sumrules}) and the values 
resulting from Monte Carlo simulations has been used to estimate the relative importance of multiparticle excitations in the kinematical regime where 
$\omega\geq |\bq|$.



%% file: Chap_Neutrino.tex
The neutrino cross sections in nuclear matter plays a central role in determining neutrino transport in both supernov\ae $ \ $ and neutron stars. Many different scattering and absorption processes can occur and , at least in principle, they should all be taken into account. 
However, the importance of a specific process mainly depends on two distinct factors: the energy of the incoming neutrino and the composition of 
nuclear matter, which in turns depends on the density regime we are interested in. 

A list of the potentially relevant neutrino processes can be found in several papers, see, e.g., Refs.  \cite{Reddy3,Reddy2,Sawyer}. They include neutral current neutrino scattering off nucleons, alpha particles and nuclei, first discussed in the pioneering work of  Freedman \cite{Freedman}, super-allowed charged-current neutrino and antineutrino absorption on nucleons, neutrino-neutrino scattering and neutrino-antineutrino absorption. The 
inverse of several neutrino production processes, such as bremsstrahlung of neutrino-antineutrino pairs and direct and modified URCA processes, 
also contribute. 

An incomplete list of the processes that may be relevant at relatively high density includes 
\[
\begin{array}{lll}
 \nu\rb{e}+n\to e^{-}+p \ , &  \bar{\nu}\rb{e}+p\to e^{+}+n \ ,   & \nu\rb{\ell}+A\to \nu\rb{\ell}+A^{*} \ ,  \\
 \nu\rb{\ell}+n\to \nu_{\ell}+n \ , &  \nu\rb{e}+n+n\to \ell^{-}+p+n \ , & \nu\rb{\ell}+A\to\ell^{-}+A^{*} \ .     
\end{array}
\]
It should be noted, however, that in the density region corresponding to the neutron star core, where the ground state of the matter
is likely to be a neutron liquid, the occurrence of many processes may be inhibited by the absence of nuclei, or of a significant
proton fraction. Typically,  purely leptonic processes, such as neutrino-lepton scattering, are negligible in neutron stars. It has been 
shown that for neutrino with energy $E_\nu \sim10$ MeV the charged-current cross section is two orders of magnitude larger than 
the $\nu\rb{e}-e$ scattering cross section \cite{Reddy2}. Nevertheless, this process may be important for the thermalization
of emergent muon neutrinos \cite{Thompson}. 

Changing the initial conditions gives rise to several different processes. For example, at low temperature ($T\leq 3-5 \ {\rm MeV}$) 
and relatively low density ($\rho\simeq 10^{12}-10^{13} \ {\rm g/cm}^{3} $) heavy nuclei are expected to be present, and 
dominate the neutrino cross section due to coherent scattering. In these conditions, the interparticle distance is large 
$d \simeq 20-40 $ fm compared to the range of nuclear forces, the 
contribution of which is small, and correlations between particles are driven by the Coulomb interaction $F\rb{C}\simeq Z^{2}e^2/d$, with 
$Z\simeq 25$. For low-energy neutrinos these coherent scattering processes are far more important than those involving nucleons \cite{Freedman}. 

As the density increases, the nuclei get larger, and their separation distance descreases. Novel heterogeneous phases of matter, dubbed 
``pasta phases'', have been predicted to occur: the nuclear surface begins to provide a significant contribution to the energy, and nuclei 
deform progressively from spherical to rod-like and slab-like shapes, in order t arrange themselves in a more energetically favorable 
configuration\cite{Ravenhall}. The occurrence of the pasta phase has been recently confirmed by a quantum molecular-dynamic inspired 
model \cite{Watanabe}. A further increase of the density leads to the appearance of a homogeneous nucleon liquid. 
In this regime, low-energy neutrino interactions are strongly affected by nuclear matter dynamics, leading to the 
occurrence of many-body excitations, such as  resonances and collective modes \cite{BS}. This work is focused on this density region. In the 
following Sections we will discuss the interaction of low-energy neutrino with neutron matter. 
Since the timescale of nuclear dynamic is much shorter than that associated to weak processes, we will assume that matter be in equilibrium
and describe many-body effects within the framework of linear response theory using the effective interaction described in Chapter \ref{CB}. 

\section{Neutrino-neutron interactions}
The interactions of  low-energy (typically few MeV) neutrinos can be described within the low-energy limit of the standard model of 
electroweak interactions. The corresponding charged- and neutral-current interaction lagrangian densities read
\begin{eqnarray}
\mathcal{L}^{{\rm cc}} = \frac{G\rb{F}}{\sqrt{2}} l\rb{\mu} j\rb{W}\rp{\mu} & \textrm{for}  & \nu\rb{\ell}+ n\to \ell+X  \nonumber \\
\mathcal{L}^{{\rm nc}} = \frac{G\rb{F}}{\sqrt{2}} l\rb{\mu} j\rb{Z}\rp{\mu} & \textrm{for}  &\nu\rb{\ell}+ n\to \nu\rb{\ell}+X , \nonumber
\end{eqnarray}
where $G\rb{F}\simeq 1.436\times 10^{-49}$ erg cm$^{-3}$ is the Fermi coupling constant. While the leptonic current is well defined in terms
of the degrees of freedom of the fundamental theory, the hadronic current involves composite particles and its form depends on the 
four momentum transfer associated with the process. However, when the latter is very small compared to the nucleon mass of \footnote{And also smaller then 1/$L$, $L$ being the length scale specifying the size of the target particle, as pointed out in Ref.  \cite{Reddy2}.    } the hadronic current can be approximated with the simple form first proposed by Fermi to explain neutron $\beta$-decay. Hence, we can write 
\[
\begin{array}{cc}
l\rb{\mu}=\bar{\ell}(1-\gamma_5)\gamma\rp{\mu} \nu\rb{\ell} & \ \ \ , \ \ \  j\rb{W}\rp{\mu}=\bar{X}(x)(g\rb{V}-g\rb{A}\gamma_5)\gamma\rb{\mu} n(x) \nonumber \\
l\rb{\mu}=\bar{\nu}\rb{\ell}(1-\gamma_5)\gamma\rp{\mu} \nu\rb{\ell} & \ \ \ , \ \ \  j\rb{Z}\rp{\mu}=
\bar{X}(x)(c\rb{V}-c\rb{A}\gamma_5)\gamma\rb{\mu} n(x) \  , \nonumber
\end{array}
\]
where $X$ labels the final hadronic state and  c\rb{V}, g\rb{V}, c\rb{A} and g\rb{A} denote the vector and axial coupling constant
  
\subsection{Neutrino-nucleon cross section in vacuum}

Lets us start by analyzing in detail the inclusive cross section of only process we need to consider in pure neutron matter
\[
\nu\rb{\ell}(k)+ n(p) \to \nu\rb{\ell}(k')+n(p') \ .
\]

We will first focus on the description of the above reaction in free space, and include medium modifications at a later stage. 
At leading order in perturbation theory, the invariant amplitude reads
\begin{equation}
\label{amplitude}
\mathcal{M}=\frac{G\rb{F}}{\sqrt{2}}\bar{u}\rb{\nu}({\bf k}',s)\gamma\rp{\mu}(1-\gamma_5)u\rb{\nu}({\bf k},s)\langle n(p')|J_{Z}^{\mu}|n(p)\rangle  ,
\end{equation}
the four momenta of the participating particles being defined as
\[
p=(E,{\bf p}), k=(\epsilon,{\bf k}), p'=(E',{\bf p}'),k'=(\epsilon',{\bf k}')  \ .
\]
Using standard techniques (see, e.g. Ref.  \cite{Mandl}) one can write the differential cross section in the Lab frame in the form
\begin{equation}
\label{xsec1}
d\sigma=\frac{4M^2}{4 [(k\cdot p)^2]^{1/2}}\overline{|\mathcal{M}|^2}(2\pi)^4\delta^4(k'+p'-k-p) \frac{d^3 \bp'}{2E' (2\pi)^3}\frac{d^3\bk'}{2\epsilon'(2\pi)^3} \ ,
\end{equation}
where $M$ denotes the neutron mass and the bar over $\mathcal{M}^2$ refers to fact that the squared transition amplitude is averaged over the 
spins of the initial state particles, $s$ and $\sigma$ and summed over the spins of the final state particles, $s^\prime$ and $\sigma^\prime$.

Substitution of eq. \eqref{amplitude} into eq. \eqref{xsec1} yields the expression
\begin{align}
d^2\sigma(\epsilon,p)=&\frac{G\rb{F} \  4 M^2 }{4 [(k\cdot p)^2]^{1/2}}\int \frac{\di^3 \bp'}{2E'(2\pi)^3} 
\frac{1}{2} \sum_{\sigma \sigma^\prime} \langle n(p)|J_{Z}^{\mu}|n(p')\rangle\langle n(p')|J_{Z}^{\mu}|n(p)\rangle   \nonumber\\ 
&\times(2\pi)^4\delta^4(k'+p'-k-p)\underbrace{\left( \frac{1}{2} \sum_{s s'} l^{*\mu}l^{\nu}\right)}_{L^{\mu\nu}}\frac{| {\bf k}'|^2d|{\bf k}'|d\Omega}{2\epsilon'(2\pi)^3} \  ,\nonumber
\end{align}
that can be rearranged in the concise form
\begin{equation}\label{vacuum_cross}
\frac{d^2\sigma(\epsilon,p)}{d\epsilon' d\Omega}=\frac{G\rb{F}^2}{(2\pi)^2}\frac{2M \epsilon'}{16 [(k\cdot p)^2]^{1/2}}L^{\mu\nu}W_{\mu\nu} ,
\end{equation}
where $L^{\mu \nu}$ and $W_{\mu \nu}$ are the leptonic and hadronic tensor, respectively, defined as
\be\label{leptontensor}
L^{\mu\nu}=8[k'^{\mu}k^{\nu}+k^{\mu}k'^{\nu}-g^{\mu\nu}(k'\cdot k)-i\epsilon^{\mu\alpha\nu\beta}k'_{\alpha}k_{\beta}] ,
\ee
and 
\[
W^{\mu\nu}=(2\pi)^3 \ M \sum_{\sigma \sigma^\prime} \int \frac{\di^3 \bp'}{2E'(2\pi)^3} \langle n(p)|J_{Z}^{\mu}|n(p')\rangle \langle n(p')|J_{Z}^{\mu}|n(p)\rangle \delta^4(k'+p'-k-p) .
\]
\subsection{Medium effects}
\subsubsection{Non relativistic Fermi gas model}
We now want to improve upon the differential cross section (\ref{vacuum_cross}), describing neutral current neutrino-nucleon scattering in vacuum, by including the effects of matter. In the region of low momentum transfer we can rely on the non relativistic approximation for nucleons, implying that 
the term arising from the flux simplifies to $ [(k\cdot p)^2]^{1/2}\simeq M\epsilon$ 

Within the Fermi gas model, neutron matter is seen as a collection of non interacting neutrons, the momentum of which is distributed 
according to the Fermi distribution. This model can be easily implemented in the formalism described in the previous Section, since the 
system can still be described in terms of single-particle states. 

The the hadronic current operator is applied to a state describing 
a neutron carrying momentum ${\bf p}$, and the cross section is obtained collecting the contributions of all target particles, weighted with 
the Fermi distribution $f({\bf p})$ according to
\[
d^2\sigma(\epsilon,\Omega)=\frac{V}{2N}\sum_{s_n}\int \frac{\di^3{\bf p}}{(2\pi)^3} f({\bf p})d^2\sigma(\epsilon,\Omega,{\bf p})  \ ,
\]
or, more briefly
\[
\frac{d^2\sigma(\epsilon)}{d\epsilon' d\Omega}=\frac{G\rb{F}^2}{(2\pi)^3}\frac{\epsilon'}{4\epsilon}L^{\mu\nu}W\rp{Matt}_{\mu\nu} ,
\]
the nuclear tensor being given by
\begin{align}
W\rp{Matt}_{\mu\nu}=\frac{1}{4\rho}\sum_{\sigma \sigma^\prime}\int  & \frac{\di^3{\bf p}}{(2\pi)^3} f( {\bf p}) \int \di^3  {\bf p}' [1-f( {\bf p}')] \langle n(p)|J_{Z}^{\mu}|n( p')\rangle \langle n(p')|J_{Z}^{\mu}|n(p)\rangle \nonumber\\
&\times  (2\pi)\delta(\varepsilon_{{\bf p}+{\bf q}}-\varepsilon_{{\bf p}}-w)\delta^3({\bf k}'+{\bf p}'-{\bf k}-{\bf p}) ,
\end{align}
where $\rho=N/V$ and the factor $[1-f({\bf p}')]$ describes the phase space above the Fermi level, available to the final state neutron. 
Rewriting the above expression in terms of the momentum and energy transfer, ${\bf q}={\bf k}-{\bf k'}$ and $w=\epsilon-\epsilon'$ we find
\begin{align}
W\rp{Matt}_{\mu\nu}=\frac{1}{4\rho}\sum_{\sigma \sigma^\prime} \int  \frac{\di^3{\bf p}}{(2\pi)^3} f( {\bf p}) & [1-f( {\bf p}+ {\bf q})] \langle n( p)|J_{Z}^{\mu}|n( p+q)\rangle \langle n( p+q)|J_{Z}^{\mu}|n( p)\rangle \nonumber\\& \times(2\pi)\delta(\varepsilon_{{\bf p}+{\bf q}}-\varepsilon_{{\bf p}}-\omega)  \ .
\end{align}
In the non relativistic limit, 
\[
J_{Z}^{\mu}=(n^{\dag} n) g^{\mu 0}-C\rb{A}(n^{\dag}\sigma_i n) g^{\mu i} \  ,
\]
and the only surviving components of the matter tensor are the density-density $(00)$ and spin-spin $(ii)$ ones
\[
W_{00}=S({\bf q},\omega) \ \ , \ \  W\rp{Matt}_{i j}=S({\bf q},\omega)\rp{A}\delta_{ij}\quad W\rp{Matt}_{0 i}=0  \ ,
\]
while all off diagonal contributions vanish. Let us focus on the density channel first. Defining 
\[
\rho\rb{p}(\bq)=\langle\bp|n^{\dag} n|\bp-\bq\rangle ,
\]
the expression of $S({\bf q},\omega)$ can be written in the form
\begin{align}\label{corr}
S({\bf q},\omega)=\frac{1}{4\rho}\sum_{\sigma \sigma^\prime}\int & \frac{\di^3{\bf p}}{(2\pi)^3} f( {\bf p})[1-f( {\bf p}+ {\bf q})] 
\rho\rb{p}({\bf q})\rho\rb{p}(-{\bf q})(2\pi)\delta(\varepsilon_{{\bf p}+{\bf q}}-\varepsilon_{{\bf p}}-\omega)\nonumber\\
=&\frac{1}{\rho} \int \frac{dt}{2 \pi} \ {\rm e}^{i \omega t} \langle  \rho({\bf q},t)\rho(-{\bf q},0)\rangle_{FG}  \ ,
\end{align}
where $\rho({\bf q},t)$ is time-dependent operator in the Heisenberg picture and $\langle \ldots \rangle_{FG}$ denotes the expectation 
value in the Fermi gas ground state.

In the spin-spin channel we find a similar result (note that as all diagonal terms are the same, we can choose $i=3$)
\[
S\rp{A}({\bf q},\omega)=\frac{1}{\rho} \int \frac{dt}{2 \pi} \ {\rm e}^{i \omega t}  \langle  \sigma_3({\bf q},t)\sigma_3(-{\bf q},0)\rangle_{FG}  \ .
\]

\subsubsection{Including interactions}
Once we have obtained the expression of the cross section in term of a ground state expectation value, including the interactions  is formally easy. 
One has to make the replacement $\langle\ldots\rangle_{FG}$ with $\langle\ldots\rangle_{INT}$, meaning that the expectation value is 
now in the ground state of the interacting system. We remind the reader that we are allowed to do so since neutrino physics and nuclear 
dynamics correspond to different time scale,  and can therefore be decoupled. The length scale of the perturbation induced by the neutrino 
is much larger than the mean free path of a neutron in matter. As a result,  the neutrino interacts with the system in equilibrium, and the zero 
sound mode can be excited.

Contracting of the lepton and matter tensors yields
\[
L^{\mu\nu}W^{{\rm matt}}_{\mu\nu}=\epsilon\epsilon'\Big[(1+\cos\theta)S+(3-\cos\theta)S_A\Big]  \ ,
\]
where $\theta$ is the angle between the directions of the incoming and outgoing neutrino. The resulting expression of the differential cross section is 
\[
\frac{d^2\sigma(\epsilon)}{d\epsilon' d\Omega}=\frac{G\rb{F}^2}{(2\pi)^3}\frac{\epsilon'^2}{4}(1+\cos\theta)S(\bq,\omega)+
(3-\cos\theta)S_A(\bq,\omega) \  ,
\]
and the neutrino mean free path is given by
\begin{align}
\label{sigma_corr}
\frac{1}{\lambda(\epsilon)}& = \rho\sigma(\epsilon) \\
\nonumber
 & =\frac{G\rb{F}^2\rho}{4(2\pi)^3}\int \di^3{\bf k'}[(1+\cos\theta)S({\bf k}-{\bf k}',\epsilon-\epsilon' )+(3-\cos\theta)
 S_A({\bf k}-{\bf k}',\epsilon-\epsilon' )]  \ .
\end{align}
Note that, from a purely formal point of view, the above expression can be readily generalized to the case for hot neutron matter, substituting the 
ground state expectation value entering the definitions of $S$ and $S_A$ with an ensemble average. 

The nuclear correlation function $S$ and ,$S_A$ are related to response function through the fluctuation-dissipation theorem \cite{Bay}. Consider an external probe $P({\bf x},t)$ coupling to an observable of the system, say the density, through $\int \di{\bf x} \rho({\bf x},t) P({\bf x},t)$ then the following relation between the correlation $S$ and the response function $K$  holds:
\[
S({\bf q},\omega)=-\frac{2}{\rho}\frac{1}{(1-\eu^{\beta \omega})}\textrm{Im}K({\bf q},\omega)  \ .
\]
It follows that we can obtain the mean free path at $T \neq 0$ from
\bea\label{mfp}
\frac{1}{\lambda(\epsilon)}=-\frac{G\rb{F}^2}{2}\int \frac{\di^3{\bf k'}}{(2\pi)^3(1-\eu^{-\beta \omega})}\!\!\!&\!\!\!&\!\!\!\!
\Big\{ (1+\cos\theta)\textrm{Im}\Big[K({\bf k}-{\bf k}',\epsilon-\epsilon' )\Big]+\nonumber\\
\!\!\!&\!\!\!&\!\!\!\!+(3-\cos\theta)\textrm{Im}\Big[\chi^{\sigma\sigma}({\bf k}-{\bf k}',\epsilon-\epsilon' )\Big] \Big\}.
\eea

\subsubsection{Phace space considerations}

In this Section we review briefly outline the calculation of the phase space integrations \cite{Iwamoto}. 
In an isotropic system the correlation functions only depend on the magnitude of the momentum transfer momentum $q=|{\bf k}-{\bf k}'|$ 
and the energy transfer $\omega=\epsilon-\epsilon' $. Choosing the $z$-axis along the direction of the momentum of the incoming 
neutrino, $\bk$, $\theta$ of eq. (\ref{mfp}) is the angle between $\bk$ and $\bk^\prime$, which can be written as a function of $q$ 
and $\omega$ according to
\begin{equation}\label{anglekk}
q^2=k^2+k'^2-2kk'\cos\theta \longrightarrow \cos\theta=\frac{1-\frac{\omega}{k}+\frac{1}{2}\frac{\omega^2}{k^2}-\frac{1}{2}\frac{q^2}{k^2}}{1-\frac{\omega}{k}} \  .
\end{equation}
In eq. (\ref{mfp}), after the trivial integration over the azimuthal angle we are left with $ 2\pi q^2 dq d\cos\theta_{qk}$, 
with  
\[
\cos\theta_{qk}=\frac{1}{2}\frac{q}{k}-\frac{1}{2}\frac{\omega}{k}\frac{\omega}{q}+\frac{\omega}{q}  \ ,
\]
as required by energy and momentum conservation.
Changing  variables from $(q,\cos\theta)$ to $(q,\omega)$ and taking into account the Jacobian of the transformation we find
\[
dq d\cos\theta\rb{qk}=|J(\frac{\partial q\cos\theta}{\partial qw})|dq dw
\]
with
\begin{equation}\label{J}
 |J|=|1\cdot\frac{\partial \cos\theta\rb{qk}}{\partial w}|=\left|(1-\frac{w}{k})\frac{1}{q}\right| .
 \end{equation}

 \begin{figure}[htbp]
\begin{center}
\includegraphics[scale=0.45]{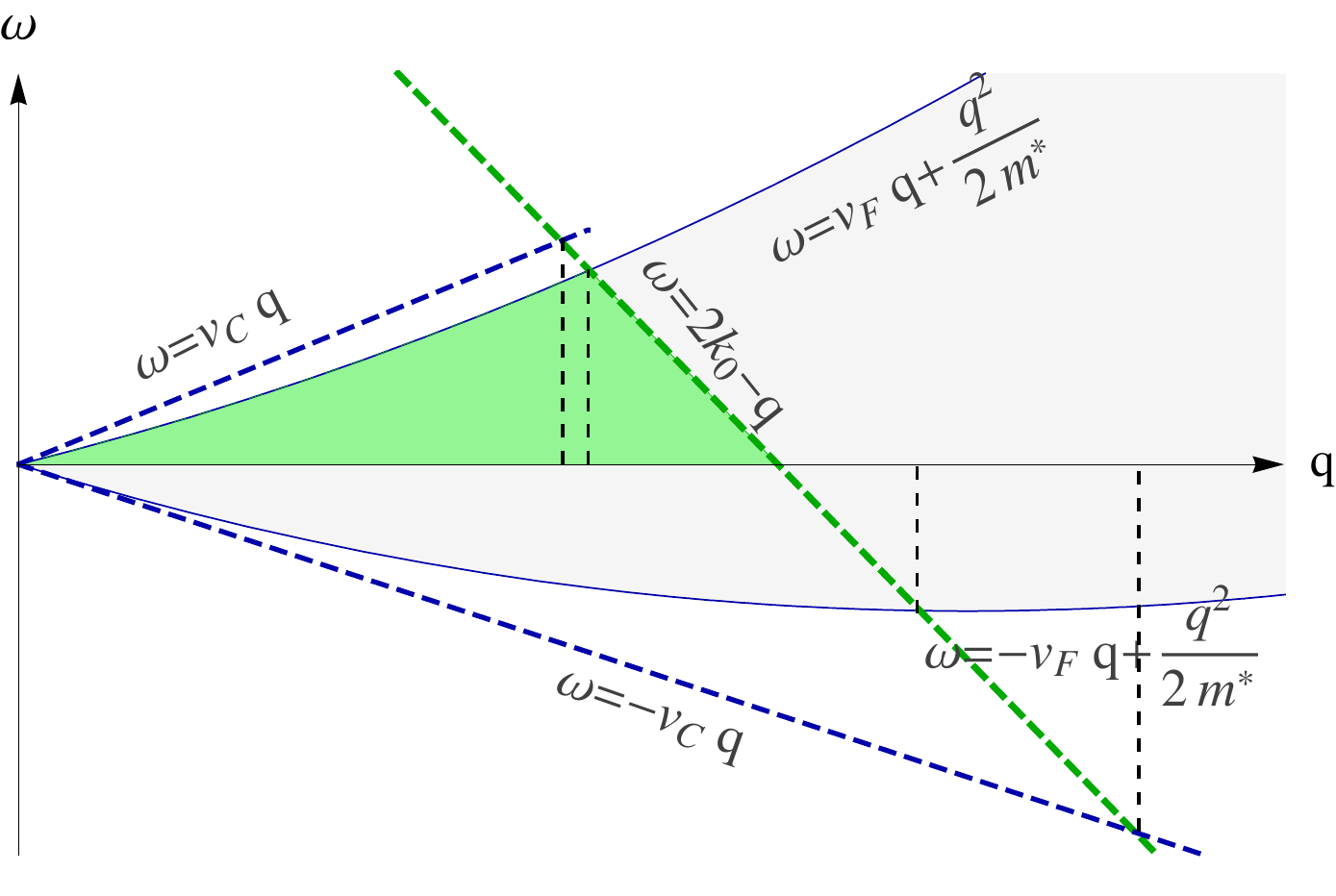}
\includegraphics[scale=0.45]{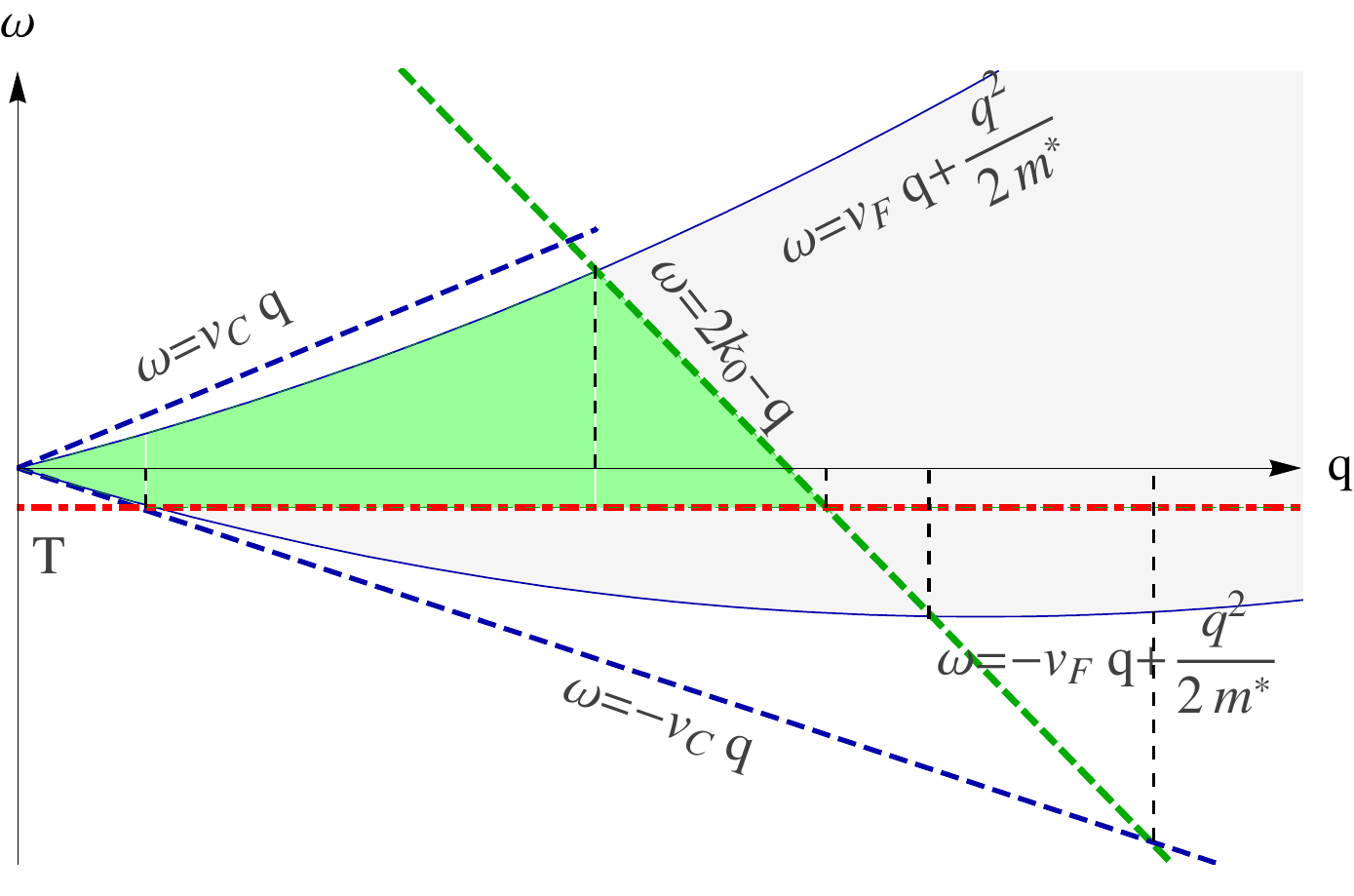}
\caption{Left panel: Qualitative sketch of the phase space available at $T=0$ (green region). The light gray area represents the region 
determined by the degenercy condition only, eq. (\ref{pS1}). Right panel: Same as the left panel, but for  $T>0$. It clearly appears 
that the available extends into the $\omega<0$ region, corresponding to de-excitation of the system. }
\label{phase_space}
\end{center}
\end{figure}

In conclusion, we obtain an expression in terms of incoming neutrino momentum, $k$, momentum transfer, $q$, and energy transfer $\omega$
 \begin{align}
\frac{1}{l(k)}&=\frac{G\rb{F}^2}{2(2\pi)^2} \ \int \frac{\di q\di \omega}{(1-\eu^{-\beta \omega})} \left| \left(1-\frac{\omega}{k}\right) \right| q \\ 
&\times \left[(1+\cos\theta(q,\omega,k))K(q,\omega )+(3-\cos\theta(q,\omega,k))\chi^{\sigma\sigma}(q,\omega )\right],
 \end{align}
where $\cos\theta(q,w,k)$ is given by eq. (\ref{anglekk}). The integration range is dictated by energy and momentum conservation. In the case of the
incoherent excitation spectrum we have
\begin{align}
\omega&=\frac{{p'}^2}{2m^{*}}-\frac{p^2}{2m^{*}}=k-k' \ , \label{pConLaw}\\
{\bf q}&={\bf p'-p}={\bf k-k'}  \ . \label{qConLaw}
\end{align}
Substitution of $p'$ obtained from eq. (\ref{pConLaw}) into eq. (\ref{qConLaw}) allows one to determine the limits of the integration 
over the energy transfer $\omega$
\begin{equation}\label{pS1}
-q v+\frac{q^2}{2m^{*}}\le \omega \le q v+\frac{q^2}{2m^{*}} \ \ \ , \ \ \  v=\frac{p}{m^{*}}  \ .
\end{equation}
For a degenerate system, the maximum value of $v$ is the Fermi velocity, $v\rb{F}$, and the term $q^2/2m^*$ is usually neglected, with respect 
to $q v_F$. A second constraint  on $\omega$,  arising from neutrino kinematics, is 
\begin{equation}\label{pS2}
|\omega|\le q\le |\omega-2k |  \ .
\end{equation}
Equations (\ref{pS1})-(\ref{pS2}) determine the integration region for incoherent single-particle excitations.  

In the spin channel we also have the contribution of collective modes. Here the integration range is determined by eq. (\ref{pS2}) only, since the 
dispersion relation of these excitations is fixed: $w=\pm v\rb{C} q$, where $\pm$ corresponds to phonon emission or absorption. 
The resulting integration limits are
\[
0\le q \le \frac{2k}{1\pm v\rb{C}} \ \ \ \ \  ,\ \ \ \ \ \omega=\pm v\rb{C} q  \ .
\]

Note that, so far, neither degeneracy nor temperature have been taken into account. 
Let us first include degeneracy at $T=0$. Now all momenta below $p\rb{F}$ are filled, and there are no quasiparticles above the Fermi level. 
Although the system can be excited, with $\omega>0$, de-excitation is suppressed by Pauli principle. In this case we should add the 
condition (see fig. \ref{phase_space}) :
\be\label{w>0}
\omega>0 \ .
\ee
Turning on the temperature, the energy transfer is bounded from below, as de-excitation is possible only in a range of energies $\sim kT$ around 
the Fermi energy. As a consequence, 
\begin{equation}\label{negEn}
-kT<\omega  \ .
\end{equation}
The integration region is determined by the intersection of the above conditions. 

Figure \ref{phase_space} qualitatively illustrates the integration regions for both cold and hot neutron matter. 
A non zero temperature results in the extension of the available space to the  $\omega<0$ region,  corresponding to 
de-excitation of the system. As the temperature increases, the lower limit can change appreciably, depending on whether or not the 
condition $-kT > -q v_{F}+q^2/2m^{*}$ is fulfilled. When $-kT<-q v_{F}+q^2/2m^{*}$, degeneracy can also play an important 
role in limiting the phase space.

\section{Numerical results}
This Section is devoted to the discussion of the numerical results of our analysis of the neutrino mean-free-path, $\lambda$. 

\begin{figure}[htbp]
\begin{center}
\includegraphics[scale=0.6]{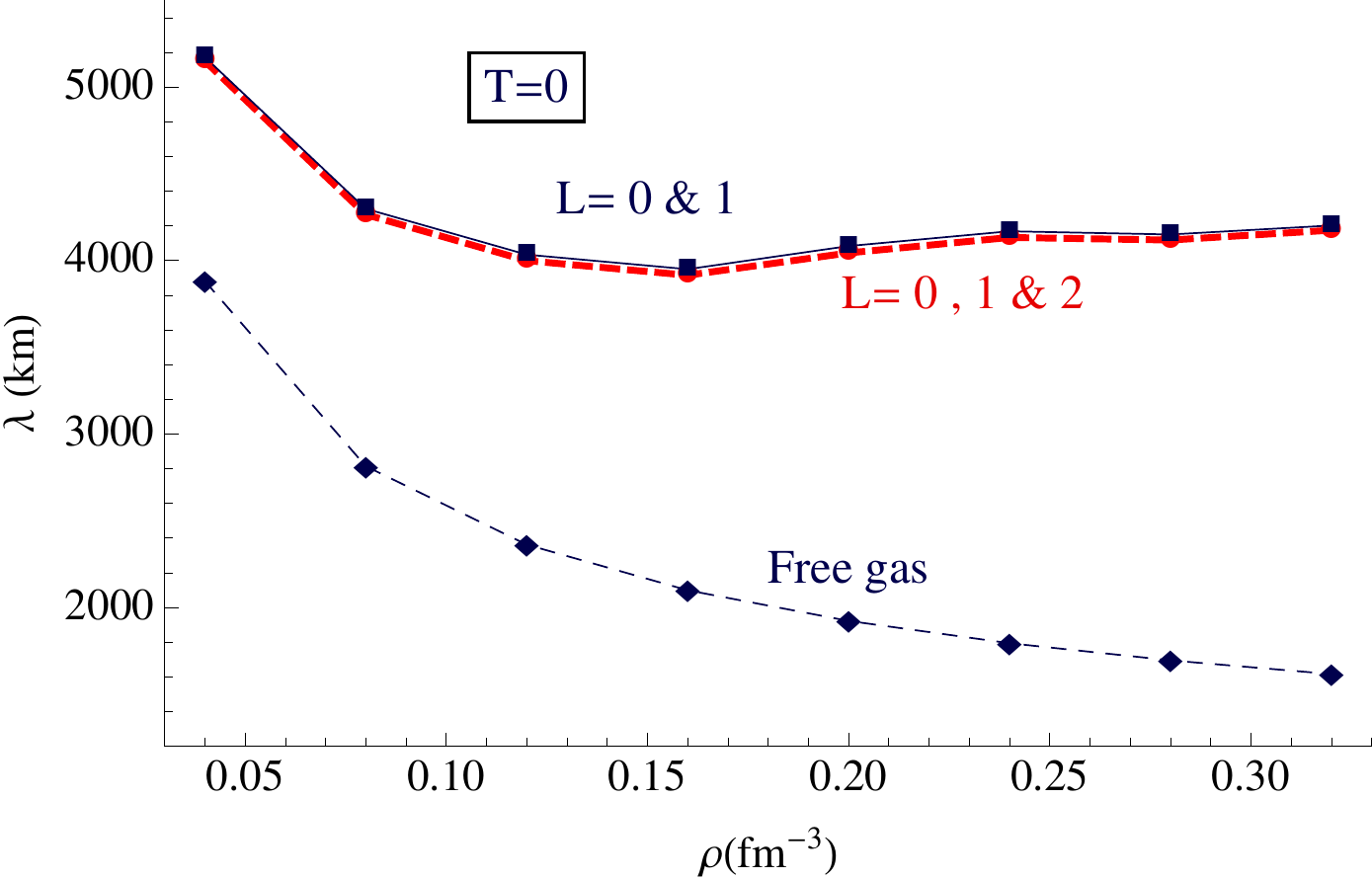}
\caption{Density dependence of the mean free path of a non degenerate neutrino with an energy $\epsilon = 1$ MeV in neutron matter at $T=0$.
The density and spin-density structure functions have been computed using the Landau parameters
$F_\ell$ and $G_\ell$ of Table \ref{LT1} with $\ell =0, \ 1$ (open dots) and $\ell =0, \ 1$ and $2$ (solid line). For comparison the dot-dash line shows
the mean free path in the free neutron gas.}
\label{T0}
\end{center}
\end{figure}

Figure \ref{T0} shows the density dependence of the mean free path of a non degenerate neutrino with an energy $\epsilon = 1$ MeV in neutron matter at $T=0$. The results have been  obtained from Eq.(\ref{sigma_corr}), using the density and spin-density structure functions computed 
using the Landau parameters $F_\ell$ and $G_\ell$ listed in Table \ref{LT1} with $\ell =0, \ 1$ (solid line) and $\ell =0, \ 1$ and $2$ (thick dashed line). Comparison with the mean free path in a
free neutron gas, displayed by the dashed line, shows that inclusion of interaction effects leads to a large enhancement of $\lambda$ 
over the whole density range. This behavior is to be ascribed to short range correlations arising from the repulsive core of the $NN$ 
interaction, that prevent two interacting nucleons from being to close to one another. 

\begin{figure}[htbp]
\begin{center}
\includegraphics[scale=0.48]{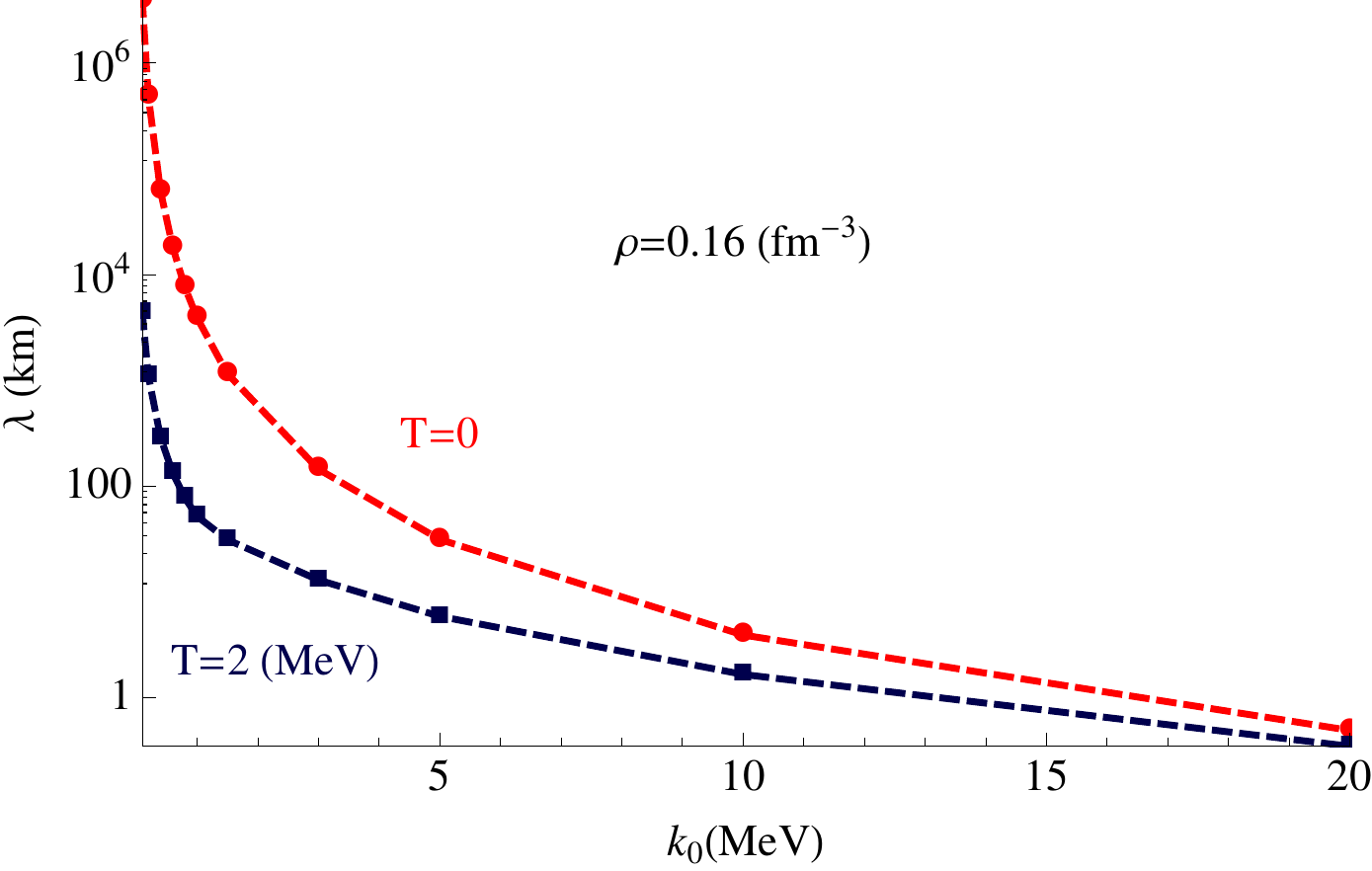}
\includegraphics[scale=0.48]{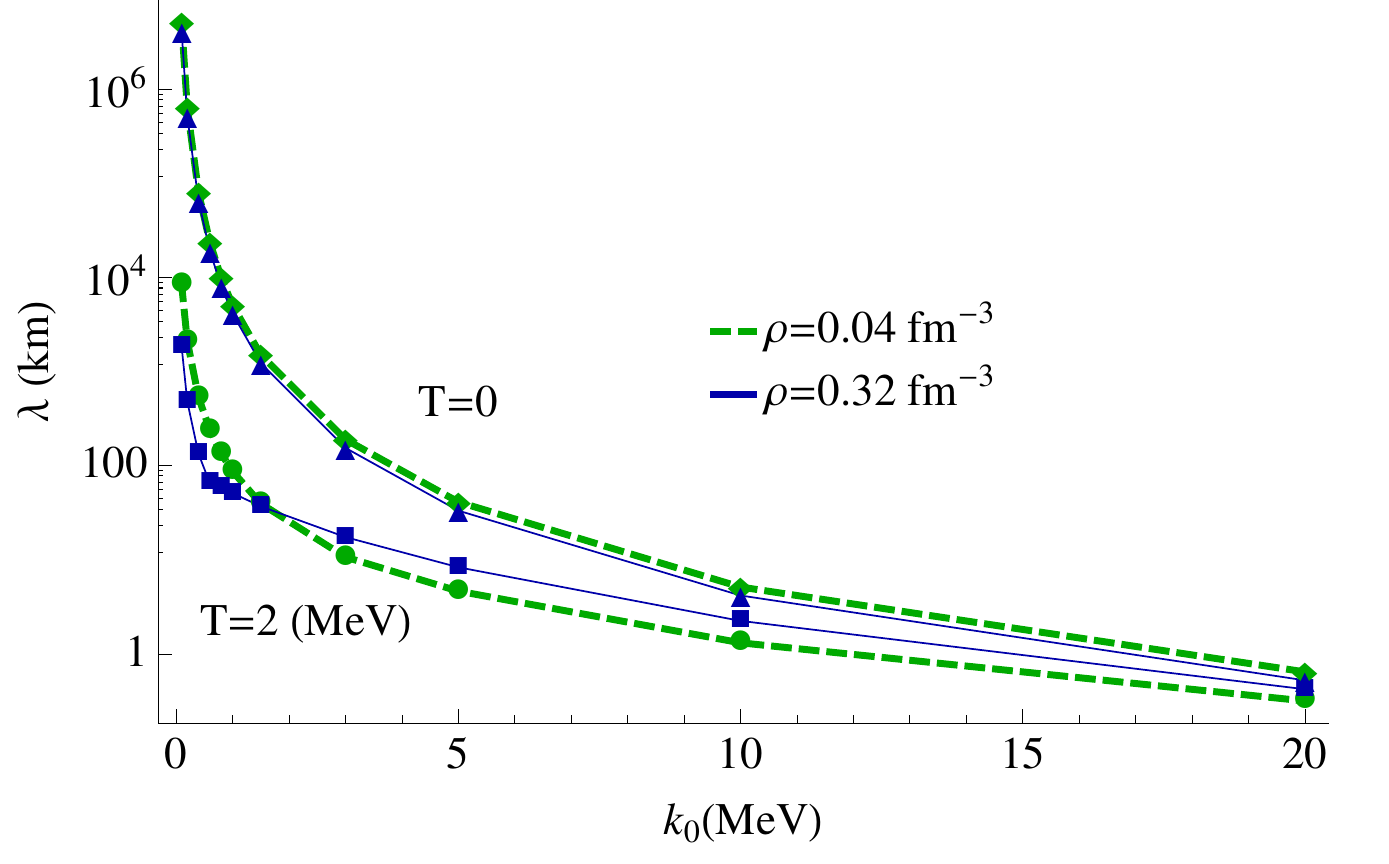}
\caption{Left panel: Energy dependence of the mean free path of a non degenerate neutrino in neutron matter 
at different temperatures. Right panel : Same as in the left panel, but for different neutron matter densities.  }
\label{qDEP}
\end{center}
\end{figure}

The left panel of Fig. \ref{qDEP} shows the energy dependence of the mean free path in neutron matter at density $\rho=0.16$,
corresponding to the Fermi temperatures $T_F = 35$ MeV.
The upper  and lower curves have been obtained setting the temperature to $T=0$ and $2$ MeV, respectively.
The right panel also displays the density dependence, in the range $0.04 \leq \rho \leq 0.32 \ {\rm fm}^{-3}$.
It appears that the cross section exhibits a $\s\approx q^3$ dependence on the energy of the incoming neutrino, 
and that at approximately $ \epsilon \sim T$ thermal effects begin to be less important. 
At $\epsilon \gg T$ the curves corresponding to $T=0$ and $T=2$ MeV are very close to each other, implying that 
the temperature dependence is negligible. Moreover, due to the stiffness of the EOS,  thermal effects appear 
to be less pronounced in the high-density region.

\begin{figure}[htbp]
\hspace{-13mm}
\includegraphics[scale=0.57]{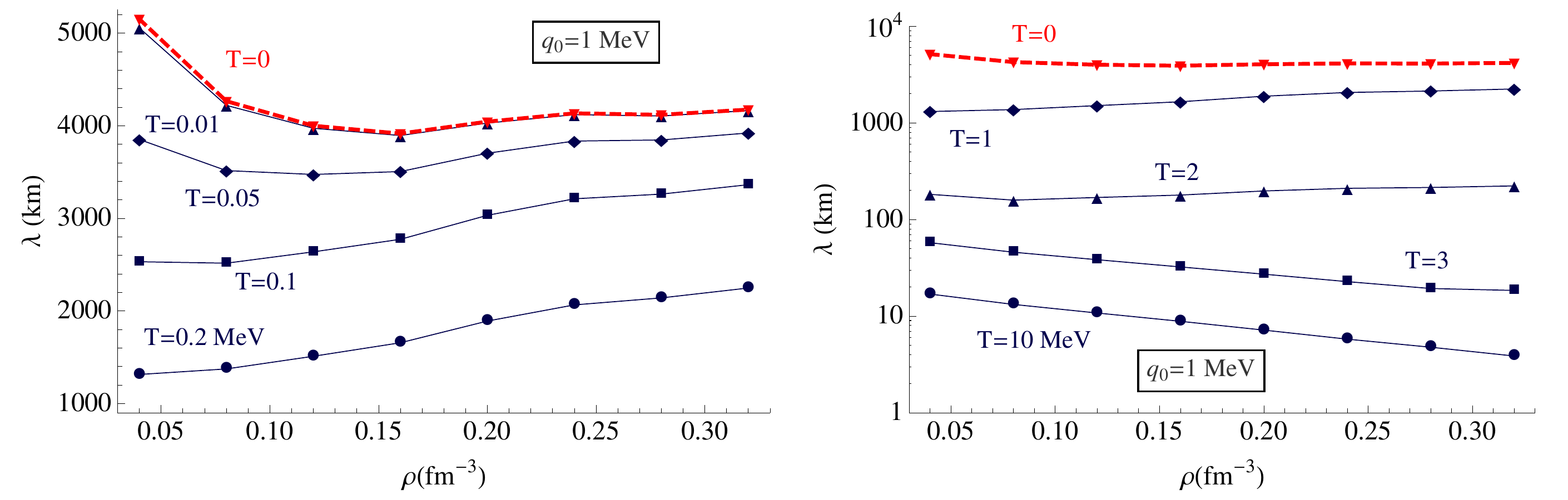}
\caption{Density dependence of the mean free path of a non degenerate neutrino with energy $\epsilon=1$ MeV.
The curves are labelled according to the values of temperature. The open dots in the left panel correspond to
$T=0.01 \ {\rm MeV}$. }
\label{F2lowT}
\end{figure}

The dependence of the mean free path of a non degenerate neutrino with energy $\epsilon=1$ MeV upon both temperature and
matter density is illustrated in Fig. \ref{F2lowT}. Note that, as the the density range $0.04 \leq \rho \leq 0.32 \ {\rm 
fm}^{-3}$ corresponds to Fermi temperatures $14 \leq T_F  \leq 55 \ {\rm MeV}$, the collisionless condition $T << T_F$ 
is always satisfied. We note again that in the high density region the EOS is very stiff and thermal effects  become important at 
larger values of $T$. However, it must be pointed out that the pattern emerging from Fig. \ref{F2lowT} results from 
the combined effects of several different factors,  such as stiffness, phase space and degeneracy.

Nevertheless the rate of depletion depends on density as shown in fig (\ref{F2lowT}). In the high density region, the EOS is very stiff and particles are very packed one each other. As a consequence the effect of the temperature will start at higher value of $T$ while we expect a deeper effect at low density. Several different factors like stiffness, phase space, degeneracy etc, inter-play at this level for $T\leq 0$ and the trend is shown in the left panel of (\ref{F2lowT}). 

As the temperature increase $T\geq q$, the available phase space saturates both for low and high density. The effect of the temperature depend now only on the factor in eq. (\ref{mfp}), no sharp change are possible and $\lambda$ turns again to decrease with density. The right panel in fig (\ref{F2lowT}) indicates the trend at high temperature.

%% file: Chap_TBF.tex
Over the past two decades, thanks to the availability of interaction models based on effective field theories, combined with the progress of many-body 
techniques, striking advances have been made towards the solution of nuclear Schr\"oedinger equation. Several \emph{ab initio} studies have been successfully carried out for light nuclei. Green's function Monte Carlo and No Core Shell Model calculations, performed using realistic 
potentials, provided exact solutions for $A\leq 12$ . For heavier nuclei, methods like the Self Consistent Green Function (SCGF)\footnote{Which is also applicable to study spectral functions and optical potential in a broad range of energies.}, Coupled Cluster or sn-medium Similarity Renormalization 
Group (SRG) yield results that are still restricted to closed shell plus the addition/removal of one or two nucleons at most.  
G-matrix perturbation theory and the variational approaches based of  CBF, such as the Fermi Hypher-Netted Chain (FHNC) scheme, have been 
extensively applied to infinite nuclear matter. 

The generalization of the available theoretical approaches to describe open shell nuclei, take into account the continuum and use realistic hamiltonians,
including three-nucleon forces (3NF) is now regarded as a most prominent issue in nuclear theory. Recently, a technique to successfully attack the first two 
problems has been developed in Ref. \cite{Soma}. This Chapter is devoted to the the discussion of the third one.  We will provide a brief outline 
of the effects of 3NF illustrated in Fig. \ref{3NF}, and focus on how their inclusion affects the theoretical neutron drip line \cite{Otsuka,Hagen}.

Three nucleon interactions are a needed element of any refined calculations of both atomic nuclei and infinite nuclear matter. Phenomenological 
potential models, which are the most widely used in literature, allow to explain several important properties, from saturation of isospin symmetric 
matter to several features of nuclear spectroscopy. A very effective procedure  to include the effects of 3NF is based on a density dependent 
modification of the two-body potential,  inspired by the well known Fujita-Mijazawa two pion exchange process.

In most of this Thesis work, we have followed the above scheme, as described in Section \ref{Eff_int},  taking into account both the 
attractive and the repulsive contributions to the 3NF in the definition of the effective interaction in a largely phenomenological fashion. 
In this Chapter we discuss a more microscopically rooted approach, in which the density-dependent correction is obtained in a fully-consistent 
background provided by by the Green function formalism. In the second quantization picture, the hamiltonian driving the dynamical interaction 
can be expanded in terms of creation/annihilation operators according to
 \begin{figure}[!h]
 \hspace{-6mm}
 \begin{center}
\includegraphics[scale=.78]{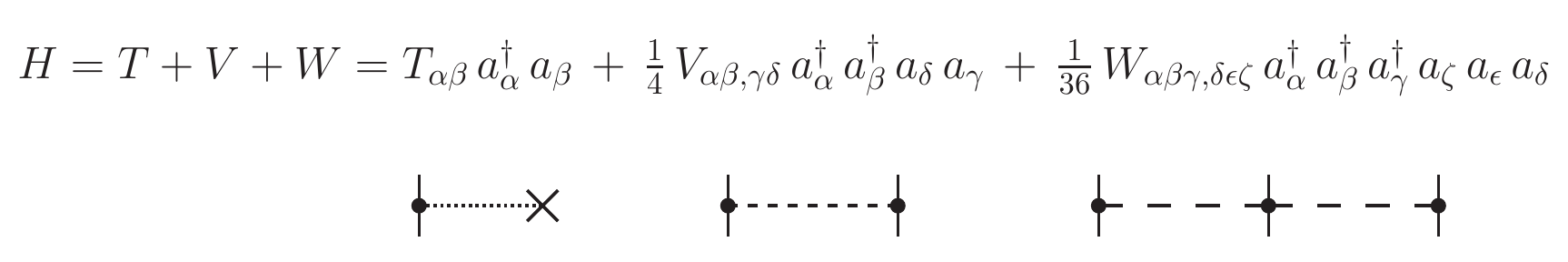}   
\end{center}
\vspace{-4mm}  
\end{figure}
\\
where $T$ , $V$ and $W$ represent the kinetic energy and the \emph{bare} two-body and three-body interactions, respectively. 
All the indexes run over the single particle states which are created/destroyed, depending on  whether they are associated to $a^{\dag}/a$. 
The coefficients are completely defined imposing the full anti-symmetry of the matrices $V$, $W$. Note that the description in terms 
of creation/annihilation operators is the framework in which the Green function formalism has been originally developed. Here we use a state-of-the-art potential derived through a similarity renormalization inspired transformation (SRG) \cite{Bogner}. The SRG evolution is an unitary rotation that allows to transform a given hamiltonian into a softer, low-momentum,  interaction, suitable for use in a perturbative approach. Two different initial hamiltonians 
based on $\chi$PT have been used: a two-body potential calculated up to N$^3$LO and a N$^3$LO two-body plus a N$^2$LO three-body potential.The flow parameter characterizing the evolution has been fixed to $\alpha=0.8$ fm$^{-4}$ and we have retained all the evolved terms up to three-body level, getting an induced (labelled with \emph{ind}) $2+3$-body force in the former case and a full (labelled with \emph{full}) $2+3$-body force in the latter \cite{Roth}.

\section{Green function formalism}

Green's function techniques are well established as a very powerful tool to describe quantum systems in a variety of  fields of Physics. 
This formalism proved very effective both in vacuum and in matter, as well as in and out equilibrium. 
In this section, the Self Consistent Green Function (SCGF) approach is applied to many-body theory. In this context, it is capable to 
provide a clear description of the dynamics,  including a wide range of observables, such as binding energies and spectral functions, which  
are analyzed using a diagrammatic technique. In the case of non-perturbative dynamics, a self-consistent approach is still possible within the 
equation of motion method. The fundamental quantity of this formalism is the single particle Green function,  that describes the propagation 
of a particle, or a hole, within the system
\begin{equation}\label{green}
g_{\alpha\beta}(t-t')=-i\langle\Psi_0^N|T[a_{\alpha}(t)a_{\beta}^{\dag}(t')]|\Psi_0^N\rangle \ .
\end{equation}
In the above equation, 
$\Psi^N_0$ is the ground-state wave function, $T$ is the time-ordering operator and $a_{\alpha}/a_{\alpha}^{\dag}$ denote the particle 
annihilation/creation operator in the state $\alpha$. The quantity defined by eq.  (\ref{green}) describes to the creation of a single particle state $\beta$ at time $t'$ together with the destruction of a single particle state $\alpha$ at time $t$ for $t-t'<0$, and the destruction of a state $\alpha$ at time $t$ together with the creation of a single particle state $\beta$ at $t'$ for $t-t'<0$. Thanks to energy conservation, this function depends on $(t-t')$ only.  It is remarkable that once this function is known, all one-body can be described\footnote{This is also true for some \emph{peculiar} multi-particle operators.}. This can be easily seen in the second quantization formalism, in which a one-body operator can be expanded in terms of creation/annihilation, $a,a^{\dag}$, operators
according to $O^{1B}=\sum_{\alpha\beta}o_{\alpha\beta}\,a^{\dag}_{\alpha}a_{\beta}$. As a results, the expectation value is given by
\begin{equation}\label{O1}
\langle O^{1B}\rangle  =  o_{\alpha\beta}\,\,\rho_{\beta,\alpha}\qquad \rho_{\alpha\beta}=\langle \Psi_0^N|a^{\dag}_{\beta}a_{\alpha} |\Psi_0^N\rangle=\iu\lim_{t'\to t^{+}}g_{\alpha\beta}(t-t').
\end{equation}
A clear picture of the information contained in $g_{\alpha\beta}$ is provided by the the Lehmann representation, obtained transforming to Fourier 
space and using the completeness relation fulfilled by the eigenstates of the $N\pm 1$-body system:
\begin{align}\label{full_green}
g_{\alpha\beta}(\omega)&=\int \,d\tau\, e^{i\omega \tau}g_{\alpha\beta}(\tau)=\nonumber\\
&=\sum_n\frac{\langle\Psi_0^N|a_{\alpha}|\Psi_n^{N+1}\rangle\langle\Psi^{N+1}_n|a_{\beta}^{\dag}|\Psi_0^{N}\rangle}{\omega-(E_{n}^{N+1}-E_0^{N})+i\eta}+\frac{\langle\Psi_0^N|a_{\beta}^{\dag}|\Psi_n^{N-1}\rangle\langle\Psi^{N-1}_n|a_{\alpha}|\Psi_0^{N}\rangle}{\omega-(E_{0}^{N}-E_n^{N-1})-i\eta}\nonumber\\
&=g_{\alpha\beta}^h(\omega)+g_{\alpha\beta}^p(\omega) \ .\nonumber
\end{align}
The poles of the above equation provide for the excitation spectra $\pm(E_{n}^{N\pm1}-E_0^{N})$ asociated with addition/removal of a particle 
to/from the ground state. The corresponding residues, reflecting the transition amplitudes, go under the name of \emph{spectroscopic amplitudes}. 
The hole part of the propagator gives information on the process of particle emission, the poles being the exact energy absorbed in the process.
Note that this analytic structure is completely general and
that these energies and amplitudes are obtained solving a Schr\"oedinger-like equation, that in diagrammatic language is usually referred to as Dyson 
equation
\begin{equation}\label{Dy}
g_{\alpha\alpha'}(\omega)=g^0_{\alpha\alpha'}(\omega)+g^0_{\alpha\beta}(\omega)\Sigma_{\beta\beta'}^{\star}(\omega)g_{\beta'\alpha'}(\omega) \ .
\end{equation}
In the above equation, $g^{0}$ is the unperturbed, \emph{free} propagator, corresponding to nucleons moving without experiencing ``dynamical'' interactions. However, statistical correlations, indiced by Pauli exclusion principle, are always present. The matrix $\Sigma^{\star}(\omega)$ embodies all the dynamical information stored in the Green function, and the star refer to the fact that only \emph{irreducible} diagrams, i.e.  diagrams that cannot be obtained combining lower-order diagrams already present in $\Sigma^{\star}(\omega)$, must be included. Equation (\ref{Dy}) can be 
written  in diagrammatic form as
\begin{minipage}[l]{.5\textwidth}
\vspace{-5mm}
\hspace{-10mm}
\includegraphics[scale=0.6]{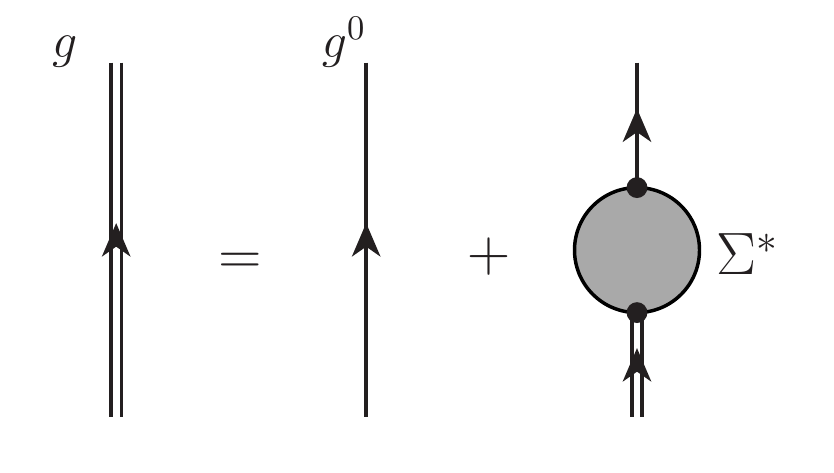}
\end{minipage}
\hspace{-20mm}
\begin{minipage}[l]{.5\textwidth}
\includegraphics[scale=0.6]{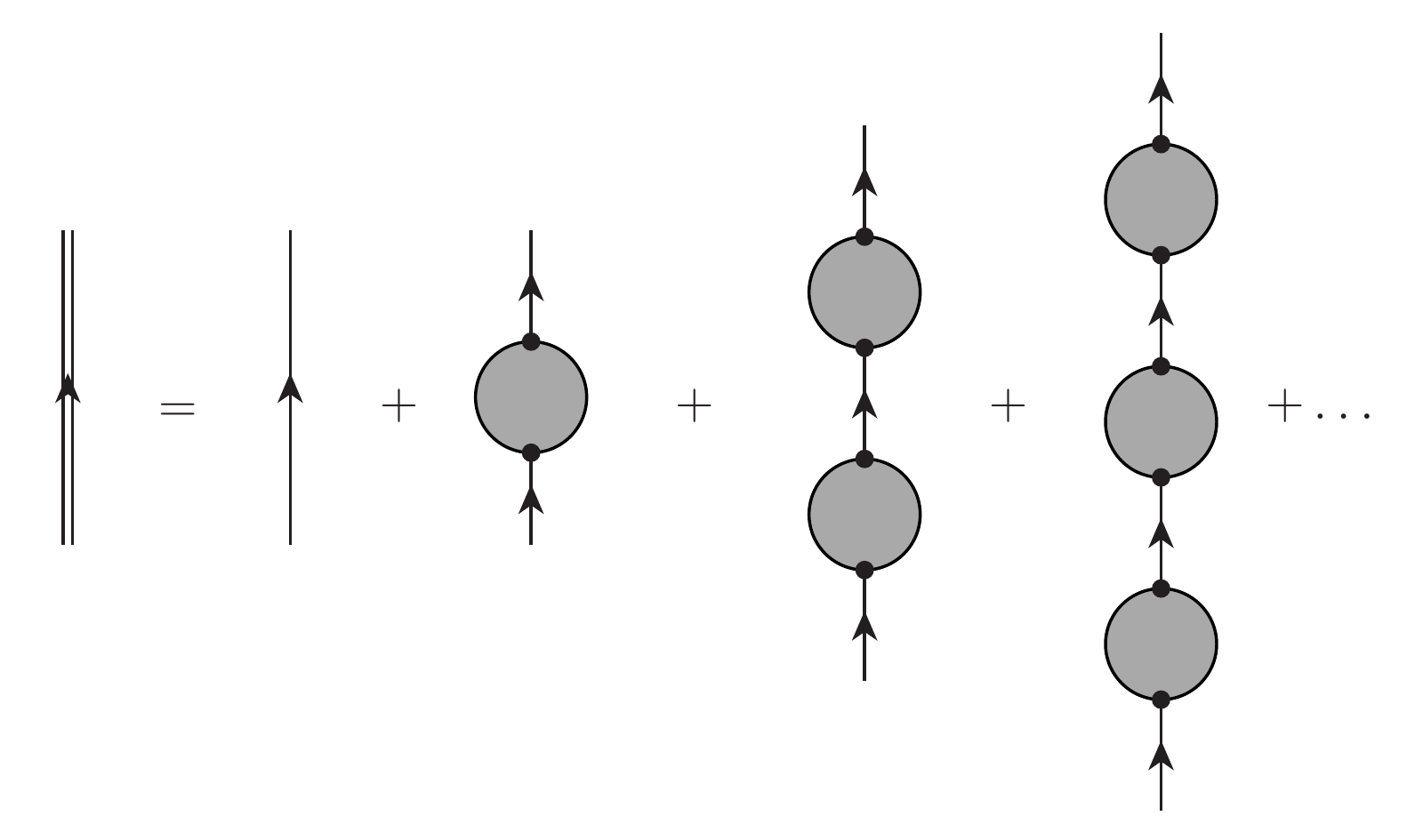}
\end{minipage}
\\
Diagrammatic representation of Dyson equation (\ref{Dy}) (left), and its perturbative expansion (right).
\\
\\
It is wort pointing out that the full knowledge of $\Sigma^{\star}(\omega)$ would be equivalent to solve the Schr\"oedinger equation without any approximations. In actual calculations, we need to select a restricted number of irreducible diagrams, which amounts to limiting many-body correction 
to a maximum number of particles. As already stated, dynamical effects come from the self-energy $\Sigma^*$, which,  in principle, includes all the diagrams that can be seen as a one-body correction arising from the \emph{bare} hamiltonian. This is what has been usually done with hamiltonians 
which include two-body interactions only. Switching on three-body interactions,  the convergence of the Dyson equation becomes very slow, 
since the thee-body term is actually transformed into a two-body effective interaction. The corresponding diagrams can appear at any order of the 
expansion, and heavily affect the results. However, thanks to the implicit form of the Dyson equation one can find an algorithm to sum up these diagrams at 
all orders. The basic idea is rearranging the diagrams 
with the help of an effective hamiltonian $\widetilde{H}$,  
defined as 
\begin{figure}[!h]
\includegraphics[scale=.85]{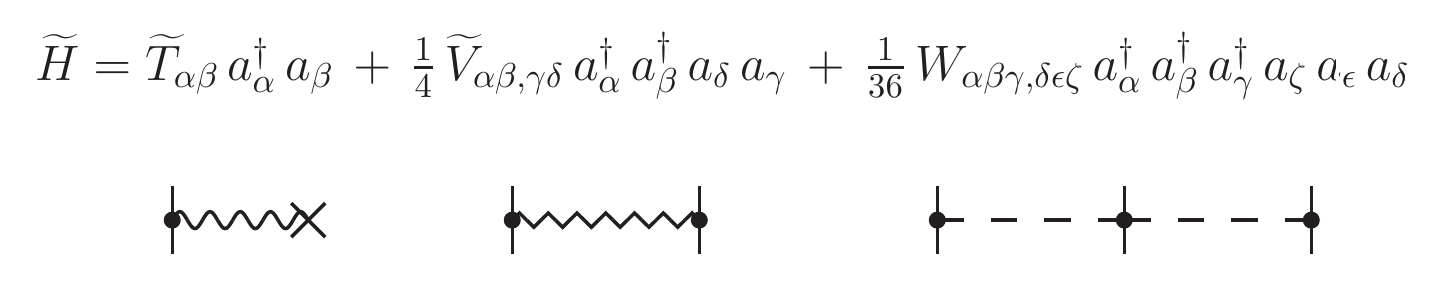}    
\end{figure}
\\
The new hamiltonian has a clear physical interpretation: $\widetilde{T},\widetilde{V}$ are effective operators which include the \emph{bare} $T$ 
and $V$ terms as well as the screening arising from the three-body force. Note that $\widetilde{H}$ has the same dynamical content as $H$,  with 
the constraint that now $W$ must act only as a pure three-body interaction, in order to avoid double counting of  the screening effect. 
In Ref. \cite{Hagen} it has been shown that the dominant effect of the three-body force can is in fact included using  $\widetilde{T}$ and 
$\widetilde{V}$,  while the pure three-body term plays a minor role. In the following, we will disregard the last term, the importance of which has been critically reviewed, up to third order, in Ref. \cite{Arianna}. 

In diagrammatic language the two operators $\widetilde{T}$ and 
$\widetilde{V}$ take  a transparent form, in which $g^{pp/hh}$ is the two-body propagator, that can be defined through a straightforward 
extension of eq. (\ref{green}), and the three-body propagator also appears.

\begin{figure}[!h]
\begin{center}
\hspace{6mm}\includegraphics[scale=.9]{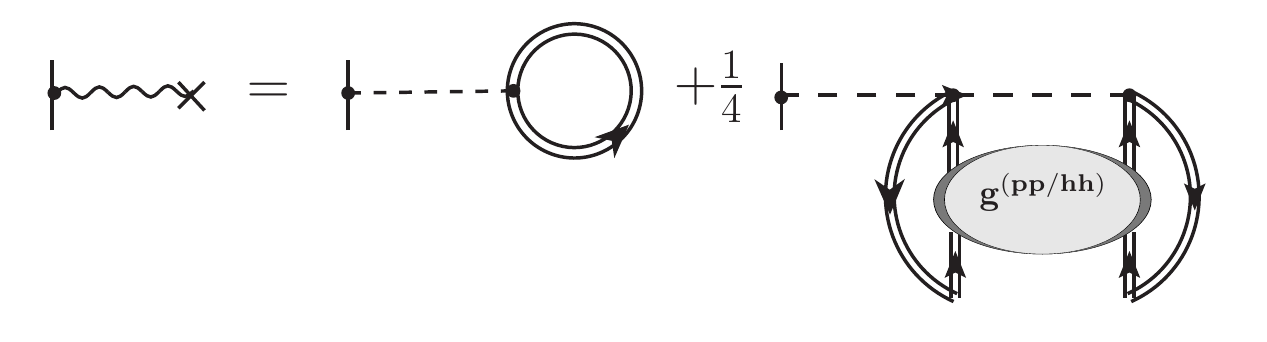}  
\includegraphics[scale=.9]{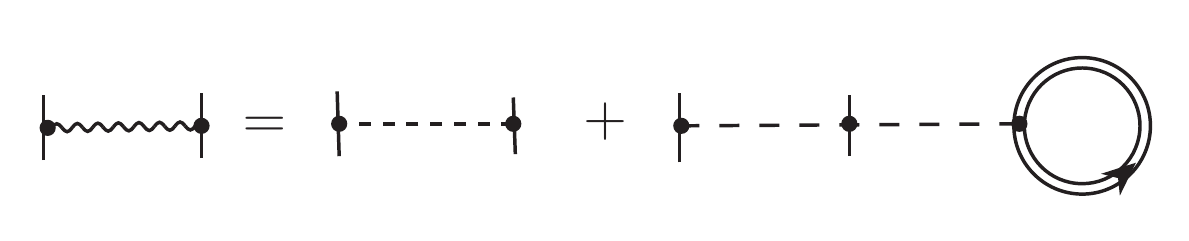}   
\caption{Diagrammatic representation of the one and two-body contributions to the \emph{effective} hamiltonian $\widetilde{H}$. }
\end{center}
\end{figure}
\begin{align}
&g^{2B}_{\alpha\beta,\alpha'\beta'}(t-t')=-\iu\langle \Psi^{N}_0|T[a_{\alpha'}(t)a_{\beta}(t)a^{\dag}_{,\alpha'}(t')a_{\beta'}^{\dag}(t')]|\Psi_0^{N}\rangle\nonumber\\
&g^{3B}_{\alpha\beta\gamma,\alpha'\beta'\gamma'}(t-t')=-\iu\langle \Psi^{N}_0|T[a_{\gamma}(t)a_{\beta}(t)a_{\alpha}(t)a^{\dag}_{\alpha'}(t')a^{\dag}_{\beta'}(t')a_{\gamma'}^{\dag}(t')]|\Psi_0^{N}\rangle
\end{align}
Proceeding as in the derivation of eq. (\ref{O1}),  we can define the average of two- and three-body operators
\begin{align}
& \langle O^{2B}\rangle   =   O^{2B}_{\alpha\beta,\alpha'\beta'}\,\,\rho^{2B}_{\alpha'\beta',\alpha\beta}\qquad \rho^{2B}_{\alpha\beta\alpha'\beta'}=\langle \Psi_0^N|a^{\dag}_{\alpha'}a^{\dag}_{\beta'}a_{\beta}a_{\alpha} |\Psi_0^N\rangle \nonumber\\ 
 &\langle O^{3B} \rangle =  O^{3B}_{\alpha\beta\gamma,\alpha'\beta'\gamma'}\rho^{3B}_{\alpha'\beta'\gamma',\alpha\beta\gamma} \qquad\rho^{3B}_{\alpha\beta\gamma\alpha'\beta'\gamma'}=\langle \Psi_0^N|a^{\dag}_{\alpha'}a^{\dag}_{\beta'}a^{\dag}_{\gamma'}a_{\gamma}a_{\beta}a_{\alpha} |\Psi_0^N\rangle  \ , \nonumber
\end{align}
with 
\[
\rho^{nB}=\iu \lim_{t'\to t^+} g^{nB}(t-t')=\iu g^{nB}(0^-)  \ .
\]
The diagrammatic representations of the corresponding expectation values are 
\begin{figure}[!h]
\hspace{-5mm}
\includegraphics[scale=.9]{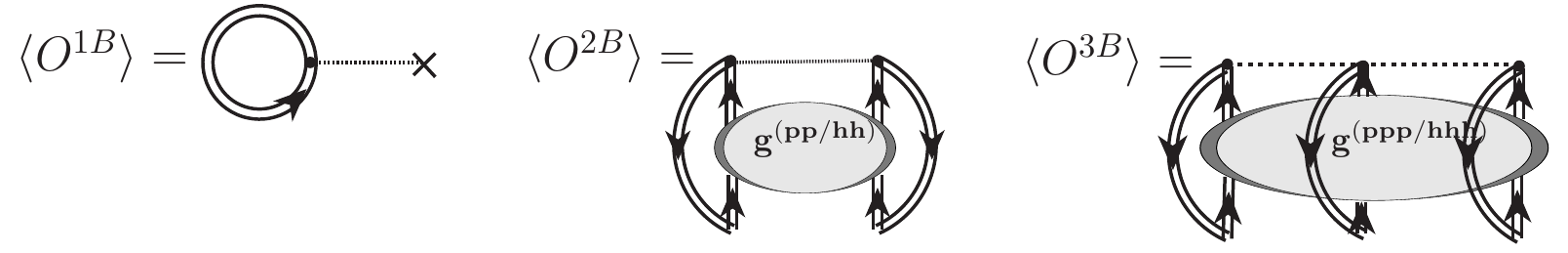}  
\caption{Diagrammatic representation of the expectation values of many-body operators evaluated through the \emph{full} Green functions.}
\label{onetwothree}
\end{figure}
\section{Sum rule for the binding energy}
In principle, the calculation of the binding energy of a nucleus binding energy within the Green function formalism is not an easy task . The 
evaluation of $\lan H \ran$ directly from kinetic, $T$, and potential, $V$ and $W$ terms, involves operators acting on states describing an up to three 
particles. As shown in fig. (\ref{onetwothree}), one would have to calculate \emph{independently} three different expectation values, involving three different propagators from three different Schr\"oedigner-like equations. This would of course imply a large amount of computer time, as well as a 
huge theoretical effort.

In this Section, we will show how the number of Green functions involved in the calculation can be reduced, exploiting a theoretical result 
first obtained by Koltun. The underlying idea stems from the observation that the operator we want to average is the same which drives 
the time evolution of the system. In the Heisenberg picture, we can write the time evolution equation for the annihilation operator in the form
 \begin{equation}\label{dt}
\iu\frac{d}{dt}a_{\alpha}(t)=[a_{\alpha}(t),H] \ .
\end{equation}
Defining the time-derivative of (\ref{green}) as
\begin{align}\label{der}
\left.\frac{d g_{\alpha\beta}(t)}{d t}\right|_{t\to 0^-}&=\iu\langle \Psi_0|a^{\dag}_{\alpha}(0)\left.\frac{d a_{\beta}(t)}{dt}\right|_{t\to 0^+}|\Psi_0\rangle\nonumber\\
&=\frac{1}{2\pi \iu}\int_{C\uparrow}d\omega \omega g_{\alpha\beta}(\omega),
\end{align}
and using the relations
\begin{align}\label{rules}
\sum_{\alpha}a_{\alpha}^{\dag}[a_{\alpha},T]&=T,\nonumber\\
\sum_{\alpha}a_{\alpha}^{\dag}[a_{\alpha},V]&=2\,V,\nonumber\\
\sum_{\alpha}a_{\alpha}^{\dag}[a_{\alpha},W]&=3\,W,
\end{align}
we obtain
\begin{align}\label{lim_eq_mot}
\langle T + 2 V +3 W \rangle= &\sum_{\alpha}\langle \Psi_0|a^{\dag}_{\alpha}[ a_{\alpha},H]|\Psi_0\rangle \nonumber\\
=&i\sum_{\alpha}\langle \Psi_0|a^{\dag}_{\alpha}(0)\left.\frac{d a_{\alpha(t)}}{d t}\right|_{t=0^+}|\Psi_0\rangle \nonumber\\
=&\sum_{\alpha}\frac{1}{2\pi i} \int_{C\uparrow} d\omega \,\omega\, g_{\alpha\alpha}(\omega).
\end{align} 
Equations (\ref{dt}) and (\ref{rules}) lead to eq. (\ref{lim_eq_mot}), which is the \emph{extended} Kultun sum rule, modified from the 
original result to include three nucleon forces. The $C \uparrow$ integration contour must be closed the in the upper half-plane, 
in order to extract the residue of quasi-hole pole. 

Different extrapolations of the ground state energy $\langle H\rangle=\langle T+V+W\rangle$ can be inferred from the above equations. 
The one we use is
\begin{equation}\label{best}
\langle H\rangle=\underbrace{\sum_{\alpha\beta}\frac{1}{4\pi i}\int_{C\uparrow} d\omega \left[\,T_{\alpha\beta}+\omega\,\delta_{\alpha\beta}\,\right] g_{\alpha\beta}(\omega)}_{ \varUpsilon }-\frac{1}{2} \langle W\rangle  \ ,
\end{equation}
where the average of $W$ must be calculated separately. A similar formula, involving thetwo-body operator  $V$ , can also be obtained  
\begin{equation}\label{worst}
\langle H\rangle=\sum_{\alpha\beta}\frac{1}{6\pi i}\int_{C\uparrow} d\omega \left[\,2\,T_{\alpha\beta}+w\,\delta_{\alpha\beta}\,\right] g_{\alpha\beta}(\omega)+\frac{1}{3}  \ .\langle V\rangle 
\end{equation}
In the following, we mainly refer to eq. (\ref{best}) since $\lan W\ran/\lan V\ran\sim 10\%$. Consequently the contribution of $\lan W\ran$ would have a minor impact in the final energy.
\section{Self-energy and iterative method}
The degree of accuracy of the SCGF formalism mainly depends on the number of irreducible diagrams one is able to include (and to sum up) in the 
self-energy. The approach we are going to described has been developed for infinite nuclear matter, and has been described in several papers for the 
caseof a hamiltonian involving a two-body interactions only \cite{Bar2,Bar3}. Here we extend the method taking into account three-body effects using 
the effective hamiltonian $\widetilde{H}$. 

\begin{figure}[!h]
\begin{center}
\includegraphics[scale=.75]{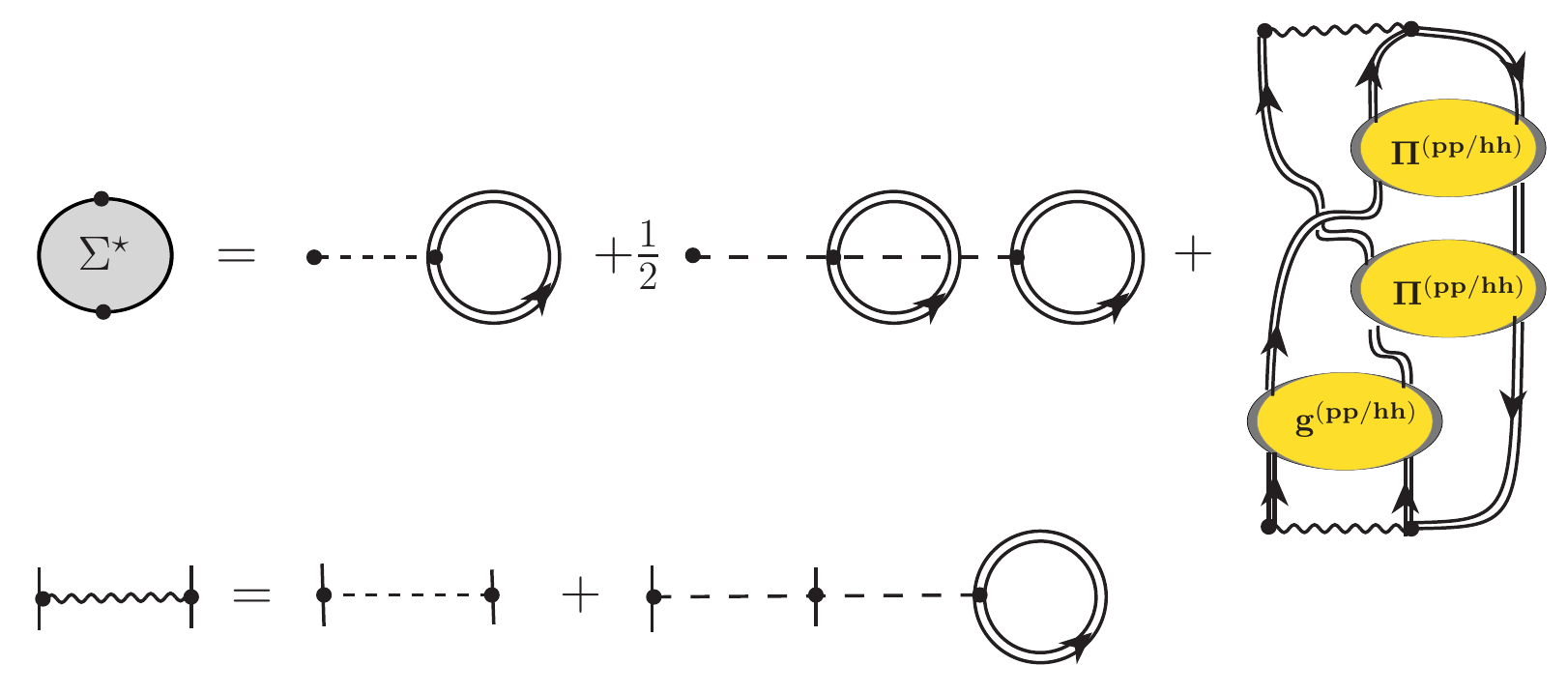}  
\caption{Diagrammatic representation of the self-energy in the Faddeev-TDA approximation. The 2h1p contribution is not represented. 
The yellow boxes correspond to the Green function in TDA approximation. In this case the double line represents the self-consistent  
Hartee-Fock propagator.}
\label{FTDA}
\end{center}
\end{figure}

The diagrams involving the purely three-body interaction $W$ have been discarded. As already stated, they should play a minor role \cite{Arianna}. 
We use a third order approximation, referred to as Fadeev-Tamm Dancoff Approximation (Faddev-TDA) approach,  in which the contribution of two particle-one hole ($2p1h$) and two hole-one particle intermediate states are taken into account self-consistently, and the interaction between $pp/hh$ or $ph$ excitations are modeled within the TDA scheme. The equations describing the polarization and particle-particle (hole-hole) propagator in TDA approximation are depicted by the yellow bubbles is fig. (\ref{FTDA}), where the effective $\widetilde{V}$ is employed. These polarization and $pp/hh$ propagators need to be added consistently to determine the $pp$-$h$,$hh$-$p$ propagator $R(\omega)$, yielding the third order contribution. 

Its contribution is sketched in fig. (\ref{FTDA}),  for the case of the $pp$-$h$ channel. It appears that it includes the effect of $ph$ and $pp/hh$ motion,  allowing for the interferences between them and giving at the same time the correct combinatorial factor to each diagrams, without the need of subtracting
spurious terms. Moreover, in the same figure we show how the new terms arising in $\widetilde{T}$ and $\widetilde{V}$ can be summed up within the same scheme. For a exhausting review, see Ref. (\cite{Barbieri}). 

In the SCGF approach the self energy matrix $\Sigma^{\star}$ is expanded in term of the dressed propagator up to the required perturbative 
order. This means that the actual degrees of freedom are the excitation of the fully correlated system, and the effects of fragmentation are already 
included in the iteration scheme for self-consistency. In applications, we need a first approximation to the propagator, e.g. the propagator obtained
within the Hartree-Fock approximation, to start the calculation. Solving for the first time the Dyson equation with this propagator we 
obtain a new $g(\omega)$, which is  then used to calculate again the self-energy. This procedure is iterated until convergence is reached.

\section{Approximations and numerical results}
Including the screening term in $\widetilde{T}$ and $\widetilde{V}$ amounts to using one- and two-body \emph{density-dependent} interactions. 
This procedure sharply increases the required computational effort, since these interactions must be rebuilt at each interaction. We adopt several approximations aimed at making the code more effective and, at the same time, at minimizing the loss of information. 
In $\widetilde{T}$ and $\lan W\ran$,  we approximate the two-body and three body propagators with two and three 
single particle propagators, which are by far the most accurately computed quantities within our approach. This approximation 
implies that the interaction between the single particles described by the Green functions are negnected. 
Moreover, we stop the evaluation of $\widetilde{T} $ and $\widetilde{V}$ at the first iteraction. 
\begin{figure}[!h]
\includegraphics[scale=.85]{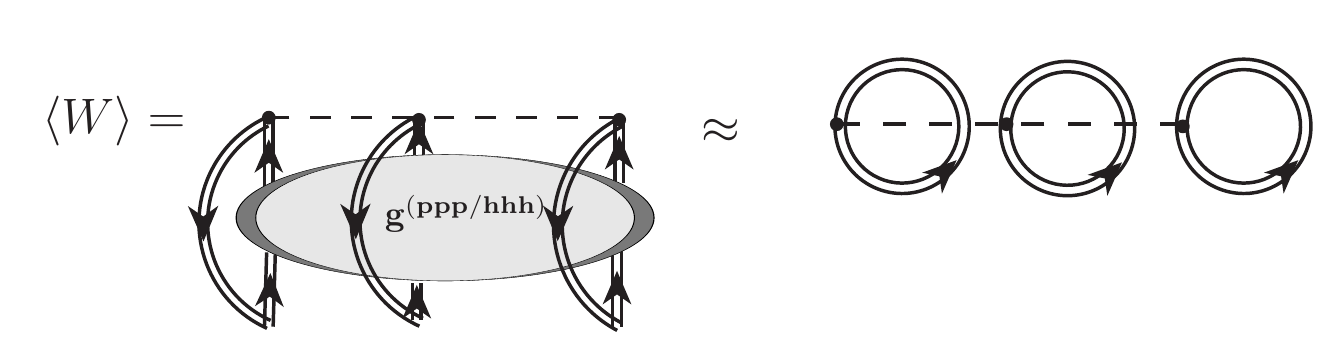}  
\caption{Diagrammatic representation of the approximation scheme employed to carry out for the three-body average. The double line represents 
the \emph{full} propagator in Faddeev-TDA approximation.}
\end{figure}

\begin{figure}[!h]
\hspace{-10mm}
\includegraphics[scale=.55]{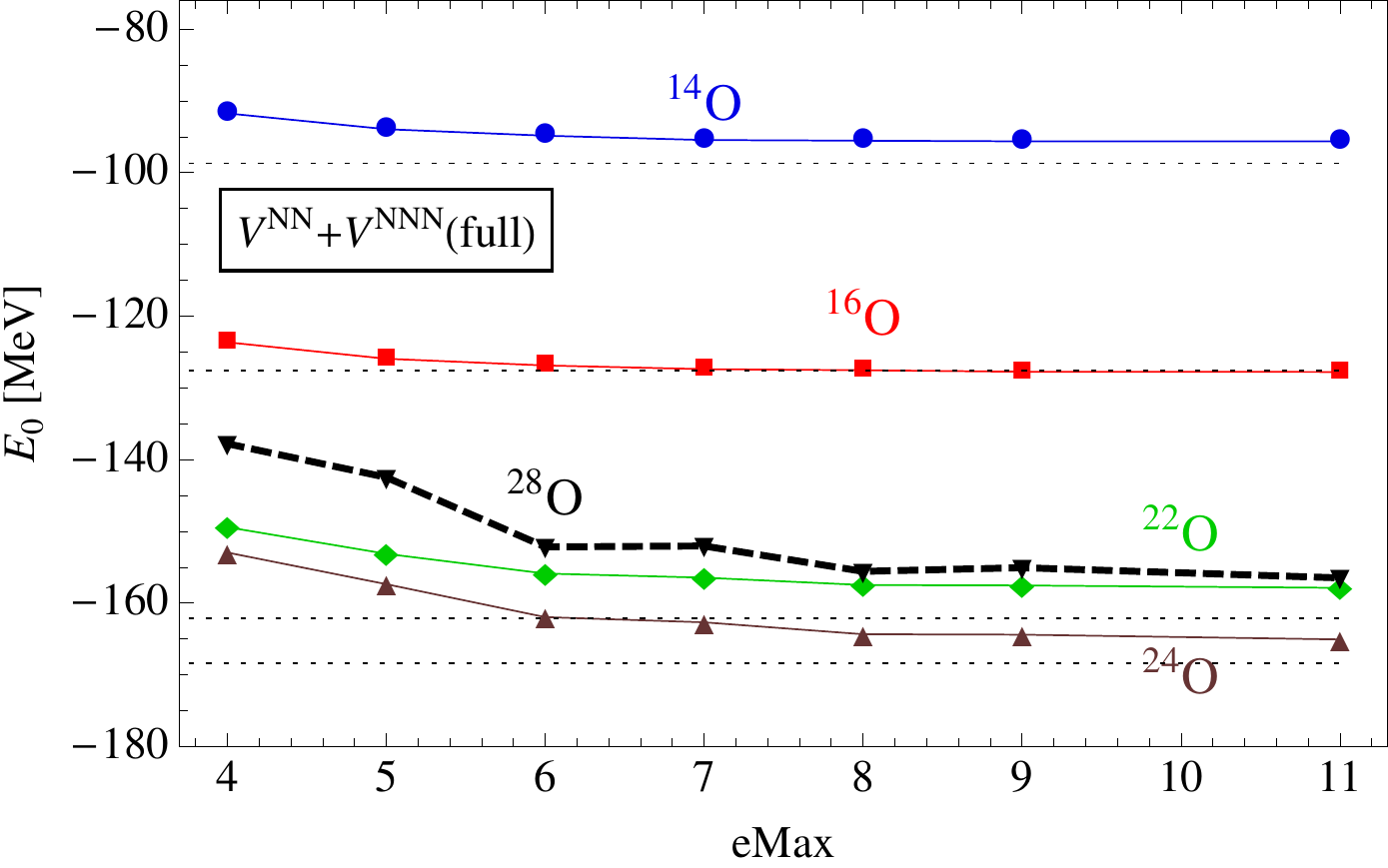}
\includegraphics[scale=.57]{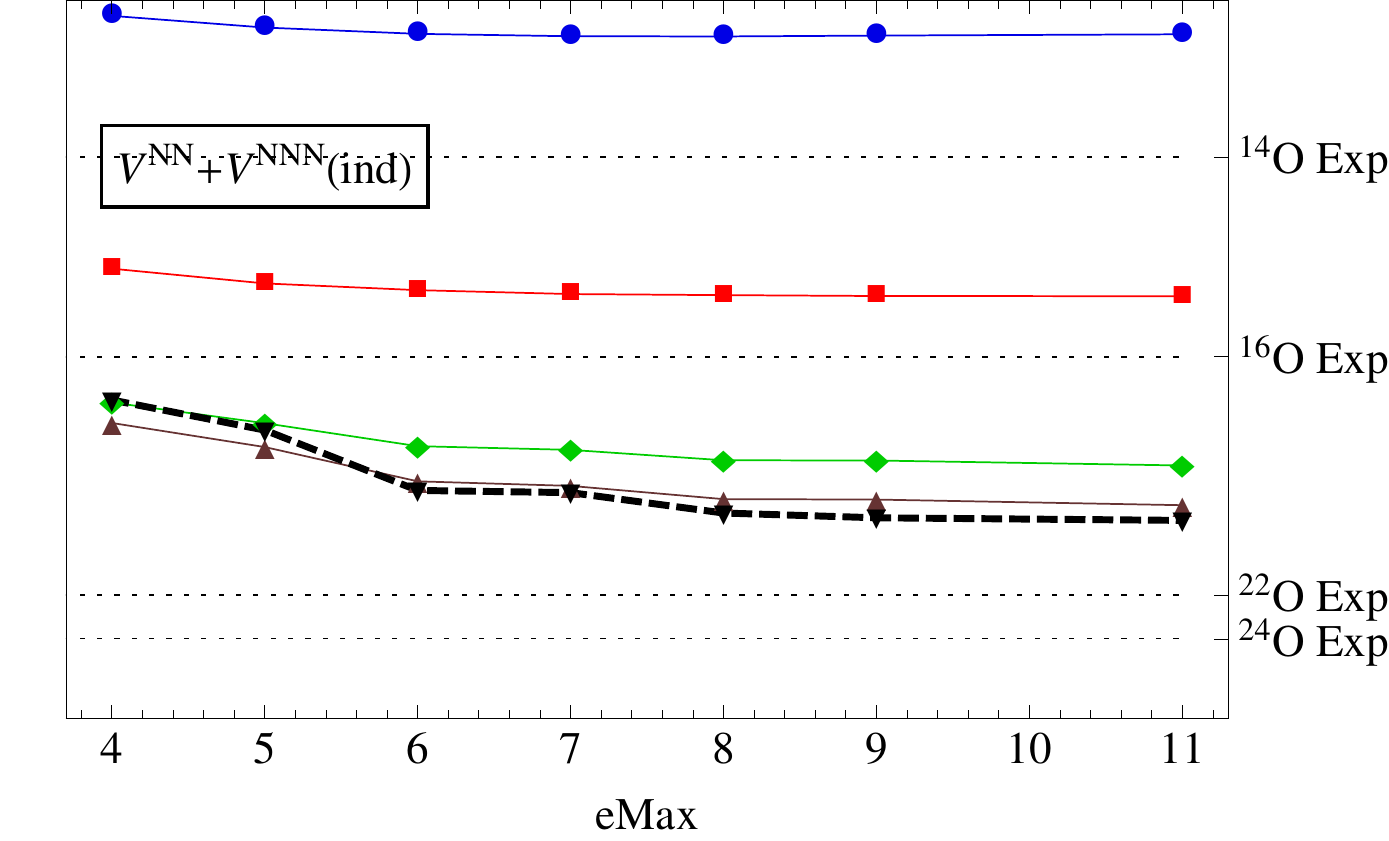}  
\caption{Behavior of the binding energies of the main oxygen isotopes, as function of the energy of the mayor shell employed, eMax. 
The results shown in the left and right panels have been obtained using the full three-nucleon force, interaction and the induced interaction, 
respectively. The thin dashed lines indicate the experimental values. The thick dashed line corresponds to  $^{28}O$, which appears to be 
unbound in the left panel and bound in the right panel. The results have been obtained using a harmonic oscillator potential with $\hbar\omega=20$ MeV.}
\label{full}

\end{figure}

The results obtained using the above procedure are collected in Fig. \ref{full}, showing the binding energy of the main oxygen isotopes. 
Consider the behavior of the energy of $^{28}O$, represented by thick dashed line. 
It appears that when we use the full three-body interaction, $^{28}O$ is unbound. On the other hand, when the induced three-nucleon force 
is employed, it seems to be bound, with a ground state energy slightly below that of $^{24}O$. 
This is to be ascribed to the fact that the latter interaction actually plays the role of a pure two-body interaction,  in which the repulsive 
component of the three-nucleon force is disregarded. The overall picture clearly indicate that a proper treatment of the three-body force 
is essential to reproduce the oxygen neutron drip line.

%% file: Chap_SO.tex
Over the past years, several refined simulations of supernovae have appeared in the literature \cite{B1,B2}. The main problem of these simulations is the description of the final stage of the evolution, that is the gravitational decoupling between the core and the mantle leading to the birth of a proto-neutron star. Many ideas have been proposed and already tested to explain the failure to achieve 
this goal. All clues seem to point to the oversimplified description of neutrino driven mechanisms in matter\footnote{Of course, there could also be "new physics" that we are completely unaware of.} \cite{Janka} . Indeed, neutrino opacity of nuclear matter is far from being understood at fully quantitative level. 

The impact of nuclear correlations on free the Fermi gas picture appears to be impressive, resulting in the appearance of a complicated pattern of excitations, and
conventional perturbative approaches are designed to fail when dealing with nuclear matter structure and dynamics. 

The aim of this Thesis has been the development of a realistic description of neutrino interactions with nuclear matter. We have restricted our 
analysis to the case of pure neutron matter, and evaluated the neutrino mean free path as function of both density and temperatures near $T\sim 0$. 
In addition, we have extended our analysis to cover a broad density range, up to twice nuclear saturation density $\rho_0$. 
In fact, there are indications that the depletion of the hard repulsive core of the $NN$ interaction allows the liquid phase to persist well beyond saturation density $\rho_0$, 
moving the threshold of the transitions to more ``exotic'' phases to densities as high as three-four times $\rho_0$.

The starting point of our work has been the two-nucleon interaction \emph{in vacuum}. We employed a phenomenological potential based on the Argonne $v_{18}$ model, yielding the 
best available fits of $NN$ scattering data. Medium effects have been included through the CBF  formalism, exploiting the expression of the energy at two-body level in the cluster expansion. 
This procedure provides an effective interaction taking into account medium effects, such as screening of the repulsive core arising from $NN$ correlations. 
The main advantages of this approach is that the effective potential can be used to carry out perturbative calculations in the Fermi gas basis, and in the 
$\rho\to 0$ limit reduces to bare Argonne potential  by construction. 

The Standard Model of Particle Physics, in the low energy limit, provides an extremely accurate description of semi-leptonic weak processes through a vector (CP conserving) and an axial (CP violating) currents,  
responsible for the so called Fermi and Gamow-Teller transitions, respectively. The description of these two transitions can be formally extended to the case on neutrino interactions in 
matter, although the relevant mechanisms strongly depend on the energy scale. In the region of low momentum transfer, many-body effects play a 
crucial role, but the self-consistent evaluation 
of the dynamic form factors within CBF involves non trivial computational problems. The authors of Ref. \cite{Farina} succeeded in carrying out a consistent CBF calculation of the 
response restricted to the density-density sector only. 

We have followed a different approach,  based on Landau theory of normal Fermi liquids. In this case the CBF effective interaction has been employed to obtain the numerical values of the Landau parameters. 

Within Landau theory the ring diagrams, which are needed to describe the response at low momentum transfer treating both the coherent and incoherent contributions on equal footing, 
 can be easily summed up. While the static properties of neutron matter obtained in the Landau scheme have the same degree of accuracy as the previously available CBF results, within our approach
we can clearly see the emergence of the zero-sound contribution. Moreover, our results are in fairly good agreement with the results of highly refined many-body calculation based on similar dynamic models. Our response functions also appear to be consistent with the existing literature. 

The evaluation of neutrino mean free path was the ultimate goal of the Thesis. We find that the dependence on neutrino energy, $\epsilon$,  is roughly 
$\epsilon^3$,  consistent with available estimates and different from the $\epsilon^2$ dependence of the free Fermi gas model. 
Our analysis also included the dependence on density and temperature, limited to 
the region $T\leq 10$ MeV, where the collisionless approximation is expected to be applicable.

%% file: Appendix1.tex
\section{$O^n$ algebra}
The six operators already introduced Chapter \ref{NMB} $[I,({\boldsymbol \sigma}_1 \cdot {\boldsymbol \sigma}_2),S_{12}({\bf \hat{r}})]
\otimes[I,({\boldsymbol \tau}_1 \cdot {\boldsymbol \tau}_2)]$ have the very important properties of forming 
an algebra with respect to multiplication, as they satisfy the relations
\[
O^{i}O^j=\sum_kK_{k}^{ij}O^K  \ .
\]
The values of $K^{k}_{ij}$ can be easily found exploiting the $SU(2)$ algebra of Pauli matrices \cite{Valli}. 
The \emph{fundamental representation} is commonly defined choosing $\s_3$ in diagonal form, 
\[
\s^1=
\left(
\begin{array}{cc}
 0 &  1 \\
 1 & 0   
\end{array}
\right)
\quad,\quad
\s^2=
\left(
\begin{array}{cc}
 0 &  -\iu \\
 \iu & 0   
\end{array}
\right)
\quad,\quad
\s^3=
\left(
\begin{array}{cc}
 1 & 0 \\
 0 & 1   
\end{array}
\right),
\]
and  the commutation and anti-commutation relations are
\begin{align}
[\s_i,\s_j]&=2\iu\epsilon_{ijk}\s^k\nonumber\\
\{\s_i,\s_j\}&=2\delta_{ij} .\nonumber
\end{align}
\section{Two-particle system}
In order to derive, and then solve, the Euler-Lagrange equations for the correlation functions, it is convenient to change basis, and use operators that 
partially decouple in the total spin-isospin space.  Let us consider two particles labelled by their spins 
$({\bf S}_1,{\bf S}_2)=({\boldsymbol \sigma}_1,{\boldsymbol \sigma}_2)/2$, respectively. The total spin states $|S,M_S\rangle$ are related to the 
single-particle spin states by
\begin{align}
|0 0\rangle&=\frac{1}{\sqrt{2}}(|\up\down\rangle-|\down\up\rangle)\nonumber\\
|1 0\rangle&=\frac{1}{\sqrt{2}}(|\up\down\rangle+|\down\up\rangle)\nonumber\\
|1 -1\rangle&=|\down\down\rangle\nonumber\\
|1  1\rangle&=|\up\up\rangle .
\end{align}
In operatorial language we can write
\[
{\bf S}^2={\bf S}_1^2+{\bf S}_2^2+2( {\bf S}_1\cdot {\bf S}_2) ,
\]
implying $\langle S=0,1|{\boldsymbol \s}_1\cdot {\boldsymbol \s}_2|S=0,1 \rangle=-3,1$, respectively. 

The projection operators on  states 
of definite total spin are 
\begin{align}
P\rb{S=0}\equiv P_1&=\frac{1-({\boldsymbol \s}_1\cdot {\boldsymbol \s}_2)}{4}  \\
P\rb{S=1}\equiv P_3&=\frac{3+({\boldsymbol \s}_1\cdot {\boldsymbol \s}_2)}{4}  \ ,\nonumber
\end{align}
and satisfy the standard properties of a projector
\[
P\rb{2S+1}^2=P\rb{2S+1} \ \ , \ \  P\rb{1}+P\rb{3}=\mathcal{I} \ \ , \ \  P\rb{3}P\rb{1}=P\rb{1}P\rb{3}=0  \ ,
\]
where $\mathcal{I}$ is the identity. 

We can easily construct the spin-exchange operator $P_{\sigma}\equiv P_3-P_1$ such that,
\[
P_{\s}|S M_S\ran=(-)\rp{S+1}|S M_S\ran  \ .
\]
The same relation can be obviously used in isospin space, so that we can define a total spin-isospin exchange operator 
$P_{\s\tau}\equiv P_{\s}P_{\tau}$ such that 
\[
P_{\s\tau}|S M_S, T M_T\ran=(-)\rp{S+T}|S M_S, T M_T\ran .
\]
\section{Tensor term}
Let us now turn to the tensor operator $S_{12}$, defined as
\[
S_{12}\equiv \sab ,
\]
where $\rh$ is a unit vector along the direction of the relative coordinate of particles 1 and 2,  while $r=|{\bf r}|$. Using the properties of Pauli 
matrices, it can be easily show that
\[
S_{12}\st=\st S_{12}=S_{12}.
\]
This relation implies that $S_{12}$ acts on $S=1$ state only and annihilates $S=0$ state\footnote{Recall that the $S=0$ state 
is an eigenstate of $\st$ with eigenvalue $-3$}. Moreover,
\[
S_{12}^2=6-2S_{12}+2\st=8-2S_{12}  \ ,
\]
and
\begin{align}
\bm{\nabla}S_{12}&=\frac{3}{r}\Big[\bm{\s}_1\sb+\bm{\s}_2\sa-2\rh\sa\sb \Big]\nn\\
\bm{\nabla}^2 S_{12}&=-\frac{6}{r^2}S_{12} \ .
\end{align}
The above relations imply that for a generic function $u(r)$ depending on the radial coordinate $r$
\[
\big(\bm{\nabla}u\big)\cdot\big(\bm{\nabla} S_{12}\big)=\left( \frac{d u}{d r} \right) \st\cdot\big(\bm{\nabla}S_{12}\big)=0  \ .
\]
Further useful properties of the tensor operator are
\begin{align}
\big(\bm{\nabla}S_{12}\big)^2&=\frac{6}{r^2}(8-S_{12})\nonumber\\
\big[S_{12},\big(\bm{\nabla}S_{12}\big)\big]&=\frac{36}{r}\iu\big({\bf S}\times \rh\big)\nonumber\\
\big[S_{12},\big(\bm{\nabla}S_{12}\big)\big]\bm{\nabla}&=-\frac{36}{r^2}\big({\bf L}\cdot{\bf S}\big)\nonumber\\
\big[S_{12},\bm{\nabla}^2 S_{12}\big]&=0\nonumber\\
\big(\bm{\nabla} S_{12}\big)\big[S_{12},\bm{\nabla}\big]&=-\big(\bm{\nabla} S_{12}\big)^2 .
\end{align}
\section{Total spin-isospin representation}
In this section we give the explicit expressions of the matrices needed to transform from the basis of the $O^n$ to the total spin-isospin 
representation. Consider a generic operator $x$ written in the form
\begin{align}
\nonumber
x &= \sum_{n=1}^6x^n(r) O^n  \\ 
\nonumber
& = x_c+x_{\tau}\tt+x_{\s}\st+x_{\s\tau}\st\tt+x\rb{t}S_{12}+x\rb{t\tau}S_{12}\tt  \ .
\end{align}
In the representation appropriate to describe channels of fixed spin and isospin it reads
\be\label{repST}
x=\sum_{TS}\Big[x_{TS}+\delta_{S1}x_{tT}S_{12}\Big]P_{2S+1}\Pi_{2T+1} ,
\ee
with
\[
\left(
\begin{array}{cccc}
 1 & -3  & -3  &  9\\
 1 &  1 &  -3 &  -3\\
 1 & -3 &  1 & -3\\
 1 &  1 &  1 &1
\end{array}
\right)
\left(
\begin{array}{c}
 x_c\\
x_{\tau}\\
x_{\s}\\
 x_{\s\tau}
\end{array}
\right)=
\left(
\begin{array}{c}
 x_{00}\\
x_{10}\\
x_{01}\\
 x_{11}
\end{array}
\right)\quad,\quad
\left(
\begin{array}{cc}
 1 & -3 \\
 1 &  1
\end{array}
\right)
\left(
\begin{array}{c}
 x\rb{t}\\
x\rb{t\tau}
\end{array}
\right)=
\left(
\begin{array}{c}
 x\rb{t0}\\
 x\rb{t1}
\end{array}
\right) ,
\]
implying
\[
\left\{
\begin{array}{rl}
x_{TS}&=x_c+(4T-3)x_{\tau}+(4S-3)x_{\s}+(4S-3)(4T_3)x_{\s\tau}     \\
 x_{tT}&=x_t+(4T-3)x_{t\tau}        
\end{array}\right.  \ .
\]
Finally, the inverse transformation is 
\[
\frac{1}{16}\left(
\begin{array}{cccc}
 1 & 3  & 3  &  9\\
 -1 &  1 &  -3 &  3\\
 -1 & -3 &  1 & 3\\
 1 &  -1 &  -1 &1
\end{array}
\right)
\left(
\begin{array}{c}
 x_{00}\\
x_{10}\\
x_{01}\\
 x_{11}
\end{array}
\right)=
\left(
\begin{array}{c}
 x_c\\
x_{\tau}\\
x_{\s}\\
 x_{\s\tau}
\end{array}
\right)
\,,\,
\frac{1}{4}\left(
\begin{array}{cc}
 1 & 3 \\
 -1 &  1
\end{array}
\right)
\left(
\begin{array}{c}
 x\rb{t0}\\
 x\rb{t1}
\end{array}
\right)=\left(
\begin{array}{c}
 x\rb{t}\\
x\rb{t\tau}
\end{array}
\right) .
\]

%% file: appendix2.tex
The energy per particle at two-body cluster level can be written (see Eqs.(\ref{Delta2}) and (\ref{eq:veff}))
\be
(\Delta E)_2 = \sum_{i<j}\:\langle
ij|\:\frac{1}{2}\bigg[f_{12},\left[\;t_1+t_2,\:f_{12}\right]\bigg] +
f_{12}v_{12}f_{12}\:|ij-ji\rangle \ ,
\ee
with
\be
t_i = -\frac{1}{2m}\bm{\nabla}^2_i  \ \ , \ \ 
t_1 + t_2= -\frac{1}{m}\bm{\nabla}^2_{\phantom{R}} -
\frac{1}{4m}\bm{\nabla}^2_R \ , 
\ee
where ${\bm \nabla}$ acts on the relative coordinate ${\bf r}$, while ${\bm
  \nabla}_R$ acts on the center of mass coordinate ${\bf R}$, defined as
\be
{\bf r}={\bf r}_1-{\bf r}_2 \ \ , \ \ {\bf R}= \frac{1}{2}({\bf r}_1+{\bf r}_2) ,
\ee
 respectively.

Using the static part of the interaction, both the correlation function $f_{12}$ 
and the two-nucleon potential $v_{12}$ are written as
\be
f_{12}=\sum^6_{p=1}f^p(r_{12})O^p_{12} \ \ , \ \
v_{12}=\sum^6_{p=1}v^p(r_{12})O^p_{12} \ ,
\ee
with the six operator $O^{n}_{12}$ the properties of which properties are discussed in 
Appendix \ref{On}.

The Fermi gas (FG) two-nucleon state is given by
\bea
\vert ij \rangle & = & \frac{1}{V}\;{\rm e}^{i({\bf k}_i\cdot{\bf r}_1+{\bf
    k}_j\cdot{\bf r}_2)}\;\vert S\,M_S,\,T\,M_T \rangle \nonumber \\
 &  = &  \frac{1}{V}\;{\rm e}^{i({\bf k}\cdot{\bf r}+{\bf K}\cdot{\bf
     R})}\;\vert S\,M_S,\,T\,M_T \rangle \ ,
\label{tns}
\eea
with
\bea
\vert {\bf k}_i\vert,\;\vert{\bf k}_j\vert & \leq & p_F \nonumber \\
{\bf k} = \frac{1}{2}({\bf k}_i-{\bf k}_j)  \  & , &  \ {\bf K} = {\bf k}_i+{\bf
  k}_j \ .
\eea

We will discuss the potential and kinetic energy term separately.

\section{Potential energy}

Consider the operator
\be
w_{12}=f_{12}v_{12}f_{12} \ ,
\ee
and the decomposition of $f_{12}$ in the $TS$-representation (see
Eq.(\ref{repST}))
\be
f_{12}=\sum_{ST}\bigg[f_{ST}+\delta_{S1}f_{tT}S_{12}\bigg]
P^{\phantom{2}}_{2S+1}\Pi^{\phantom{2}}_{2T+1} \ .
\ee
In the above equation, $P^{\phantom{2}}_{2S+1}$ and
$\Pi^{\phantom{2}}_{2T+1}$ are spin and isospin projection operators,
whose properties are given in Appendix \ref{On}. Performing the
decomposition for $w_{12}$ and $v_{12}$ we find,
\bea
w_{12} & = & \sum_{TS}
\Bigg\{\delta^{\phantom{2}}_{S0}f^2_{T0}v^{\phantom{2}}_{T0}+ \delta_{S1}
\Big\{\;v^{\phantom{2}}_{T1}\bigg[\;f^2_{T1}+8f^2_{tT}+
2\left(f^{\phantom{2}}_{T1}f^{\phantom{2}}_{tT}-f^2_{tT}\right)S_{12}\bigg] +  \nonumber\\
 & + &
 \!\!v^{\phantom{2}}_{tT}\bigg[16\left(f^{\phantom{2}}_{T1}f^{\phantom{2}}_{tT}-f^2_{tT}\right)
+\left(f^2_{T1}-4f^{\phantom{2}}_{T1}f^{\phantom{2}}_{tT}+12f^2_{t1}\right)S_{12}\bigg]\Big\}\Bigg\}
P^{\phantom{2}}_{2S+1}\Pi^{\phantom{2}}_{2T+1} \ , \nonumber
\eea
and then we can identify
\bea
w^{\phantom{2}}_{T0} & = & \phantom{2}v^{\phantom{2}}_{T0}\phantom{\Big(}f^2_{T0}\phantom{\Big)} \nonumber \\
w^{\phantom{2}}_{T1} & = & \phantom{2}
v^{\phantom{2}}_{T1}\Big(f^2_{T1}+8f^2_{tT}\Big)+16v^{\phantom{2}}_{tT}
\Big(f^{\phantom{2}}_{T1}f^{\phantom{2}}_{tT}-f^2_{tT}\Big)  \\
w^{\phantom{2}}_{tT} & = & 2v^{\phantom{2}}_{T1}\Big(f^{\phantom{2}}_{T1}
f^{\phantom{2}}_{tT}-f^2_{tT}\Big)+v^{\phantom{2}}_{tT}\Big(f^2_{T1}
-4f^{\phantom{2}}_{T1}f^{\phantom{2}}_{tT}+12f^2_{t1}\Big)  \ . \nonumber
\eea
Now, after replacing 
\be
\sum_{i<j} \longrightarrow \frac{1}{2}\,\sum_{i j} \ ,
\ee
the potential energy contribution to $(\Delta E)_2$ reads
\bea
\langle w \rangle & = &
\frac{1}{2}\frac{1}{V^2}\sum_{SM_S}\sum_{TM_T}\sum_{{\bf k}_i{\bf
    k}_j}\sum_{S^\prime T^\prime}\Bigg\{\int \!d^3r_1d^3r_2 
 \bigg[w_{S^\prime T^\prime}(r)\langle P_{2S^\prime+1}\Pi_{2T^\prime+1}\rangle+ \nonumber \\
 & & \delta_{S^\prime 1}w_{tT^\prime}(r)\langle
 S_{12} P_{2S^\prime+1}\Pi_{2T^\prime+1}\rangle\bigg]- \int
 \!d^3r_1d^3r_2\,{\rm e}^{i({\bf k}_i\cdot{\bf r}-{\bf k}_j\cdot{\bf r})} \\
 & & \bigg[w_{S^\prime
   T^\prime}(r)\langle P_{2S^\prime+1}\Pi_{2T^\prime+1} P_{\sigma\tau}\rangle+\delta_{S^\prime1}w_{tT^\prime}(r)\langle S_{12}P_{2S^\prime+1}\Pi_{2T^\prime+1}P_{\sigma\tau}\rangle\bigg]\Bigg\} \ , \nonumber
\eea
where $P_{\sigma\tau}$ is the spin-isospin exchange operator defined in
Appendix \ref{On} and the expectation values $\< O \>$ are taken over
two-nucleon states of definite total spin and isospin $\vert S\, M_S,\, T\, M_T
\rangle$. Using
\be
\int d^3r_1d^3r_2  =  \int d^3r\,d^3R = V\int d^3r \ ,
\ee
the definition of the Slater function,
\be
\sum_{\vert{\bf k}\vert \leq p_F}{\rm e}^{i{\bf k}\cdot{\bf
    r}}=\frac{V}{(2\pi)^3}\int_{\vert {\bf k}\vert \leq p_F}d^3k\,e^{i{\bf
    k}\cdot{\bf r}}=\frac{N}{\nu}\ell(p_Fr) \ ,
\ee
and the results of Appendix \ref{On}, we finally obtain
\be
\label{pot:en}
\langle w \rangle=
\frac{1}{2}\frac{1}{V^2}\frac{N^2}{\nu^2}V\sum_{ST}\left(2S+1\right)\left(2T+1\right)\int
d^3r\,w_{ST}(r)\left[1-(-1)^{S+T}\ell^2(p_Fr)\right] \ ,
\ee
for example, in the case of symmetric nuclear matter ($\nu=4$),
\bea
\frac{1}{N}\langle w \rangle & = & \frac{\rho}{32}
\int d^3r \Big\{\big[w_{00}(r)+9w_{11}(r)\big]a_{-}(p_Fr)+ \nonumber \\
& & \phantom{\frac{\rho}{32}\int d^3r }\!\!+
\big[3w_{01}(r)+3w_{10}(r)\big]a_{+}(p_Fr)\big]\Big\} \ ,
\label{tbe:p}
\eea
where $\rho=N/V$ is the density and
\be
a_{\pm}(x)=1\pm\ell^2(x) \ .
\ee

\section{Kinetic energy}

Let us now discuss the kinetic contribution to the energy, given by
\be
\frac{1}{2}\bigg[f_{12},\left[\;t_1+t_2,\:f_{12}\right]\bigg]=-\frac{1}{2m}\bigg[f_{12},\left[\;\nabla^2,\:f_{12}\right]\bigg]
\ . \label{en:kin}
\ee

We consider spin-zero and spin-one channels separately.

\paragraph{Spin-zero channels}
In these channels, the relevant part of the correlation function is given by
\be
f_{12} = \sum_T f_{T0}(r)\;P_1\Pi_{2T+1} \ .
\ee

Making use of the results of Appendix \ref{On}, as well as the relation
\be
\Big[f_{T0},\,\nabla^2 f_{T0}\Big]=0 \ \ , \ \
 \Big[f_{T0},\,\big({\bm \nabla}f_{T0}\big){\bm \nabla}\Big]=-\big({\bm
   \nabla}f_{T0}\big)^2 \ ,
\ee
we find
\bea
\Big[f_{12},\left[\;\nabla^2,\:f_{12}\right]\Big] & = & \sum_{TT'}\Big[f_{T0}
\;P^{\phantom{2}}_1\Pi^{\phantom{2}}_{2T+1},\left[\nabla^2,\;f_{T0}\right]\;
P^{\phantom{2}}_1\Pi^{\phantom{2}}_{2T'+1}\Big] \nonumber \\
 & = & \sum_{TT'}\Big[f_{T0},
 \left[\nabla^2,\;f_{T0}\right]\Big]\;P^2_1\Pi^{\phantom{2}}_{2T+1}
\Pi^{\phantom{2}}_{2T'+1} \nonumber \\
 & = & \sum_T\Big[f_{T0},\, \big(\nabla^2 f_{T0}\big)+2\big({\bm
   \nabla}f_{T0}\big){\bm \nabla}\Big]\;P^{\phantom{2}}_1\Pi^{\phantom{2}}_{2T+1} \nonumber \\
 & = & 2\sum_T\Big[f_{T0},\,\big({\bm \nabla}f_{T0}\big){\bm \nabla}\Big]
\;P^{\phantom{2}}_1\Pi^{\phantom{2}}_{2T+1} \nonumber \\
 & = & -2\sum_T\big({\bm
   \nabla}f_{T0}\big)^2\;P^{\phantom{2}}_1\Pi^{\phantom{2}}_{2T+1} \ .
\eea

Finally,
\be
-\frac{1}{2m}\bigg[f_{12},\left[\;\nabla^2,\:f_{12}\right]\bigg] = \frac{1}{m}\sum_T\big({\bm
   \nabla}f_{T0}\big)^2\;P^{\phantom{2}}_1\Pi^{\phantom{2}}_{2T+1} \ .
\label{tbe:ks}
\ee

\paragraph{Spin-one channels}

In these channels, the correlation function is given by
\be
f_{12}=\sum_T\bigg[f_{T1}(r)+f_{tT}(r)S_{12}\bigg]P^{\phantom{2}}_3\Pi^{\phantom{2}}_{2T+1}
\ .
\ee
Relying once more on Appendix \ref{On}, we calculate
\[
\sum_{T'}\Big[\nabla^2,\,
\big(f_{T'1}+f_{tT'}S_{12}\big)P^{\phantom{2}}_3\Pi^{\phantom{2}}_{2T'+1}\Big]
  =  \sum_{T'}\Big\{\big[\nabla^2,\, f_{T'1}\big]+\big[\nabla^2,\,
f_{tT'}S_{12}\big]\Big\}
P^{\phantom{2}}_3\Pi^{\phantom{2}}_{2T'+1}
\]
\[
 =\sum_{T'}\Big\{\big(\nabla^2 f_{T'1}\big)+2\big({\bm \nabla} f_{tT'}\big){\bm
  \nabla}+\big(\nabla^2 f_{tT'}S_{12}\big)+2\big({\bm
  \nabla}f_{tT'}S_{12}\big){\bm
  \nabla}\Big\}P^{\phantom{2}}_3\Pi^{\phantom{2}}_{2T'+1}
\]
\[
  =  \sum_{T'} \Big\{ \big(\nabla^2 f_{T'1}\big) + 2\big({\bm \nabla}
f_{tT'}\big){\bm \nabla} + \big(\nabla^2 f_{tT'}\big)S_{12} +
\big(\nabla^2S_{12} \big)f_{tT'} 
\]
\be
+ 2\big({\bm \nabla}f_{tT'}\big)\big({\bm \nabla}S_{12}\big) +2S_{12}\big({\bm
   \nabla}f_{tT'}\big){\bm \nabla} +2f_{tT'}\big({\bm \nabla}S_{12}\big){\bm
   \nabla}\Big\} 
P^{\phantom{2}}_3\Pi^{\phantom{2}}_{2T'+1} \ .
\label{graffe}
\ee

Hence, the commutator in Eq.(\ref{en:kin}) can be rewritten as
\bea
\Big[f_{12},\left[\;\nabla^2,\:f_{12}\right]\Big] & = &
\sum_{TT'}\Big[\big(f_{T1}+f_{tT}S_{12}\big)
P^{\phantom{2}}_3\Pi^{\phantom{2}}_{2T+1},\left\{\ldots\right\}
P^{\phantom{2}}_3\Pi^{\phantom{2}}_{2T'+1} \Big] \nonumber \\
 & = & \sum_T\Big[f_{T1}+f_{tT}S_{12}, \left\{\ldots\right\}\Big]
P^{\phantom{2}}_3\Pi^{\phantom{2}}_{2T+1} \nonumber \\
 & = & \sum_T \Big( F^{(1)}_T + F^{(2)}_T \Big)
 P^{\phantom{2}}_3\Pi^{\phantom{2}}_{2T+1} \ ,
\eea
with
\be
F^{(1)}_T = \Big[f_{T1}, \left\{\ldots\right\}\Big] \ \ , \ \ 
F^{(2)}_T = \Big[f_{tT}S_{12}, \left\{\ldots\right\}\Big] \ ,
\ee
and
\bea
\Big\{ \ldots \Big\} & = & \Big\{ \big(\nabla^2 f_{T'1}\big) + 2\big({\bm \nabla}
f_{tT'}\big){\bm \nabla} + \big(\nabla^2 f_{tT'}\big)S_{12} +
\big(\nabla^2S_{12} \big)f_{tT'} \nonumber \\
 & + & 2\big({\bm \nabla}f_{tT'}\big)\big({\bm \nabla}S_{12}\big) +2S_{12}\big({\bm
   \nabla}f_{tT'}\big){\bm \nabla} +2f_{tT'}\big({\bm \nabla}S_{12}\big){\bm
   \nabla}\Big\} \ .
\eea
We find
\be
F^{(1)}_T = -2\big({\bm \nabla}f_{T1}\big)^2 -2\big({\bm \nabla}f_{T1}\big)
\big({\bm \nabla}f_{tT}\big)S_{12} \ ,
\ee
and
\bea
F^{(2)}_T & = & \Big[f_{tT}S_{12},\, 2\big({\bm \nabla}f_{T1}\big){\bm
  \nabla}\Big]  + \Big[ f_{tT}S_{12},\, 2S_{12}\big({\bm \nabla}f_{T1}\big){\bm \nabla}\Big] + \nonumber \\
&  & +\Big[ f_{tT}S_{12},\, 2f_{T1}\big({\bm \nabla}S_{12}\big){\bm \nabla}\Big] = \nonumber \\
 & = & -2\big({\bm \nabla}f_{T1}\big)\big({\bm
   \nabla}f_{tT}\big)S_{12}-2\big({\bm \nabla}f_{tT}\big)^2S^2_{12} + \nonumber \\
 & & +2f^2_{tT} \Big[S_{12}, \,\big({\bm \nabla}S_{12}\big){\bm \nabla}\Big]
 \nonumber \\
& = & -2\phantom{\Bigg[}\big({\bm \nabla}f_{T1}\big)\big({\bm
  \nabla}f_{tT}\big)S_{12}-2\big({\bm \nabla}f_{tT}\big)^2\big(8-2S_{12}\big)\phantom{\Bigg]}+ \nonumber \\
 & - &
 2f^2_{tT}\Bigg[\frac{36}{r^2}\;\big({\bf L}\cdot{\bf S}\big)+\frac{6}{r^2}\;\big(8-S_{12}\big)\Bigg] \ .
\eea

Collecting all pieces togheter, we find for the spin-one channels
\bea
-\frac{1}{2m}\bigg[f_{12},\left[\;\nabla^2,\:f_{12}\right]\bigg] & = & \frac{1}{m}\sum_T \Bigg\{
\big({\bm \nabla}f_{T1}\big)^2 +  \big({\bm \nabla}f_{T1}\big)\big({\bm
  \nabla}f_{tT} S_{12}\big)
 + \nonumber \\
 & & + \phantom{\Bigg[}\big({\bm \nabla}f_{tT}\big)^2\big(8-2S_{12}\big)\phantom{\Bigg]} +  f^2_{tT}\Bigg[\frac{36}{r^2}\;\big({\bf L}\cdot{\bf
   S}\big)+\frac{6}{r^2}\;\big(8-S_{12}\big)\Bigg]
\Bigg\}P^{\phantom{2}}_3\Pi^{\phantom{2}}_{2T+1} \nonumber \\
& = & \frac{1}{m}\sum_{TS}\Bigg\{\big( {\bm
  \nabla}f_{TS}\big)^2+\delta_{S1}\Big[2\big({\bm \nabla}f_{TS}\big)
\big({\bm \nabla}f_{tT}\big)S_{12}
+ \nonumber \\
 & & + \big({\bm \nabla}f^{\phantom{2}}_{tT}\big)^2
 S^2_{12}+f^{\phantom{2}}_{tT}\frac{36}{r^2}
\;\big({\bf L}\cdot{\bf S}\big)+\frac{6}{r^2}\;\big(8-S_{12}\big)\Big]\Bigg\}
P^{\phantom{2}}_{2S+1}\Pi^{\phantom{2}}_{2T+1} \nonumber \\
 & = & \sum_{TS} \Big\{t_{TS}(r) + \delta_{S1}\bigg[t_{tT}(r)S_{12} +
 t_{bT}(r)\big({\bf L}\cdot{\bf S}\big)\bigg]\Big\}
P^{\phantom{2}}_{2S+1}\Pi^{\phantom{2}}_{2T+1} \ ,
\eea
with
\bea
t_{T0} & = & \frac{1}{m} \big({\bm \nabla}f_{T0}\big)^2 \nonumber \\
t_{T1} & = & \frac{1}{m} \Big[\big({\bm \nabla}f_{T1}\big)^2 + 8\big({\bm
  \nabla}f_{tT}\big)^2 + \frac{48}{r^2}\;f^2_{tT} \Big] \nonumber \\
t_{tT} & = & \frac{1}{m} \Big[2\big({\bm \nabla}f_{T1}\big)\big({\bm \nabla}f_{tT}\big) 
- 2\big({\bm \nabla}f_{tT}\big)^2 -\frac{6}{r^2}\;f^2_{tT} \Big] \nonumber \\
t_{bT} & = & \frac{1}{m} \frac{36}{r^2}\;f^2_{tT} \ . \nonumber
\eea

\section{Explicit form of $\left(\Delta E\right)_2$}
We can rewrite
\be
\left(\Delta E\right)_2 = \sum_{i<j} \langle ij \vert W_{12} \vert ij-ji
\rangle \ ,
\ee
with
\[
W_{12}  =  -\frac{1}{m}\bigg[f_{12},\left[\;\nabla^2,\:f_{12}\right]\bigg] +
f_{12}v_{12}f_{12}
\]
\[
  =  \sum_{TS} \Big\{W_{TS}(r) + \delta_{S1}\bigg[W_{tT}(r)S_{12} +
 W_{bT}(r)\big({\bf L}\cdot{\bf S}\big) \bigg]\Big\}
P^{\phantom{2}}_{2S+1}\Pi^{\phantom{2}}_{2T+1} \ ,
\]
where
\bea
W_{T0} & = & \frac{1}{m}\big({\bm \nabla}f^{\phantom{2}}_{T0}\big)^2 + v_{T0}f^2_{T0} \nonumber \\
W_{T1} & = & \frac{1}{m}\Big[\big({\bm \nabla}f_{T1}\big)^2 + 8\big({\bm
  \nabla}f_{tT}\big)^2 
+ \frac{48}{r^2}\;f^2_{tT} \Big]+ \nonumber \\
& & + v^{\phantom{2}}_{T1}\Big(f^2_{T1}+8f^2_{tT}\Big)+16v^{\phantom{2}}_{tT}
\Big(f^{\phantom{2}}_{T1}f^{\phantom{2}}_{tT}-f^2_{tT}\Big) \nonumber \\
W_{tT} & = & \frac{1}{m}\Big[2\big({\bm \nabla}f_{T1}\big)\big({\bm
  \nabla}f_{tT}\big)
 -2\big({\bm \nabla}f_{tT}\big)^2 -\frac{6}{r^2}\;f^2_{tT} \Big]+ \nonumber \\
 & &
 +2v^{\phantom{2}}_{T1}\Big(f^{\phantom{2}}_{T1}f^{\phantom{2}}_{tT}-f^2_{tT}\Big)
+v^{\phantom{2}}_{tT}\Big(f^2_{T1}-4f^{\phantom{2}}_{T1}f^{\phantom{2}}_{tT}+12f^2_{t1}\Big) \nonumber \\
W_{bT} & = & \frac{1}{m}\;\frac{36}{r^2}\;f^2_{tT} \ . \nonumber 
\eea

Making use of the expression for the expectation values given in Appendix
\ref{On}, we finally obtain for symmetric matter (compare to Eq.(\ref{tbe:p}))
\bea
\frac{(\Delta E)_2}{N} & = & \frac{\rho}{32} \int\;d^3r
\Bigg\{\Big[W_{00}(r) + 9W_{11}(r)\Big]a_-(p_Fr)+ \nonumber \\
 & & +\Big[3W_{01}(r) + 3W_{10}(r)\Big]a_+(p_Fr)\Bigg\} \ .
\label{tbe:k}
\eea

%% file: appendix3.tex
\section{Spin singlet channels: uncoupled equations}
In the spin-zero channels, the energy per particle of SNM, evaluated at two-body
cluster level, reads (compare to Eqs.(\ref{tbe:p}) and (\ref{tbe:ks}))
\bea
\frac{(\Delta E)_2}{N} & = &
\frac{\rho}{32}\;\left(2T+1\right)\int\;d^3r\left[\big({\bm \nabla}f^{\phantom{2}}_{T0}\big)^2 + v_{T0}f^2_{T0}\right]a_{T0}(p_Fr) \nonumber \\
 & = & \frac{\rho}{32}\;\left(2T+1\right)\;4\pi\int\;r^2dr\left[\big(f^\prime_{T0}\big)^2 + v_{T0}f^2_{T0}\right]a_{T0}(p_Fr) \nonumber \\
 & = &
 \textrm{const}\;\int^{\infty}_{0}dr\;F\left[f^{\phantom{^\prime}}_{T0},\;f^\prime_{T0}\right] \ , 
\eea
where $a_{TS}(x)=1-(-)^{T+S} \ell^2(x)$ and
\be
F\left[f^{\phantom{^\prime}}_{T0},\;f^\prime_{T0}\right]  =  \Bigg[\big(f^\prime_{T0}\big)^2 + 
 \;v_{T0}f^2_{T0}\Bigg] \phi^2_{T0} \ ,
\ee
with
\be
\phi_{T0}  =  r\sqrt{a_{T0}} \ .
\label{eq:phi}
\ee
The corresponding Euler-Lagrange (EL) equations for the unknown functions $f_{T0}$
are given by
\be
\frac{d}{dr}\frac{\partial F}{\partial f^\prime_{T0}}-\frac{\partial F}{\partial
  f^{\phantom{^\prime}}_{T0}}=0 \ .
\ee

From
\bea
\frac{\partial F}{\partial f^{\phantom{^\prime}}_{T0}} & = &
2\;\;v_{T0}\;f^2_{T0}\;\phi^2_{T0} \ , \nonumber \\
\frac{\partial F}{\partial f^\prime_{T0}} & = & 2\;f^\prime_{T0}\;\phi^2_{T0} \ , \nonumber \\
\frac{d}{dr}\frac{\partial F}{\partial f^\prime_{T0}} & = &
2\;f^{\prime\prime}_{T0}\;\phi^2_{T0} + 4\;f^\prime_{T0}\;\phi^\prime_{T0}\;\phi_{T0} \ ,
\eea
we obtain
\be
f^{\prime\prime}_{T0}\;\phi^2_{T0} + 2\;f^\prime_{T0}\;\phi^\prime_{T0}
-\;v_{T0}\;f^2_{T0}\;\phi^2_{T0}=0 \ .
\label{EL0:f}
\ee

Introducing
\be
g_{T0} \equiv f_{T0}\;\phi_{T0} \ ,
\ee
we can put Eq.(\ref{EL0:f}) in the form
\be
g^{\prime\prime}_{T0} - \Bigg(\frac{\phi^{\prime\prime}_{T0}}{\phi_{T0}} + 
\;v_{T0}\Bigg)\;g_{T0}=0 \ .
\ee

Now we introduce a Lagrange multiplier, in order to fulfill the requirement
\be
\left. g^\prime_{T0} \right|_{r=d}   = \left.  \phi^\prime_{T0} \right|_{r=d} \ .
\ee
The resulting equation is Eq.(4) of \cite{BCFR}
\be
g^{\prime\prime}_{T0} - \Bigg(\frac{\phi^{\prime\prime}_{T0}}{\phi_{T0}} +
\;\left(v_{T0}+\lambda\right)\Bigg)\;g_{T0}=0 \ ,
\ee
to be integrated with the boundary conditions
\bea
\left. g_{T0} \right|_{r=0} & = & 0 \ \ , \\ 
\left. g_{T0} \right|_{r=d} & = & \left. \phi_{T0} \right|_{r=d} \ . 
\eea

\section{Spin triplet channels: coupled equations}

In the spin-one channels, the contribution to the energy is given by (see
Eqs.(\ref{tbe:p}) and (\ref{tbe:k}))
\bea
\frac{(\Delta E)_2}{N} & = & \frac{\rho}{32}\;\left(2T+1\right)
\int\;d^3r\Bigg\{\left[\big({\bm \nabla} f^{\phantom{2}}_{T1}\big)^2 
+8\big({\bm \nabla} f^{\phantom{2}}_{tT}\big)^2 + \frac{48}{r^2}\;f^2_{tT}\right] + \nonumber \\
 & & + v_{T1}\bigg(f^2_{T1}+8f^2_{tT}\bigg)
 +16v_{tT}\bigg(f^{\phantom{2}}_{T1}f^{\phantom{2}}_{tT}
-f^2_{tT}\bigg)\Bigg\}\;a_{T1}(p_Fr) \nonumber \\
 & = & \textrm{const}\;\int^{\infty}_{0}dr
\;F\left[f^{\phantom{^\prime}}_{T1},\;f^{\phantom{^\prime}}_{tT};\;f^\prime_{T1},\;f^\prime_{tT}\right] \ ,
\eea
where
\bea
F\left[f^{\phantom{^\prime}}_{T1},\;f^{\phantom{^\prime}}_{tT};\;f^\prime_{T1},\;f^\prime_{tT}\right] 
& = &
\left(f^\prime_{T1}\right)^2\phi^2_{T1}+8\left(f^\prime_{tT}\right)^2\phi^2_{T1}+\frac{48}{r^2}
\;f^2_{tT}\phi^2_{T1}+ \ \ \ \ \ \ \ \ \ \ \ \ \ \ \ \ \ \nonumber \\
 & + & \Big[v_{T1}\left(f^2_{T1}+8f^2_{tT}\right) 
+16v_{tT}\left(f^{\phantom{2}}_{T1}f^{\phantom{2}}_{tT}-f^2_{tT}\right)\Big] \ .
\eea

In this case we have two coupled EL equations
\be
\left\{
\begin{array}{l}
\frac{d}{dr}\frac{\partial F}{\partial f^\prime_{T1}}-\frac{\partial F}{\partial
  f^{\phantom{^\prime}}_{T1}} 
 =  0 \nonumber \\
\ \ \ \ \ \\
\frac{d}{dr}\frac{\partial F}{\partial f^\prime_{tT}}-\frac{\partial F}{\partial
  f^{\phantom{^\prime}}_{tT}}  =  0 \ .
\end{array}
\right.
\ee

Carrying out the derivativees as in the spin-zero channels and defining
\be
g_{T1} \equiv f_{T1}\phi_{T1} \ \ , \ \  g_{tT} \equiv \sqrt{8}f_{tT}\phi_{T1}
\ ,
\ee
we find
\be
\left\{
\begin{array}{l}
g^{\prime\prime}_{T1} - \left(\frac{\phi^{\prime\prime}_{T1}}{\phi_{T1}} +
\;v_{T1}\right)\;g_{T1}-\sqrt{8}v_{tT}g_{tT}
 =  0  \\
\ \ \ \ \ \\
g^{\prime\prime}_{tT} - \left[\frac{\phi^{\prime\prime}_{T1}}{\phi_{T1}} + \;\left(v_{T1}-
2v_{tT}\right)+\frac{6}{r^2}\right]\;g_{tT} - \sqrt{8}v_{tT}g_{T1} = 0
\ .
\end{array}
\right.
\ee

Finally, inclusion of the Lagrange multipliers needed to guarantee
\bea
\left. g^\prime_{T1}\right|_{r=d_1} & = & \left.\phi^\prime_{T1}\right|_{r=d_1} \ , \\
\left. g^\prime_{tT}\right|_{r=d_2} & = & \left.\phi^\prime_{T1}\right|_{r=d_2} \ ,
\eea
with, in general, $d_1\neq d_2$, leads to (compare to Eq.(5) of \cite{BCFR})
\be
\left\{
\begin{array}{l}
g^{\prime\prime}_{T1} - \left[ \frac{\phi^{\prime\prime}_{T1}}{\phi_{T1}} +
\;\left(v_{T1}+\lambda_1\right) \right] 
\;g_{T1}-\left(\sqrt{8}v_{tT}+\lambda_2\right)g_{tT}  =  0 \\ 
 \ \ \ \\ 
g^{\prime\prime}_{tT} - \left[\frac{\phi^{\prime\prime}_{T1}}{\phi_{T1}} + \;\left(v_{T1}-
2v_{tT}+\lambda_1\right)+\frac{6}{r^2}\right]\;g_{tT}
- \left(\sqrt{8}v_{tT}+\lambda_2\right)g_{T1} = 0 \ , 
\end{array}
\right.
\ee
with the boundary conditions
\bea
\left. g_{T1}\right|_{r=0}   & = &  0  \ , \\
\left. g_{T1}\right|_{r=d_1} & = &  \left. \phi_{T1}\right|_{r=d_1}  \ ,
\eea
and
\bea
\left. g_{tT}\right|_{r=0}    & = & 0 \ , \\
\left. g_{tT}\right|_{r=d_2}  & = & 0 \ .
\eea

%% file: App_ex.tex
In \cite{FT} the Fourier transform (FT) of all the eighteen operators of Argonne potential are performed exploiting the equivalence between 
${\bf q}$ and $-\iu\bm{\nabla}_{\br}$. Here we use a different approach,  in which the FT is obtained carrying out the integrations through the decomposition over the orbital wave functions. 

We will start from the $S_{12}(\hat{\br})$ operator as it appears in the expression of the energy at two-body cluster level
\bea
\delta\mathcal{E}=\ldots+&&\!\!\!\!\!\!\!\!\!\!\!\!\frac{1}{2V}\sum_{\bk_i,\bk_j}\sum_{i,j}\delta n_{i}(\bk_i)\delta n_j(\bk_j)\Bigg\{\nonumber\\ 
+\ldots+&\!\!\!\!\!\! \Big[&\!\!\!\!\!\!\underbrace{\int d^{3}\br\Big(w^5(r)+w^6(r)\Big)(1-\eu\rp{-\iu({\bf q \cdot r})})(3(\bm{\s}_1\cdot\hat{\br})\,(\bm{\s}_2\cdot\hat{\br})-\bm{\s}_1\cdot\bm{\s}_2)}_{\mbox{\textcircled{c}}} \Big]\Bigg\}  \ , \nonumber
\eea
and show that 
\[
\textcircled{c}=\Big[\underbrace{-\frac{4\pi}{5}\int dr r^2 \Big(w_5(r)+w_6(r)\Big)g_2(r,q)}_{\mbox{$h(q)$}}\Big](3(\bm{\s}_1\cdot\hat{\bq})(\bm{\s}_2\cdot\hat{\bq})-\bm{\s}_1\cdot\bm{\s}_2) .
\]
\\
\subsubsection{Fourier transform of $S_{12}(\hat{{\bf r}})$}
The term in $\textcircled{c}$ can be written in the form:
\be
\propto\left[\tilde{V}_{12}(0)-\tilde{V}_{12}(\bk_i-\bk_k)\right] ,
\ee
where $\bq=\bk_i-\bk_j$ and:
\bea
&\tilde{V}\!\!\!&\!\!\!_{12}(0)=\int d^3\br \Big(\ldots\Big)(3(\bm{\s}_1\cdot\hat{\br})\,(\bm{\s}_2\cdot\hat{\br})-\bm{\s}_1\cdot\bm{\s}_2)=0\nn\\
&\tilde{V}\!\!\!&\!\!\!_{12}(\bq)=\int d^3\br \Big(\ldots\Big)e^{-i\bq\cdot\br}(3(\bm{\s}_1\cdot\hat{\br})\,(\bm{\s}_2\cdot\hat{\br})-\bm{\s}_1\cdot\bm{\s}_2) .\nn
\eea
The quantity in the first line vanishes, since the round brackets contain a function of $|{\bf r}|$ only. 
The one surviving term is the exchange contribution,  which is proportional to the FT of the tensor operator. The integration is performed 
fixing the $z$-axis and denoting
$x=\cos\th=\hat{z}\cdot \hat{r}$, $x'=\cos\th'=\hat{z}\cdot \hat{q}$ and $\hat{q}\cdot\hat{r}=\cos\g=y$ with $q=|\bq|$. 

We expand  the exponential in Legendre polynomials according to
\be
\eu\rp{-\iu q r y}=\sum_\ell g_\ell P_\ell(y)\quad g_\ell(r,q)=\frac{2\ell+1}{2}\int_{-1}^{1}\eu\rp{-\iu q r y}P_\ell(y)dy ,
\ee
and use the standard decomposition of a scalar product ${\bf v}\cdot \hat{{\bf r}}$ in spherical harmonics, as well as the addition theorem for Legendre polynomials
\bea
&&\bm{\s}_1\cdot \hat{\br}=\sum_{\n=-1}^1(-)^{\n}\sigma^{-\n}\sqrt{\frac{4\pi}{3}}Y_{1\n}(x,\phi)\nn\\
&&P_\ell(y)=\frac{4\pi}{2\ell+1}\sum_{p=-\ell}^\ell Y^*_{\ell p}(x,\phi)Y_{\ell p}(x',\phi ') ,\nn
\eea
where $\sigma^{-\n}$ denotes the spin rising and the lowering operator:
\[
\sigma^{+}=-\sqrt{2}\left(\begin{array}{cc}
0 & 1\\
0 & 0
\end{array}\right)\quad
\sigma^{-}=\sqrt{2}\left(\begin{array}{cc}
0 & 0\\
1 & 0
\end{array}\right)\quad
\sigma^{0}=\left(\begin{array}{cc}
1 & 0\\
0 & -1
\end{array}\right).
\]
It follows that
\begin{align}
\tilde{V}_{12}=&\int  \rmd r \,r^2 w_{5}(r)\sum_l g_\ell(r,q) \int \rmd \O P_\ell(y)\times\nonumber\\
&\times\Big[\overbrace{(4 \pi)\sum_{\n \n'}(-)^{\n+\n'}\s_1^{-\n}\s_2^{-\n'}Y_{1\nu}(x,\phi)Y_{1\nu'}(x,\phi)}^{\mbox{\textcircled{a}}}-\underbrace{\,\bm{\s}_1\cdot\bm{\s}_2\,}_{\mbox{\textcircled{b}}}\Big] .\nonumber
\end{align}
Using the above property of $P_\ell(y)$, the first term can be rewritten
\begin{align}
\textcircled{a}=&\int  \rmd r \,r^2 w_{5}(r)\sum_l \frac{g_\ell(r,q)}{2l+1}(4\pi)^2\sum_{\n \n'}(-)^{\n+\n'}\s_1^{-\n}\s_2^{-\n'}\times\nonumber\\
 &\times\overbrace{\left[\int \rmd \O Y_{1\nu}(x,\phi)Y_{1\nu'}(x,\phi)Y^*_{l p}(x,\phi)\right]}^{\mbox{$(I)$}}Y_{l p}(x',\phi') ,\nonumber
\end{align}
and, substituting
\[
Y^*_{l p}(x,\phi)=(-)^pY_{l-p}(x,\phi) ,
\]
we finally obtain the expression
\bea
(I)&=&\sum_p(-)^p\left[ \frac{9(2 l+1)}{4\pi}\right]^{1/2}\left(\begin{array}{ccc}
1 & 1 & l\\
0 & 0 & 0
\end{array}\right)\left(\begin{array}{ccc}
1 & 1 & l\\
\n & \n' & -p
\end{array}\right)\nn\\
&=& \sum_p\left[ \frac{9}{4\pi(2 l+1)}\right]^{1/2}\underbrace{\ll\,1\, 1\, 0\, 0\,|\, l\, 0\,\rr}_{\mbox{$\neq0$ for $l=0,2$}}\ll\,1\, 1\, \nu\, \nu'\,|\, l\, p\,\rr  \ , \nn
\eea 
where the amplitude $\ll l_1 l_2 m_1 m_2| l m \rr$ is the usual Clebsch-Gordan coefficient describing the addition of $\ell_1, m_1$ and $\ell_2, m_2$ 
in the $\ell,m$ channel. Note that the first coefficient $\ll1 1 0 0| \ell 0\rr$ is different from zero only for $\ell=0,2$.
\subsubsection*{$\ell=0$ term}
For $\ell=0$ the contribution denoted \textcircled{a} turns out to be
\bea
\textcircled{a}_{\ell=0} &=&\int  \rmd r \,r^2 \Big(\ldots\Big)g_0(r,q)(4\pi)^{3/2}\sum_{\n \n'}(-)^{\n+\n'}\s_1^{-\n}\s_2^{-\n'} 3\underbrace{\ll1 1 0 0| \ell 0\rr}_{\mbox{$-\frac{1}{\sqrt{3}}$}}\underbrace{\ll\,1\, 1\, \nu\, \nu'\,|\, \ell \, p\,\rr}_{\mbox{$(-)^{1+\n}\frac{1}{\sqrt{3}}\delta_{\n-\n'}$}} Y_{0 0}(x',\phi')\nn\\
&=&\int  \rmd r \,r^2 w_{5}(r) g_0(r,q)(4\pi)^{3/2}\underbrace{\left[\sum_{\n}(-)^{\n}\s_1^{-\n}\s_2^{\n}\right]}_{\mbox{$\bm{\s}_1\cdot\bm{\s}_2$}}\underbrace{Y_{0 0}(x',\phi')}_{\mbox{$\frac{1}{\sqrt{4\pi}}$}}\nn\\
&=&4\pi(\bm{\s}_1\cdot\bm{\s}_2)\int  \rmd r \,r^2 w_{5}(r) g_0(r,q)\nn\\
&=&\textcircled{b} .\nn
\eea
As a consequence, $\textcircled{a}_{\ell=0}-\textcircled{b}=0$ in $\tilde{V}_{12}$ and
\[
\tilde{V}_{12}=\textcircled{a}_{\ell=2}.
\]
\subsubsection*{$\ell=2$ term}
In this case 
\[
\tilde{V}_{12}=\int  \rmd r \,r^2 \Big(\ldots\Big) g_2(r,q)\left(\frac{4\pi}{5}\right)^{3/2}\bigg[ 3 \sum_{\n \n' p}(-)^{\n+\n'}\s_1^{-\n}\s_2^{-\n'}\underbrace{\ll1 1 0 0| 2 0\rr}_{\mbox{$\sqrt{\frac{2}{3}}$}}\ll\,1\, 1\, \nu\, \nu'\,|\, \ell \, p\,\rr\bigg] Y_{2 p}(x',\phi') ,
\]
or
\[
\tilde{V}_{12}=\int  \rmd r \,r^2 \Big(\ldots\Big) g_2(r,q)\left(\frac{4\pi}{5}\right)^{3/2}\sqrt{6}\bigg[  \sum_{\n \n' p}(-)^{\n+\n'}\s_1^{-\n}\s_2^{-\n'}\ll\,1\, 1\, \nu\, \nu'\,|\, 2\, p\,\rr\bigg] Y_{2 p}(x',\phi') .
\]
In order to simplify the above result, we start from the definition 
\[
(\bm{\s}_1\cdot\hat{\bq})(\bm{\s}_2\cdot\hat{\bq})=\sum_{\n\n' }(-)^{\n+\n'}\s_1^{-\n}\s_2^{-\n'}\frac{4\pi}{3}Y_{1\n}Y_{1\n'} ,
\]
and use
\[
Y_{1\n}Y_{1\n'}=\sum_{\ell p}\frac{3}{\sqrt{4\pi(2\ell+1)}}\ll1 1 0 0| \ell 0\rr \ll\,1\, 1\, \nu\, \nu'\,|\, \ell \, p\,\rr Y_{2 p}  \ .
\]
In order to eliminate the $\ell=0$ contribution we add
\[
\bm{\s}_1\cdot\bm{\s}_2 ,
\]
to obtain the final result
\[
3(\bm{\s}_1\cdot\hat{\bq})(\bm{\s}_2\cdot\hat{\bq})-\bm{\s}_1\cdot\bm{\s}_2=\sqrt{\frac{6(4 \pi)}{5}} \sum_{\n\n' p}(-)^{\n+\n'}\s_1^{-\n}\s_2^{-\n'}\ll\,1\, 1\, \nu\, \nu'\,|\, 2\, p\,\rr Y_{2 p} ,
\]
that has been employed in the calculation of the energy from 
\bea
\delta\mathcal{E}&&\!\!\!\!\!\!=\ldots+\frac{1}{2V}\sum_{\bk,\bk'}\sum_{i,j}\delta n_{i}(\bk)\delta n_j(\bk')\Bigg\{+\ldots+\nonumber\\ 
-&&\!\!\!\!\!\!S_{12}(\hat{\bf q}) \frac{4\pi}{5}\int dr r^2 g_2(r,q)\Big(w_5(r)+w_6(r)\Big)\Bigg\} .\nonumber
\eea

%% file: Main_thesis.bbl
\begin{thebibliography}{200}
\addcontentsline{toc}{chapter}{Bibliography}
\bibitem{Cowan} F. Reines and C. L. Cowan Jr., Nature {\bf 178}, 446 (1956)
\bibitem{Bigongiari} C. Bigongiari (Antares Collaboration), Journal of Physics: Conferences Series {\bf 173},012024 (2009)
\bibitem{ReddyLong} M. Prakash et al., Lect. Notes Phys. {\bf 578}, 364-423 (2001)
\bibitem{Pet} C.J. Pethick, Rev. Mod. Phys. {\bf 64} (1992) 1133.
\bibitem{Hirata} K. S. Hirata  et al., Phys. Rev. Lett. {\bf 58} 1490-93 (1987). 
\bibitem{Bionta} R. M. Bionta et al., Phys. Rev. Lett. {\bf 58} 1494-96 (1987).
\bibitem{laganke}
K.H. Langanke,  Lectures delivered at the Fifth Course of the Nuclear Physics School "Raimondo Anni" (Otranto, Italy, June 2011).
\bibitem{B1} A. Burrows, E. Livne, L. Dessart, C. D. Ott, and J. Murphy,  ApJ {\bf 640}  (2006) 878.
\bibitem{B2} 
M. Ruffert and H. Th. Janka, A\&A  {\bf 380} 2 (2001) 544.
\bibitem{Otsuka} T. Otsuka et al., Phys. Rev. Lett. {\bf 105}, 032501 (2010)
\bibitem{Ish} S. Aoki et al, arXiv:1206.5088v1[hep-lat]
\bibitem{Ish1} S. Aoki et al, Prog. Theor. Phys. Vol {\bf 123} No. 1 (2010),
\bibitem{Ecker} G. Ecker, Prog. Part. Nucl. Phys. {\bf 35}, 1-80 (1995)
\bibitem{WeinBook1} S. Weinberg, \emph{The Quantum theory of field I-II} ,(1995-6)
\bibitem{CCWZ} S. Coleman et al, Phys. Rev. Vol. {\bf 177} N. 5, 2239 (1969); G. Callan et al, Phys. Rev.  {\bf 177} N. 5, 2247 (1969)
\bibitem{Nij} V.G.J. Stoks et al., Phys. Rev. C {\bf 48}, 792 (1993)
\bibitem{GolTre} M.L. Goldberger and S.B. Treiman, Phys. Rev. {\bf 111}, 354 (1958) 
\bibitem{Argonne} R.B. Wiringa  et al., Phys. Rev. C {\bf 51}, 38 (1995)
\bibitem{Andrea} A. Cipollone et al, in preparation
\bibitem{Li} Li, Lombardo et al., Phys. Rev. C {\bf 74}, 047304 (2006)
\bibitem{FujMiy} J. Fujita and H. Miyazawa, Prog. Theor. Phys. {\bf 17}, 360 (1957)
\bibitem{He} B.S. Pudliner et al, Phys. Rev. Lett. {\bf 74}, 4396 (1995) 
\bibitem{Nav}P. Navra\'til et al, Phys. Rev. Lett. {\bf 84}, 5728 (2000); Phys. Rev. C {\bf 62}, 054311 (2000)
\bibitem{SRG} S. K. Bogner et al, Phys. Rev. C {\bf 75}, 061001 (2007)
\bibitem{Baldo} M. Baldo, "Nuclear Methods and Nuclear Equation of State", Ed. M. Baldo (World ScientiÞc, 
Singapore, 1999)
\bibitem{Dick} W.H. Dickhoff and C. Barbieri, Prog. Part. Nuc. Phys. {\bf 52}, 377Ð496  (2004)
\bibitem{Martino} K. Kowaiski et al, Phys. Rev. Lett. {\ref 92}, 132501 (2004)
\bibitem{Bisconti} C. Bisconti et al., Phys. Rew. C {\bf 75}, 054302 (2007)
\bibitem{Carlson} J. Carlson et al., Phys. Rev. C {\bf 68}, 025802 (2003)
\bibitem{Gandolfi} S. Gandolfi et al, Phys. Rev. Lett. {\bf 98}, 102503 (2007)
\bibitem{feenberg} E. Feenberg, \emph{ Theory of quantum fluids} (Academic Press, New York, 1967).  
\bibitem{clark79} J.W. Clark, Prog. Part. Nucl. Phys. {\bf 2} (1979) 89.
\bibitem{Fantoni1} S. Fantoni and S. Rosati, Nuovo Cimento A {\bf 25}, 593 (1975) 
\bibitem{Lovato} A. Lovato, Ph.D. Thesis, SISSA (2012)
\bibitem{Pand} S. Cowell and V.R. Pandharipande, Phys. Rev. C {\bf 70} (2004) 035801.
\bibitem{Ben} O. Benhar and M. Valli, Phys. Rev. Lett. {\bf 99} (2007) 232501.
\bibitem{Lag} I. Lagaris and V.R. Pandharipande, Nucl. Phys. A {\bf 359}, 349 (1981)
\bibitem{Ak} A. Akmal and V.R. Pandharipande, Phys. Rev. C {\bf 56}, 2261 (1997) 
\bibitem{Valli} M. Valli, PhD Thesis, University of Rome "La Sapienza" 
\bibitem{Fantoni} A. Sarsa et al. ,Phys. Rev.C {\bf 68} (2003) 024308
\bibitem{Ak1} A. Akmal et al. , Phys. Rev. C {\bf 58}, 1804 (1998), and reference there in.
\bibitem{Abr} A. A. Abrikosov et al, \emph{Methods of Quantum Field Theory in Statistical Physics}, Dover Book.
\bibitem{Landau} L. D. Landau, Lifsits, \emph{Statistical physics}, vol 9.
\bibitem{Olss1} E. Olsson et al, Phys. Rev. C {\bf 70}, 025804 (2004)
\bibitem{Arima}  A. Arima et al, Adv. Nucl. Phys. {\bf 18}, 1 (1987)
\bibitem{Haensel} P. Haensel, Phys. Lett. B {\bf 62}, 268 (1976); P. Hansel, Nucl. Phys. A 298, 139 (1978)
\bibitem{FT} S. Veersamy and W. N. Polyzou, arXiv:1106.1934v1/nucl-th
\bibitem{Farina}  O. Benhar ant N. Farina, Phys. Lett. B {\bf 680}, 305-309 (2009)
\bibitem{Bay} G. Baym and C. Pethick, \emph{Landau Fermi-liquid theory}, WILEY-VCH
\bibitem{Leg} A. J. Leggett, Ann. Phys. {\bf 46},76-113 (1968)
\bibitem{par1} S. O. Backman, et al. 43 B, {\bf 263} (1973); S. O. Backman et al, Nucl. Phys. A {\bf 321}, 10-24 (1979) 
\bibitem{par2} C. J. Pethick and D. G. Ravenhall, Ann. phys. {\bf 183}, 131-165 (1988) and reference there in: Friedman and Pandharipande, Nucl. Phys. A {\bf 361}, 502 (1981)
\bibitem{par3} J. Nitsch, Z. Physik. {\bf 251}, 141-151 (1972)
\bibitem{friman} B. L. Friman and A. K. Dhar, Phys. Lett. B {\bf 85}, 1 (1979)
\bibitem{Haensel1} P. Haensel and A. J. Jerzak, Phys. Lett. B {\bf 112}, 285 (1982) 
\bibitem{Mer} N. D. Mermin, Phys. Rev. {\bf 159}, 161 (1967)
\bibitem{Liu} K. F. Liu, Nuovo Cimento A {\bf 70} N.4, 329 (1982)
\bibitem{Gogny} D. Gogny and R. Padjen, Nucl. Phys. A {\bf 293}, 365-378 (1977)
\bibitem{Noz} P. Nozieres, \emph{Theory of interacting Fermi system}
\bibitem{Dic} W. H. Dickhoff and D. Van Neck, \emph{Many-Body Theory Exposed!}
\bibitem{Iwamoto} N. Iwamoto et al., Phys. Rev. D {\bf 25}, (1982) 313
\bibitem{Safier} E. Safier and A. Widom, Journal of Low Temperature Physics Vol {\bf 6} No 3/4, 397 (1972)
\bibitem{Reddy4} G. Shen, et al, arXiv:1205.6499 [nucl-th]
\bibitem{Reddy3} S. Reddy, \emph{The Micro-Physics of neutrino transport at extreme density}, Compact Stars: pp. 495-509
\bibitem{Reddy2} A. Burrows et al., Nucl. Phys. A {\bf 777},  356-394 (2006)
\bibitem{Sawyer} M. Prakash et al., Ann. Rev. Nucl. Part. Sci. {\bf 51}, 295-344 (2001)

\bibitem{Freedman} D. Z. Freedman, Phys Rev D {\bf 9} 1389 (1974)
\bibitem{Thompson} T. A. Thompson et al, Phys Rev C {\bf 62} 035802 (2000)
\bibitem{Ravenhall} D. G. Ravenhall et al, Phys Rev Lett {\bf 50} 2066 (1983)
\bibitem{Watanabe} G. Watanabe et al, Phys. Rev. C {\bf 68}, 035806 (2003)
\bibitem{BS} A. Burrows and R.F. Sawyer, Phys. Rev. C {\bf 58} (1998) 554
\bibitem{Mandl} F. Mandl and G. Shaw, \emph{Quantum Field Theory} , (Mcgraw-Hill, New York,1980)
\bibitem{BCFR} O. Benhar et al, Phys. Lett. B {\bf 70}, 1 (1977) 
\bibitem{Reddy1} S. Reddy et al., Phys. Rev. D. {\bf 58}, 013009 (1998)

\bibitem{Soma} V. Som\'a \& P. Bozek , Phys. Rev. C {\bf 78}, 054003 (2008)
\bibitem{Hagen} G. Hagen et al, Phys. Rev. Lett. {\bf 108}, 242501 (2012)
\bibitem{Bogner} Bogner et al, Phys. Rept. {\bf 386}, 1 (2003)
\bibitem{Roth} R. Roth et al, Phys. Rev. Lett. {\bf 107},072501 (2011)
\bibitem{Bar2}S. J. Waldecker et al, arXiv: nucl-th/1105.4257v1
\bibitem{Bar3} C. Barbieri et al, Phys. Rev. A {\bf 76}, 052503 (2007)
\bibitem{Arianna} A. Carbone et al, in preparation
\bibitem{Barbieri} W.H. Dickhoff \& C.Barbieri, Prog. Part. Nucl. Phys. {\bf 52}, 377-496 (2004)

\bibitem{B1} C. D. Ott et al, J. Phys. {\it Conference Series} vol  {\bf 180} issue 1, 012022 (2009) or arXiv: 0907.4043
\bibitem{B2} A. Burrows et al, ApJ {\bf 640} 878 (2006)
\bibitem{Janka} R. Buras et al, arxiv: astro-ph/0303171
\bibitem{Lom} C. Shen et al, Phys. Rev. C {\bf 68}, 055802 (2003)
\bibitem{Marg} J. Margueron et al, Nucl. Pys. A {\bf 719}, 169c (2003)

\end{thebibliography}
